\documentclass[a4paper]{article}
\pdfoutput=1
\usepackage{jheppub}

\usepackage{amsmath}
\usepackage{amssymb}
\usepackage{graphicx,color,xcolor}
\usepackage{ulem}
\usepackage{enumerate}

\usepackage{amsmath,amsthm}
\usepackage{txfonts}

\usepackage{mathrsfs}

\usepackage{bm}

\newcommand{\Ref}[1]{(\ref{#1})}

\newcommand{\Z}{\mathbb{Z}}
\newcommand{\R}{\mathbb{R}}
\newcommand{\C}{\mathbb{C}}

\newcommand{\D}{\mathrm{d}}
\newcommand{\E}{\mathrm{e}}
\newcommand{\I}{{\rm i}}
\newcommand{\DD}{{\mathcal D}}
\newcommand{\ab}{{ab}}

\newcommand{\A}{{\mathcal{A}}}
\newcommand{\F}{{\mathcal{F}}}

\newcommand{\bA}{{\bar{A}}}
\newcommand{\bF}{{\bar{F}}}

\newcommand{\Tr}{\text{Tr}}

\newcommand{\Y}{{{Y}_\gamma}}

\newcommand{\SLDC}{{\mathrm{SL}(2,\mathbb{C})}}
\newcommand{\sldc}{{\frak{sl}_2\mathbb{C}}}

\newcommand{\Slc}{\mathrm{SL}(2,\mathbb{C})}
\newcommand{\slc}{\fs\fl_2\mathbb{C}}

\newcommand{\Su}{\mathrm{SU}(2)}
\newcommand{\su}{\fs\fu_2}

\newcommand{\OO}{{\mathbf{O}}}
\newcommand{\oo}{{\mathbf{o}}}

\def\be{\begin{eqnarray}}
\def\ee{\end{eqnarray}}

\newcommand{\cd}{\mathcal D}

\newcommand{\cf}{\mathcal F}

\newcommand{\ch}{\mathcal H}

\newcommand{\cj}{\mathcal J}

\newcommand{\cm}{\mathcal M}

\newcommand{\fl}{\mathfrak{l}}  
  \newcommand{\Fm}{\mathfrak{M}}
\newcommand{\fn}{\mathfrak{n}}

\newcommand{\fs}{\mathfrak{s}}  
  
\newcommand{\fu}{\mathfrak{u}}

\renewcommand{\a}{\alpha}

\newcommand{\g}{\gamma}
\newcommand{\G}{\Gamma}

\newcommand{\eps}{\varepsilon}

\newcommand{\sig}{\sigma}

\renewcommand{\l}{\lambda}
\renewcommand{\L }{\Lambda}
\renewcommand{\o}{\omega}
\renewcommand{\O}{\Omega}
\renewcommand{\t}{\tau}

\newcommand{\rmd}{\mathrm d}

\newcommand{\lt}{\left}
\newcommand{\rt}{\right}

\newcommand{\lag}{\left\langle}
\newcommand{\rag}{\right\rangle}

\newcommand{\tr}{\mathrm{tr}}

\newcommand{\sgn}{\mathrm{sgn}}

\newcommand{\rmi}{\mathrm{i}}

\renewcommand{\tilde}{\widetilde}
\newcommand{\bLambda}{{\boldsymbol\Lambda}}
\newcommand{\bR}{{\mathbf{R}}}
\newcommand{\bJ}{{\mathbf{J}}}
\newcommand{\bK}{{\mathbf{K}}}

\graphicspath{{./figs/}{../figs/}}

\title{SL(2,C) Chern-Simons Theory, a non-Planar Graph Operator, and 4D Loop Quantum Gravity with a Cosmological Constant: \\ \Large Semiclassical Geometry}

\author[1]{Hal~M.~Haggard,}
\author[2]{Muxin~Han,}
\author[3]{Wojciech~Kami\'nski,}
\author[4]{and Aldo~Riello}

\affiliation[1]{Physics Program, Bard College, Annandale-on-Hudson, NY 12504, USA. }
\affiliation[2]{Institute for Quantum Gravity, University of Erlangen-N{\"u}rnberg, Staudtstra{\ss}e 7 / B2, 91058 Erlangen, Germany  }
\affiliation[3]{Instytut Fizyki Teoretycznej, Uniwersytet Warszawski, ul. Ho\.{z}a 69, 00-681 Warszawa, Poland}
\affiliation[4]{Perimeter Institute for Theoretical Physics, 31 Caroline St N, Waterloo, Ontario, Canada, N2L 2Y5 }

\emailAdd{hhaggard[at]bard[dot]edu}
\emailAdd{muxin.han[at]fau[dot]de} %
\emailAdd{Wojciech.Kaminski[at]fuw.edu[dot]pl}%
\emailAdd{ariello[at]perimeterinstitute[dot]ca} 

\abstract{
We study the expectation value of a nonplanar Wilson graph operator in SL(2,C) Chern-Simons theory on $S^3$. In particular we analyze its asymptotic behaviour in the double-scaling limit in which both the representation labels and the Chern-Simons coupling are taken to be large, but with fixed ratio. When the Wilson graph operator has a specific form, motivated by loop quantum gravity, the critical point equations obtained in this double-scaling limit describe a very specific class of flat connection on the graph complement manifold. We find that flat connections in this class are in correspondence with the geometries of constant curvature 4-simplices. The result is fully non-perturbative from the perspective of the reconstructed geometry. We also show that the asymptotic behavior of the amplitude contains at the leading order an oscillatory part proportional to the Regge action for the single 4-simplex in the presence of a cosmological constant. In particular, the cosmological term contains the full-fledged curved volume of the 4-simplex. Interestingly, the volume term stems from the asymptotics of the Chern-Simons action. This can be understood as arising from the relation between Chern-Simons theory on the boundary of a region, and a theory defined by an $F^2$ action in the bulk. Another peculiarity of our approach is that the sign of the curvature of the reconstructed geometry, and hence of the cosmological constant in the Regge action, is not fixed \textit{a~priori}, but rather emerges semiclassically and dynamically from the solution of the equations of motion. In other words, this work suggests a relation between 4-dimensional loop quantum gravity with a cosmological constant and SL(2,C) Chern-Simons theory in 3-dimensions with knotted graph defects.
}


\arxivnumber{}

\begin{document}

\maketitle

\section{Introduction and Overview}\label{sec_intro}

In this paper we show that the $\Slc$ Chern-Simons expectation value of a particular Wilson-graph operator is related to the Regge action of discretized general relativity with cosmological constant in four spacetime dimensions and Lorentzian signature. This relation is found in the double-scaling limit in which both the representation labels associated to the Wilson-graph operator and the modulus of the complex Chern-Simons level are scaled to infinity, while keeping their ratio, as well as the phase of the complex level, fixed. As an intermediate result we also show that the critical point equations obtained in this double-scaling limit allow a full reconstruction of the geometry of a homogeneously and non-perturbatively curved 4-simplex. The sign of the curvature can be either positive or negative. To be more precise we find that for a given graph, if any, there are always two critical points related by an orientation flip. To further fix ideas and notation, let us write the expectation value we are interested in explicitly
\be
Z_\text{CS}\lt(S^3;\G_5\big|\,\vec{j},\vec{i}\rt)=\int\cd A\cd\bar{A}\ \E^{\I \text{CS}\lt[S^3\,|\,A,\bar{A}\rt]}\ \G_5\lt(\vec{j},\vec{i}\,\big|A,\bar{A}\rt), 
\label{eq_expvalue1}
\ee
where $\G_5\lt(\vec{j},\vec{i}\,\big|A,\bar{A}\rt)$ is a specific graph operator, depending in particular on the representation labels $\vec{j}$ decorating the edges of the graph and the (complex-conjugate) connections $(A,\bA)$, and where $\text{CS}\lt[S^3\,|\,A,\bar{A}\rt]$ denotes the $\Slc$ Chern-Simons action on $S^3$ with complex (inverse) coupling constants $(h,\bar{h})$
\be
\text{CS}[S^3\,|\,A,\bar{A}]=\frac{h}{8\pi}\int_{S^3}\tr\lt(A\wedge \rmd A+\frac{2}{3}A\wedge A\wedge A\rt)+\frac{\bar{h}}{8\pi}\int_{S^3}\tr\lt(\bA\wedge \rmd \bA+\frac{2}{3}\bA\wedge \bA\wedge \bA\rt).
\ee
Then, if we let $j$ refer to a uniform scaling of all of the representation labels, the double-scaling limit (d.s.l.) we will refer to is 
\be
j, \, |h|\,\rightarrow \infty
\qquad \text{while} \qquad
j/|h|\sim cnst, \; \text{and} \; \text{arg}(h) = cnst.
\label{eq_DoubScal}
\ee
and its result on the expectation value of eq. \eqref{eq_expvalue1} is (when not suppressed, and modulo an overall phase that we are not writing here for clarity, but which will be discussed at the end of the paper)
\be
Z_\text{CS}\lt(S^3;\G_5\big|\,\vec{j},\vec{i}\rt) \xrightarrow{\text{d.s.l.}} \lt[\mathscr{N_+}\E^{ \I \lt(\sum_{t=1}^{10} \text{a}_t \Theta_t -  \lambda V_4  \rt)} + \mathscr{N_-}\E^{ - \I \lt( \sum_{t=1}^{10} \text{a}_t \Theta_t -  \lambda V_4  \rt)}\rt]\lt[1 + \text{O}(j^{-1}, h^{-1})\rt],
\label{eq_asymptoticformula1}
\ee
where $\text{a}_t, \; \Theta_t$ and $V_4$ are respectively the areas of the triangular faces of the reconstructed, curved 4-simplex,\footnote{The boundary of the 4-simplex is composed by purely space-like subsimplices.} the (hyper-)dihedral angles associated to such faces, and the (non-oriented) 4-volume of the 4-simplex. All the quantities appearing on the right-hand side are functions of the representation labels and the Chern-Simons level appearing on the left-hand side of the equality. Finally, $\mathscr N_\pm$ are weights depending at most polynomially on $j$ and $h$. Precise definitions of all the elements entering these formulae are given later in the paper.

This result brings to the forefront a new relationship between $\Slc$ Chern-Simons theory and 4-dimensional geometry. Indeed, a somewhat similar relationship between $\Slc$ Chern-Simons theory (with real level) and 3-dimensional geometry is well-studied in the research surrounding the so-called ``volume conjecture''. In this context, the  Chern-Simons expectation value of most Wilson-line (knots) \cite{Kashaev1997,Murakami2001} and of some Wilson-graph \cite{Veen2008,Veen2010,Costantino2014,Nawata2014} operators has been shown to reproduce, in a double-scaling limit very similar to the one discussed here, the 3-volume of certain hyperbolic manifolds \cite{Gukov2005,Dimofte2010,Dimofte2011,Dimofte2011a,Gukov2012}. Although refined mathematical techniques are being developed to rigorously study these relationships, the result \textit{per se} might seem natural to physicists. In fact, since the work of Witten \cite{Witten1988} and others starting at the end of the 1980's \cite{Achucarro1986,Ezawa1994,Matschull1999}, we know that three-dimensional quantum gravity \cite{Carlip2003} can be formulated, modulo some important subtleties, as a Chern-Simons theory for different gauge groups, depending on the sign of the cosmological constant and on the signature of the spacetime. In particular, $\Slc$ Chern-Simons with real level is related to Euclidean 3-dimensional quantum gravity. In light of this understanding, one can interpret the Wilson lines as topological defects induced by the presence of some particles. It follows that the double-scaling limit is nothing more than a semiclassical limit in which one selects stationary trajectories of the quantum theory, i.e. the classical solutions of the 3-dimensional Einstein equations. The result is an homogeneously curved hyperbolic manifold with particular conical singularities determined by the presence of the particles \cite{Gukov2005}. Certainly, things are more complicated than this physical picture might suggest, however, it has the advantage of clarifying \textit{why} the volume conjecture is reasonable. We will argue that an intuition can also be built for our result, and will try to convey it later in this  introduction. For the moment we observe that even though our construction bears similarities to the one appearing in the context of the volume conjecture, there are also major differences. {The most relevant one is that the $\Slc$ Chern-Simons theory used in the volume conjecture is the result of an analytic continuation \cite{Witten2010} of the SU(2) theory, which results in a generic real Chern-Simons level; by contrast, we deal with a genuine $\Slc$ Chern-Simons theory whose level has integer real part. As a consequence, in the asymptotic regime, we get a purely oscillating behavior, with no exponential growth of the amplitude as in the volume conjecture framework.}

In order to explain the picture we have in mind, let us start from the result. There, we see the emergence of Einstein-Hilbert gravity (with the proper boundary terms included) via its on-shell action; this action is Hamilton's principal function for gravity evaluated on the homogeneous solution within a single, curved 4-simplex. Each homogeneous 4-simplex has to be eventually thought of as part of a large triangulation that in some continuous limit gives smooth general relativity. In this approach curvature is distributionally concentrated in the form of a conical singularities over the triangles. In three dimensions the picture is similar, except that one is using tetrahedra to triangulate the manifold and the curvature is concentrated along the sides of the triangulation. This way of dealing with gravity is known as Regge calculus \cite{Regge1961,Bahr2010}. Our result, presents a quantum version for the amplitude of a single ``building block,'' in the spirit of spinfoam models \cite{Perez2012}. Further work will be needed not only to obtain a completely mathematically well-defined 4-simplex amplitude, but  also to understand how to obtain a sensible interplay between different 4-simplices, and, of course, to take the necessary continuum limit. The latter problem is particularly complex and there are entire research programs developed to tackle it, e.g. \cite{bahr2012discretisations,Oriti2007,Dittrich2014} to mention a few.

 In this paper we shall focus on the amplitude for a single 4-simplex. In the body of the paper we will explain in detail \textit{why} Chern-Simons theory can be expected to implement the bulk homogeneity of each building block, the idea being that it is the holographic projection of the topological quantum field theory \cite{Atiyah1988,Witten1988a} given by $BF$ theory plus a cosmological term \cite{Baez1996} (once the $B$ field has been integrated out). What about the curvature defects concentrated along the triangles, what is their origin? These defects are crucial, since they allow us to convert an otherwise homogeneously curved manifold into an approximation of virtually any manifold. At the level of the topological quantum field theory, such defects must originate in the breaking of  the ``triviality'' of the topological dynamics on a given manifold. Here the defects are exactly sourced by the Wilson graph. Notice, though, that the defects are geometrically associated to 2-dimensional surfaces while the graph defect is intrinsically one-dimensional. Indeed, in our picture, the graph corresponds to the \textit{dual} to the 4-simplex boundary triangulation: each of its five vertices (hence the name $\G_5$) corresponds to a tetrahedron, and each of its ten edges corresponds to a triangular face. In a precise sense, the graph carries quanta of area and should be thought of as carrying gravitational degrees of freedom instead of matter-like ones. This is exactly the picture emerging from the kinematics of loop quantum gravity \cite{Rovelli2004,Thiemann2007,Ashtekar2004}.

In order to better contextualize our work, we shall review very briefly what a spinfoam model is \cite{Perez2012}. The prototypical spinfoam model is the Ponzano-Regge model \cite{Ponzano1968,Barrett2009} for 3-dimensional Euclidean quantum gravity. In this model one starts from a triangulated manifold, and assigns SU(2) representations to the sides of the triangulation and trivalent intertwiners among three such representations to its triangles; then one contracts all the intertwiner indices following the combinatorics of the triangulation, multiplies the resulting amplitude by some weight factors (which have a clear group theoretical, and geometrical, meaning), and finally sums on all possible assignments of representations to sides in the bulk of the triangulation. This procedure assigns to every tetrahedron a function of the six spins attached to its sides, known as a $6j$-symbol. Ponzano and Regge noticed a relation with quantum gravity when they realized that in the large spin limit $j\rightarrow\infty$, the $6j$-symbol gives the imaginary exponential of the Regge action for the tetrahedron (without cosmological constant). Moreover, 3-dimensional quantum gravity is topological, and as such can be argued to be triangulation invariant. This is a property the Ponzano-Regge model satisfies morally; it is indeed invariant under Pachner moves, but only up to some infinite volume factors that signal the fact that the model is not completely gauge fixed \cite{Freidel2004,Bonzom2010}. A way to overcome this difficulty is to consider a quantum deformation of the group theoretical ingredients appearing in the model, in particular one can substitute the $6j$-symbols with $q$-deformed $6j$-symbols, and the dimensions of the representations, $(2j+1)$, with $[2j+1]_q$, to obtain the so-called Turaev-Viro state sum model \cite{Turaev1992}. Such models cure the divergences present in the Ponzano-Regge model by cutting them off at a maximal spin, related to the parameter $q$, making its triangulation invariance not only formal but mathematically exact. However, this is not enough, the deformed model presents an even more interesting asymptotics \cite{Mizoguchi1992,Taylor2006}. In the limit in which both the spins and the cutoff are taken to be uniformly large, one finds that the $6j_q$-symbol gives the Regge action for a homogeneously curved tetrahedron augmented by the cosmological term $\Lambda V_3$, where the  cosmological constant is related to the maximal spin. Finally, it also becomes clear that a deep relationship between the Turaev-Viro state sum model and Chern-Simons theory exists, thus giving a beautifully consistent picture of all the forms in which 3-dimensional quantum gravity can be understood. 

This much simplified account of the Ponzano-Regge and Turaev-Viro state sum models is not meant to be complete, but rather aims to illustrate why many researchers have dreamt that $q$-deforming spinfoam models for 4-dimensional gravity, could lead to a mathematically well-defined version of these ideas with the added feature of automatically incorporating in it a cosmological constant. This is exactly what was tried in the context of the Barrett-Crane model \cite{Barrett1998,barrett2000lorentzian}, see \cite{Noui2003}, and more recently the Engle-Pereira-Rovelli-Livine model \cite{Engle2008,Freidel2008}, see \cite{Han2011c,Han2011,Fairbairn2012}. The latter model, often abbreviated as EPRL (or EPRL/FK, for Freidel and Krasnov, when referring to its Euclidean version) is to-date the most developed and studied spinfoam model of 4-dimensional Lorentzian quantum gravity. It will constitute also our starting point for constructing the Wilson-graph operator $\G_5$. 

Previous efforts to include the cosmological constant have been largely motivated by analogy with the 3-dimensional case rather than obtained by means of some constructive principle \textit{leading} to the quantum group structure. In this paper we aim for a constructive inclusion and provide both heuristic and formal procedures for understanding the construction of a spinfoam model including a cosmological constant, which reduces to the usual ``flat'' spinfoam model in the appropriate proper limit. As a byproduct, we understand that introducing a cosmological constant in four dimensions is intimately related to a coupling of the spinfoam model to Chern-Simons theory, and as such it may---or may not---lead to a known quantum group structure for the spin networks under investigation \cite{Podles1990,Buffenoir1999,Buffenoir2002}. 

Indeed, our analysis shows that the introduction of a cosmological constant within the EPRL model of 4-dimensional Lorentzian quantum gravity requires the use of $\Slc$ Chern-Simons theory with a general complex level $h\in\mathbb C$, to which no known quantum group structure has yet been associated. Therefore, we propose a more general approach which can \textit{explain} why in some situations (e.g. in the case of 4-dimensional Euclidean quantum gravity) quantum groups are a relevant tool, but that they are not required as an \textit{a priori} starting point. Another useful consequence of our approach is to replace the algebraic language of  quantum groups with the field-theoretical language of Chern-Simons theory. In this way we are allowed---at least at the semiclassical level we investigate in this article---to talk about quantities such as holonomies, which admit a more direct geometric interpretation. This enormously simplifies the study of the model's asymptotics, shedding light on both discrete curved geometries and the way they are encoded in the classical solutions of Chern-Simons theory on the graph-complement manifold. Nonetheless, because of the asymptotic behavior of eq. \Ref{eq_asymptoticformula1} our model, $Z_\text{CS}\lt(S^3;\G_5\big|\,\vec{j},\vec{i}\rt)$, can be viewed as a generalization of the Turaev-Viro model (and its quantum $6j$-symbol formulation) that produces 4-dimensional gravity with a cosmological constant in the double-scaling limit of a single 4-simplex. 

Our work's connection with $\Slc$ Chern-Simons theory does, however, raise a question of mathematical rigor. Chern-Simons theory with a compact gauge group $G$ and its quantization are well understood, but the non-compact case is much more involved. In the specific case in which the non-compact group is the complexification of a compact one $G_\mathbb{C}$, e.g. $\Slc = \text{SU}(2)_\mathbb{C}$, which is the relevant one for our analysis, much more is known and actively investigated. In particular, in recent years, progress has been made in this arena, see e.g. \cite{Witten1991,Bar-Natan1991,Buffenoir2002,Gukov2005,Witten2010,Dimofte2010,Dimofte2011a,Dimofte2011,Gukov2012,Andersen2014}. Incorporating these new results and techniques into our framework is one of our main goals for the near future. For the moment let us just remind the non-expert reader that the differences between the theories on the gauge groups $G$ and $G_\mathbb{C}$ are not just technical, since qualitative changes occur. For example the Hilbert space of Chern-Simons theory with a noncomapct gauge group is infinte-dimensional \cite{Buffenoir2002,Dimofte2010}, while the Hilbert space of the theory with compact group is finite-dimensional. 

The idea of an interplay between Chern-Simons theory and loop quantum gravity, or spinfoams, is not a new one. It can be traced back to the discovery of the Kodama state \cite{Kodama1988} as a (formal) solution to the quantum constraints of canonical gravity with a cosmological constant expressed in Ashtekar's variables. This perspective has been investigated in the intervening years, in particular by Smolin \cite{Smolin1995,Smolin2002,Randono2006,Randono2006a,Wieland2011,Markopoulou:1997hu} (see also \cite{Witten2003}). This approach gives the same Chern-Simons structure that we have found from a covariant perspective, when applied to canonical loop quantum gravity with the Barbero-Immirzi-Holst twist \cite{Immirzi1993,BarberoG.1995,Holst1996} and expressed in terms of complex selfdual and anti-selfdual Ashtekar variables. Indeed, our construction can also be interpreted (even if this is not the interpretation we prefer) as the projection of the Kodama state onto a particular spin-network state, i.e. as taking a particular component of the loop-transform of such a state. Interestingly, the discovery of such a relationship between Chern-Simons theory and 4-dimensional loop gravity served in the 1990's as a further---though not the only---motivation to investigate the interplay between topological quantum field theories (with defects) and quantum gravity \cite{crane1993topological,crane1993categorical,crane1995clock,barrett1995quantum,Smolin1995,Baez1996,Baez2000}. {In the context of quantum deformation of (Euclidean) spinfoam models, the coupling with Chern-Simons has been proposed by one of the authors in \cite{Han2011}.} Finally, there is another point of contact between Chern-Simons theory and loops, this is  the study of quantum black holes and their entropy in loop gravity under the quantum isolated horizons paradigm \cite{Ashtekar1998,Ashtekar2002,Beetle2010,Engle2010,Engle2011,Frodden2014,Achour2014,Han2014}. It would be interesting to further investigate how this literature relates to the present work. 

\begin{figure}[t]
\begin{center}
\includegraphics[height=4cm]{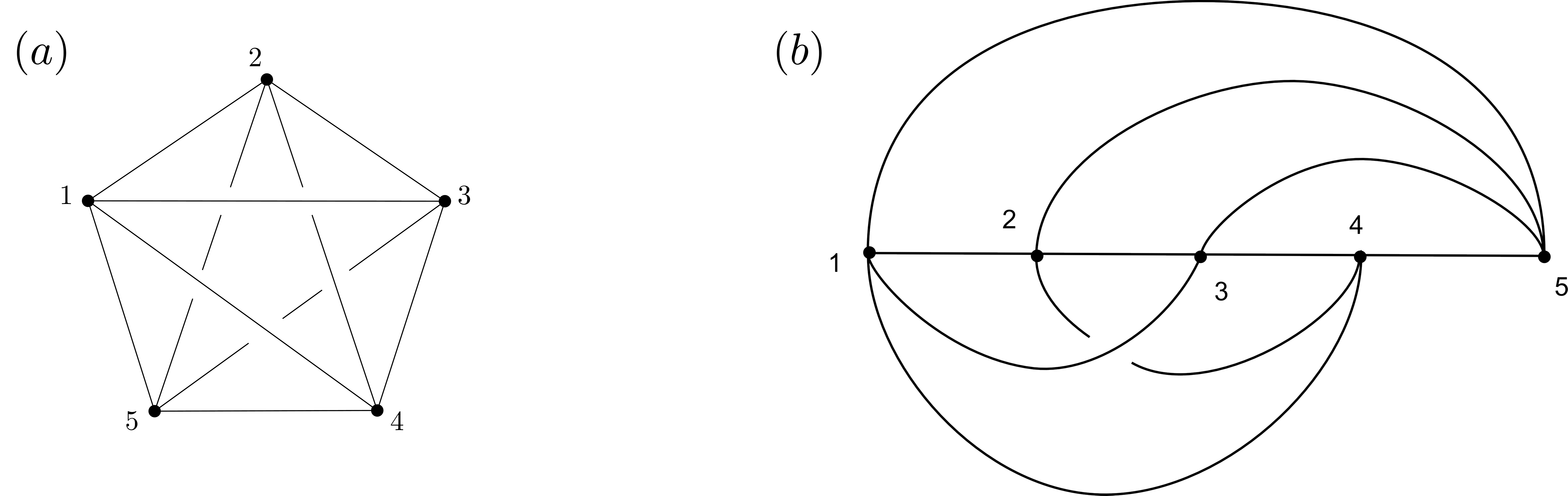}
\caption{Both panels illustrate the graph $\G_5$ with its five 4-valent vertices. (a) This panel explicitly displays the combinatorial structure of $\G_5$ as the dual to the boundary of a 4-simplex. (b) A topological deformation of $\G_5$ illustrates the single essential crossing of the graph projection.}
\label{fig_gamma5}
\end{center}
\end{figure}

After this excursus, we return to our calculation and try to give a bird's-eye view of what we will accomplish in the rest of the paper. After constructing the particular graph operator needed to implement the geometricity of the boundary of the 4-simplex, we study its asymptotic, semiclassical, properties in the double scaling regime described at the beginning of this introduction. Physically this is equivalent to sending $\hbar$ to zero, while keeping the size of the physical areas and of the cosmological constant fixed at finite values. In this way we freeze the fluctuating quantum geometries, picking out the most relevant classical solutions. For what concerns Chern-Simons theory, these classical solutions are given by flat connections. However, the graph plays the r\^ole of a source for such connections, which are hence flat everywhere but on the graph. 

To make mathematical sense of this statement one is lead to consider flat connections on the graph-complement manifold $M_3 = S^3\setminus\G_5$, obtained by removing from the 3-sphere an infinitesimal neighborhood of the $\G_5$ graph. Since $\G_5$ is dual to the boundary of a 4-simplex, see \autoref{fig_gamma5}, it is not too hard to see that $M_3$ is a 3-manifold bounded by a genus-6 surface. All the relevant information about the flat connection in $M_3$ can then be repackaged into equations for a set of holonomies in $M_3$. We divide these holonomies into two subsets, longitudinal holonomies that are computed along the length of the tubes bounding the edges of the thickened graph and transverse holonomies that cycle around these tubes. These equations encode the proper boundary conditions for the $M_3$-connection induced by the presence of a graph in the original manifold $S^3$. 

The main result of the paper, is to show that these very same holonomies can be reinterpreted as the holonomies on the boundary of a homogeneously curved 4-simplex. In a sense, we provide a translation between a connection whose curvature is concentrated along 1-dimensional defects carrying quanta of area, and a connection whose curvature is  homogeneously distributed in the 4-simplex. In the second case, the curvature ``defects'' are concentrated along the extended 2-dimensional submanifolds. (See \cite{Freidel2013} for a more precise description of this correspondence in the flat context. Somewhat similar ideas are also present in \cite{Dittrich2014a,Dittrich2014b}.) Note, that the flat connections of the graph complement we use to reconstruct the 4-simplex geometry have some very peculiar properties, inherited from the specific graph operator. Probably their main property is that the four transverse holonomies associated to a single graph vertex, when calculated infinitesimally close to it, are all in the same SU(2) subgroup of $\Slc$. This property is crucial for the interpretation of such holonomies in terms of face holonomies of a homogeneously curved tetrahedron that is \textit{flatly embedded}\footnote{A submanifold is said to be flatly embedded in a Riemannian manifold, if its extrinsic curvature vanishes. This turns out to be equivalent to the requirement that the surface is \textit{totally geodesic}, see e.g. \cite{Aminov2001}.} in the ambient de Sitter, or anti-de Sitter, space. An important ingredient of this interpretation is the fact that the representation labels (spins) associated to the graph edges are interpreted as the areas of the triangles in the 4-simplex (expressed in units of the Planck area times the Barbero-Immirzi parameter $\gamma$).

Once a 4-simplex geometry has been built from a specific class of flat connections on the graph complement, we can interpret the phase appearing in the semiclassical approximation geometrically. The semiclassical action is composed of two pieces: one coming from the Wilson-graph observable, and the other from the Chern-Simons action itself. Analogously to the flat\footnote{By flat we mean without cosmological constant.} EPRL model, one can see that the phase contributed by the graph operator corresponds to the $\I\sum_t \text{a}_t \Theta_t$ term in the Regge action. More interestingly, the Chern-Simons term contributes the cosmological term, $-\I\lambda V_4$, of the Regge action. Though not derived in this way here, this term can be seen as originating in the fact that a Chern-Simons theory on the boundary of a 4-dimensional region can be interpreted as the holographic projection of an $F^2$ theory in the bulk, where $F$ is the curvature of the Chern-Simons connection. If the curvature of $F$ is constant and equal to $\frac{\lambda}{6}e\wedge e$, this bulk theory provides exactly the sought after 4-volume term. Another feature of this asymptotic approximation is that critical solutions come in pairs of oppositely oriented 4-simplices, hence the two terms in eq. \eqref{eq_asymptoticformula1}. This is not a new feature, it was present already in the flat EPRL and EPRL/FK model, as well as in the Ponzano-Regge-Turaev-Viro state sum. It can be seen as due to the fact that one is not quantizing metric gravity, but rather first order gravity expressed in vielbein-connection variables, and in such a representation the metric can become degenerate and the orientation can flip. The physical reliability of this feature is a matter of debate \cite{Witten1988,Engle2013,Christodoulou2013,Riello2013}.  

{
Interestingly, Engle's explanation for the presence of both orientations \cite{Engle2011,Engle2013a,Engle2013} gives an explanation of why we find geometric sectors associated to both signs of the cosmological constant. He observed that the construction leading to the definition of what we call the EPRL graph operator has two sectors of solutions related by a sign flip. This sign appears in the equation that classically relates the curvature to the tetrad field when a cosmological constant is present effectively changing its sign. More technically, one can say that the linear simplicity constraints admit solutions in two Plebanski sectors, corresponding to $B=\pm e\wedge e$, respectively. Because the cosmological term is quadratic in $B$, while the Ricci term is only linear in it, the two sector effectively correspond to the Einstein-Hilbert with different signs of the cosmological constant. This is reflected in the equation for the curvature $F = \frac{\Lambda}{3}B= \pm\frac{\Lambda}{3}e\wedge e$.}

{
To conclude this overview of the work, we would like to acknowledge that other terms come out of the asymptotic analysis, which must be added to the Regge action. Their geometrical interpretation is not fully clear for the moment, though it is understood that these terms are parity invariant and therefore factorize as a phase common to the two differently oriented critical point contributions. At first sight, this may seem to be completely degenerate to an irrelevant phase choice of boundary state. However, upon closer analysis this seems not to be the right interpretation since this extra phase depends on the geometrical and dynamical variables associated to the 4-simplex (e.g. the spins), and therefore will \textit{a priori} superpose in a complicated way with the dynamics of the model for any 4-simplex in the bulk of the triangulation, where the boundary state phase choices are irrelevant. Nonetheless, there are some indications that these contributions should add up to zero for a given triangle sitting in the bulk of the triangulation, where the sum over its spins is relevant. Therefore, the r\^ole of these extra phases has not yet been completely clarified. We leave the analysis of this issue for future work, when we will study the amplitude of simplicial complexes.}

The paper is structured as follows. In \autoref{sec_CSandGraph} we formally define the expectation value $Z_\text{CS}\lt(S^3;\G_5\big|\,\vec{j},\vec{i}\rt)$, while in \autoref{sec_relationLQG} we discuss its relation with 4-dimensional gravity and LQG. In \autoref{sec_largej}, we introduce the semiclassical (double-scaling) limit and the zero-cosmological-constant limit of  $Z_\text{CS}\lt(S^3;\G_5\big|\,\vec{j},\vec{i}\rt)$. From \autoref{sec_lEPRL_explicit} to \autoref{sec_equations}, we derive the critical point equations to study the semiclassical limit just introduced. From \autoref{sec_idea} to \autoref{sec_curved4simplex} we relate these equations to the 3- and 4-dimensional simplicial geometries of constant curvature. We dedicate \autoref{sec_criticalaction} to the evaluation of the graph operator $\Gamma_5\lt( \vec j, \vec i \big| A,\bA\rt)$ and the Chern-Simons action functional $CS\lt[S^3\big| A,\bA\rt]$ at the critical points. In \autoref{sec_parity} we comment on the fact that to any semiclassical solution there always corresponds a second, opposite spacetime with reversed orientations. In \autoref{sec_NRterms} we discuss the role of the parity-invariant non-Regge contribution to the action, while in \autoref{sec_2pi-area} we discuss a subtlety in the geometrical interpretation of the spin variables and show that it entails no consequences for the final result. We finally conclude in \autoref{sec_conclusions}. In the appendices, we fix notations and conventions, e.g. those of the general relativistic action and others in the context of the selfdual and anti-selfdual split of $\Slc$ (\autoref{app_GRconv} and \autoref{app_notations}); we perform explicit calculations not spelled out in the text in \autoref{app_variations}; we present some relevant details about 4-dimensional discrete geometries in \autoref{app_Lorentziangluing}; and we calculate the Chern-Simons functional at the critical point perturbatively around a flat-geometry solution in \autoref{app_perturbative}. 
We hope that our effort to make the paper self-contained and pedagogical enough to be understandable to readers with different backgrounds has been successful. The reader may want to skip sections containing familiar material.



\section{$\Slc$ Chern-Simons Theory and the $\G_5$ Graph Operator} \label{sec_CSandGraph}

In this section, we give a brief overview of $\Slc$ Chern-Simons theory and introduce knotted graph operators. We study the expectation value of a particular nonplanar, knotted graph operator $\G_5(\vec{j},\vec{i}|A,\bA)$, defined below.

Given a compact oriented 3-dimensional manifold $\Fm_3$, the SU(2) Chern-Simons functional is\footnote{We use $\Fm_3$ for a general 3-manifold; the graph complement of central importance in this paper will always be denoted $M_3 = S^3\setminus\G_5$. }
\be
W[A]:=\frac{1}{4\pi}\int_{\Fm_3}\tr\lt(A\wedge \rmd A+\frac{2}{3}A\wedge A\wedge A\rt),
\ee
here $A=A^j \t_j$ is a \textit{real} $\su$-valued connection on the 3-manifold $\Fm_3$. Each of the components $A^j$ is an $\R$-valued 1-form on $\Fm_3$ and $\t_j=-\frac{i}{2} \sig_j $ are anti-Hermitian $2\times2$ generators of $\su$, with $\{\sig_j\}_{j=1,2,3}$ the Pauli matrices.

This action can be analytically continued to a holomorphic action for the complexified connection $A^\C$ with value in $\su^\C$, which is in turn isomorphic to $\slc$. Thus, in order to define Chern-Simons theory for the $\slc$-connection $\A$ on the manifold $\Fm_3$, one can first decompose $\A$ into its holomorphic and anti-holomorphic parts, $A^\C$ and $\bA^\C$ respectively, and then combine their Chern-Simons functional with a complex weight (inverse coupling) $h$:
\be
\text{CS}\lt[\Fm_3 \big| \A=(A^\C,\bA^\C) \rt] := \frac{h}{2}W[A^\C] + \frac{\bar h}{2} W[\bA^\C].
\ee
Henceforth, we refer to $\A$ as the $\slc$ connection, and---dropping the $\C$ superscript---to $A$ and $\bA$ as the holomorphic and anti-holomorphic $\slc$ connection, respectively. Sometimes we will even drop the adjective (anti-)holomorphic if this is not confusing.

Finally, quantum Chern-Simons theory on $\Fm_3$ is defined via the functional integral\footnote{The path integral measure should be  understood to contain all the gauge fixing (ghost) terms needed to make this expression meaningful. Since, for the purpose of this paper, we are interested only in the \textit{phase} resulting from the semiclassical approximation of the path integral, these terms are not going to play any r\^ole, and are therefore not considered explicitly. 
}
\be
Z_\text{CS}(\Fm_3)=\int\cd A\cd\bar{A}\ \E^{\I \text{CS}[\Fm_3\,|\,A,\bar{A}]}.\label{qcs}
\ee
It is expected that the partition function gives an interesting topological invariant of the 3-manifold. Also, recent progress towards its rigorous definition can be found e.g. in \cite{Dimofte2010,Witten2010}.

It is common to parametrize the complex Chern-Simons couplings as
\be
h=k+\I s\qquad\text{and}\qquad \bar{h}=k-\I  s,
\ee 
with $k$ and $s$ initially arbitrary complex numbers.\footnote{An analytic continuation of Chern-Simons theory has been proposed in \cite{Witten2010}, where both $k$ and $s$ can, in principle, be extended to arbitrary complex numbers (so $h,\ \bar{h}$ become independent complex numbers). The analytically continued $\Slc$ Chern-Simons theory has gauge group $\Slc\times\Slc$ (as a complexification of $\Slc$). Thus, the connections $A$ and $\bar{A}$ are treated as independent variables, although the integration contour in eq. \eqref{qcs} is usually chosen to be real with $\bA$ the complex conjugate of $A$. Non-integral values of $k$ in the analytically continued theory imply that the integration cycle should belong to a covering space to the space of connections, where two connections are equivalent if they are related by an infinitesimal (rather than a finite) gauge transformation. In this paper we restrict ourselves to the case with $k\in\Z$ and $s\in\R$. A study of analytic continuation will appear in \cite{3dblocks}.} However, requiring the invariance of $\exp\lt(\I\ \text{CS}[A,\bar{A}]\rt)$ under finite gauge transformations, restricts $k\in\Z$; moreover, $s$ is also constrained, by the requirement of unitarity  \cite{witten,Buffenoir2002}. There are two possibilities for a unitary $\Slc$ Chern-Simons theory corresponding to $s$ real or purely imaginary, these are: (\textit{i}) if  $s\in\mathbb R$, then $\bar{h}$ is the usual complex conjugate of $h$, and $A$ is invariant under the reversal of orientation; the theory is unitary in the Lorentzian sense;\footnote{That is, such that its action $\text{CS}_L$ is real and appears in the partition function in the form $\exp (\I \text{CS}_L)$.} and (\textit{ii}) if $s\in i\R$, on the other hand, $A\mapsto\bar{A}$ under orientation reversal, and the theory is unitary in the Euclidean sense;\footnote{That is, such that its action $\text{CS}_E$ is real, but appears in the partition function in the form $\exp( - \text{CS}_E)$.} Indeed, Chern-Simons theory relates to (2+1)-dimensional quantum gravity in de Sitter spacetime in case (\textit{i}), and to Euclidean quantum gravity in 3-dimensional Anti-de Sitter space in case (\textit{ii}).\footnote{For details on the relation between Chern-Simons theory and 3-dimensional gravity, see e.g. \cite{Witten1988,Achucarro1986,Ezawa1994,Matschull1999,Carlip2003,Gukov2005}.} Here, we stick to the choice (\textit{i}). Hence, in the rest of the paper, $\bar{h}$ is the usual complex conjugate of $h$.

It is well known that quantum Chern-Simons theory with compact gauge group $\mathrm{SU}(N)$ and level $k$ i related to the representation theory of the quantum group $\mathrm{SU}_q(N)$, where the quantum group deformation parameter $q$ is the root of unity $q=q(k,N)=\exp\lt(\frac{4\pi \I}{k+N}\rt)$. For Chern-Simons theory with non-compact gauge group $\Slc$ and couplings $k$ and $s$ the situation is more complicated. In \cite{Buffenoir2002} it is shown that for the Lorentzian unitary case (\textit{i}), with $s\in\R$, and under the restriction $k=0$, the quantum $\Slc$ Chern-Simons theory is related to the unitary irreps of the quantum group $\mathrm{SL}_q(2,\C)$ with real deformation $q=\exp\lt(\frac{2\pi}{s}\rt)$. Moreover, a quantum group deformation of $\Slc$ is only known for such a real deformation parameter $q$ \cite{Podles1990,Buffenoir1999}. In this paper, however, we focus on $\Slc$ Chern-Simons with \textit{non-vanishing} $k$ due to the interesting relation it bares with 4-dimensional geometry and quantum gravity; we do this despite the fact that the theory with non-vanishing $k$ has no known quantum group structure behind it. The $\Slc$ Chern-Simons theories with general values of $k$ and $s$, which exist as quantum field theories, might lead to a generalization of quantum group structures to $\Slc$. The present work provides additional motivation for a generalization in this direction.

We consider $\Slc$ Chern-Simons theory on $S^3$, the 3-sphere, with Wilson-line operator
\be
G_{\ell}[A]=\mathbb P \exp \int_{\ell}A, \ \ \ \ G_{\ell}[\bar{A}]=\mathbb P \exp\int_{\ell}\bar{A}
\ee 
along the (piecewise differentiable) curve $\ell$ embedded in $S^3$. We focus on Wilson-line operators in $S^3$ carrying unitary irreps of $\Slc$,
\be
D^{(j,\rho)}_{l,n;l',n'}\lt(A,\bar{A}\rt)=\lag(j,\rho);l,n\lt|\ D^{(j,\rho)}\lt(G_{\ell}[A],G_{\ell}[\bar{A}]\rt)\ \rt|(j,\rho);l',n'\rag.\label{WL}
\ee
The (infinite-dimensional) unitary irreps of $\Slc$ are classified by two parameters $(j,\rho)$ with $j\in\frac{\Z}{2}$ and $\rho\in\R$ \cite{Ruhl1970}. A canonical basis in the unitary irrep $\ch^{(j,\rho)}=\oplus_{l\in j +\mathbb N}\ch_l$ is denoted $\lt|(j,\rho);l,n\rag$, where $\ch_l$ is the spin-$l$ irrep of $\mathrm{SU(2)}\subset\Slc$. Note that $D^{(j,\rho)}\lt(A,\bar{A}\rt)$ depends non-trivially on both $A$ and $\bar{A}$, this will appear explicitly in the formulae of \autoref{sec_lEPRL_explicit} (e.g. eq. \eqref{IG5}, where both $G[A]$, and $G^\dagger[\bA]$ appear).

In a major part of the literature on Chern-Simons theory, the curve $\ell$ is taken to be a knot, where the Wilson-line operator is a Wilson-loop (e.g. \cite{Witten1989a,Dimofte2011a,Dimofte2011,Gukov2012,Gukov2005}). However, we are interested in knotted graphs that admit vertices with valences greater than 2. In particular, the knotted graph operator investigated here is the $\G_5$ graph of \autoref{fig_gamma5}; this is a non-planar, 4-valent graph with a single crossing.\footnote{A few quantum group spin networks based on the $\G_5$ graph have been proposed and studied in \cite{Han2011c,Fairbairn2012,Han2011,Noui2003}.} We adopt the same framing convention for the graph as in \cite{Witten1989}.

The 4-valent knotted graph operator $\G_5(\vec{j},\vec{i}\,|A,\bar{A})$ in $\Slc$ Chern-Simons theory is constructed through a series of four steps: 
\begin{enumerate}[(i)]

\item \textit{$\Slc$ unitary irreps:} An $\Slc$ unitary irrep $(j_\ell,\rho_\ell)$ is associated to each edge $\ell$ in the knotted graph. We define the ratio $\g=\rho_\ell/j_{\ell}\in\R$ and restrict $\g$ to be a constant independent of $\ell$. This restriction\footnote{In the flat case such a restriction is redundant in the semiclassical limit, because it happens to be one of the critical point equations \cite{Barrett2010}. This is not the case in the present setting.} is important for later geometrical interpretations in 4-dimensions, where $\g$ defines a fundamental unit for surface areas. In the context of LQG, $\g$ corresponds to the Barbero-Immirzi parameter \cite{Immirzi1993,BarberoG.1995,Engle2008,Ding2011}. We will label the edge connecting the vertices $a$ and $b$ ($a,b\in\{1,\cdots,5\}$) by $\ell_{ab}=\ell_{ba}$, and fix its orientation from $b$ to $a$ when $a<b$. Thus, the $\Slc$ unitary irrep associated to the edge $\ell_{ab}$ is often denoted by $(j_{ab},\g j_{ab})$.

\item \textit{Intertwiners:} All the vertices of the $\G_5$ graph are 4-valent. To maintain gauge invariance, an $\Slc$ intertwiner 
\be
I^{(j_1,\g j_1)\cdots (j_v,\g j_v)}\in \mathrm{Inv}_{\Slc}\lt(\ch^{(j_1,\g j_1)}\otimes\cdots\otimes\ch^{(j_v,\g j_v)}\rt),
\ee
is associated to each vertex. Here $v$ labels the valency of the vertex ($v=4$ for us) and $\ch^{(j,\g j)}$ is a carrier space for the unitary irrep of $\Slc$. The space of intertwiners $\mathrm{Inv}_{\Slc}\lt(\ch^{(j_1,\g j_1)}\otimes\cdots\otimes\ch^{(j_v,\g j_v)}\rt)$ is infinite-dimensional when $v\geq4$.\footnote{When $v=3$, the space of intertwiners is 1-dimensional, \cite{Naimark1964}.} However, we restrict attention to a finite-dimensional subspace of intertwiners that lead to a nice geometrical interpretation. This subspace of $\Slc$ intertwiners is determined via an $\Slc$  diagonal action on the SU(2) intertwiners $i^{j_1\cdots j_v}$, followed by group averaging\footnote{The tensor components of the intertwiner are all finite for $v>2$, \cite{Engle2009}.}
\be
I^{(j_1,\g j_1)\cdots (j_v,\g j_v)}_{j_1'm_1';\cdots;j_v'm_v'}(i)=\int_{\Slc}\rmd g \sum_{m_1\cdots m_v} i^{j_1\cdots j_v}_{m_1\cdots m_v}\prod_{l=1}^v D^{(j_l,\g j_l)}_{j_lm_l,j_l'm_l'}(g).\label{intertwiner}
\ee
This defines an embedding map from the space of SU(2) intertwiners to the space of $\Slc$ intertwiners. Equation \eqref{intertwiner} can be written abstractly as $I(i)=P_{\Slc}\circ {\Y}(i)$. In this expression, the EPRL map ${\Y}$ is an injection $\ch_j\hookrightarrow\ch^{j,\g j}$ given by the identification of the SU(2) irrep $\ch_j$ with the lowest subspace in the tower $\ch^{j,\g j}=\oplus_{l\in j +\mathbb N}\ch_l$, i.e. 
\be
{\Y}|j,m\rangle=|(j,\g j);j,m\rangle , \label{Ymap}
\ee 
and $P_{\Slc}$ is a projector onto the space of $\Slc$ intertwiners. The subspace of $\Slc$ intertwiners given by the image $I(i)$ was first introduced by Engle, Pereira, Rovelli, and Livine \cite{Engle2008} and further developed by Dupuis and Livine \cite{Dupuis2010}. The classical spin networks with intertwiners in this image have been shown to relate to simplicial (piecewise-flat) geometry in 4-dimensions by \cite{Barrett2010,Han2012,Han2013a}.

\item \textit{Contraction:} The knotted graph operator $\G_5(\vec{j},\vec{i}\,|A,\bar{A})$, as a gauge invariant observable of $\Slc$ Chern-Simons theory, is defined by contraction of the Wilson-line operators of eq. \eqref{WL} with the intertwiners of eq. \eqref{intertwiner} at each vertex:
\be
\G_5\lt(\vec{j},\vec{i}\,\big|A,\bar{A}\rt)=\bigotimes_{a<b}D^{(j_{ab},\g j_{ab})}\lt(A,\bar{A}\rt) \bullet \bigotimes_{a=1}^5 I\lt(i_a\rt),\label{graphoperator}
\ee
where $\bullet$ stands for contraction of the indices at each vertex following the appropriate 4-simplex combinatorics. This knotted graph operator is, in the terminology of \cite{Dupuis2010}, an $\Slc$ projected spin-network function. 

\item  \textit{Coherent basis:} For the concrete computation we choose a particular class of SU(2) intertwiners, the coherent intertwiners introduced by Livine and Speziale in \cite{Livine2007}. Given the SU(2) unitary irrep $\ch_j$, a coherent state $|j,\xi\rangle\in\ch_j$ is defined by an SU(2) action on the highest-weight state \cite{Perelomov1971},\footnote{See, e.g. \cite{Freidel2008} for a compact introduction to SU(2) coherent state.}
\be
|j,\xi\rangle:=g(\xi)|j,j\rangle,\ \ \ \ g(\xi)\equiv \begin{pmatrix}
  \xi^1 & -\bar{\xi}^2  \\
  \xi^2 & \,\bar{\xi}^1
 \end{pmatrix},
\ee 
where $\xi$ is a normalized 2-spinor according to the Hermitian inner product $\lag\xi,r\rag=\delta_{\dot{\a}\a}\bar{\xi}^{\dot{\a}}r^\a$, i.e. $\lag\xi,\xi\rag=1$. The $\mathrm{SU(2)}$ group element $g(\xi)$ rotates the 3-vector $\hat{z}=(0,0,1)$ into the unit vector $\hat{n}(\xi)=\lag\xi,\vec{\sig}\xi\rag$, where $\vec{\sig}$ is the vector of Pauli matrices. The coherent states $|j,\xi\rangle$ form an over-complete basis of $\ch_j$ and provide a resolution of the identity
\be
\mathbb{I}_j=(2j+1)\int_{S^2} \rmd\mu( \xi)\ |j,\xi\rangle\langle j,\xi |.
\ee
Since $|j,\xi\rangle\mapsto \E^{i\phi}|j,\xi\rangle=|j,\E^{i\phi}\xi\rangle$ leaves the integrand invariant, the domain of integration is the coset $S^2=\mathrm{SU(2)}/\mathrm{U(1)}$, on which $\rmd\mu(\xi)$ is the uniform measure. The phase of $\xi$ must be fixed conventionally to complete the definition of these coherent states.

A coherent intertwiner $i_{\vec{\xi}}\in \mathrm{Inv_{SU(2)}}\lt(\ch_{j_1}\otimes\cdots\otimes\ch_{j_v}\rt)$ can be defined by group averaging the projected tensor product of coherent states, \cite{Livine2007},
\be
i_{\vec \xi}= i^{j_1\cdots j_v}_{\xi_1\cdots \xi_v}:=\int_{\mathrm{SU(2)}}\rmd h\ \otimes_{l=1}^v D^{j_l}(h)|j_l,\xi_l\rangle.
\ee
These form an over-complete basis in the space of SU(2) intertwiners and relate to the quantization of three-dimensional flat polyhedral geometry \cite{Barbieri1997,Baez1999,Conrady2009,Bianchi2011PRD, BianchiHaggard2011,BianchiHaggard2012,Haggard2013}. These coherent intertwiners are mapped by $I$, eq. \eqref{intertwiner}, to $\Slc$ intertwiners $I(i_{\vec{\xi}})$; the latter enter the definition of the knotted graph operator $\G_5\lt(\vec{j},\vec{i}_{\vec{\xi}}\,\big|A,\bar{A}\rt)$ in eq. \eqref{graphoperator}. In the analysis below we often employ the following convention for labeling the $\xi$ variables. At the vertex $a$ of the graph $\G_5$ we denote the SU(2) coherent intertwiner by $i_a=i_{\{\xi_{ab}\}_{b\neq a}}=\int_{\mathrm{SU(2)}}\rmd h\ \otimes_{b\neq a} D^{j_{ab}}(h)|j_{ab},\xi_{ab}\rangle$ and the corresponding $\Slc$ intertwiner by $I(i_{a})$. The $\xi_{ab}$ label the coherent intertwiner at the vertex $a$, while the $\xi_{ba}$ label the coherent intertwiner at vertex $b$. Thus, we have in total 20 spinors $\xi_{ab}$ with $\xi_{ab}\neq \xi_{ba}$; these two distinct spinors are located at the opposite ends of the edge $\ell_{ab}$. The knotted graph operator is finally denoted $\G_5\lt({j}_{ab},{\xi}_{ab}\,\big|A,\bar{A}\rt)$.

\end{enumerate}

The Chern-Simons expectation value of the knotted graph operator on $S^3$ is the central object studied here:\footnote{The expectation value of eq. \eqref{Zgamma5} has been normalized by the partition function of Chern-Simons theory on $S^3$. This normalization procedure is understood throughout the paper.} 
\be
Z_\text{CS}\lt(S^3;\G_5\big|j_{ab},\xi_{ab}\rt)=\int\cd A\cd\bar{A}\ \E^{\I \text{CS}\lt[S^3\,|\,A,\bar{A}\rt]}\ \G_5\lt(j_{ab},\xi_{ab}\,\big|A,\bar{A}\rt). \label{Zgamma5}
\ee
In particular, we are interested in the asymptotic behavior of this expectation value when the double-scaling limit of eq. \eqref{eq_DoubScal} is taken uniformly in every spin $j_{\ab}$. As discussed in the introduction, in the asymptotic regime of the limit, a relation emerges between the data $j_{ab},\xi_{ab},A,\bar{A}$ of Chern-Simons theory and four-dimensional, constant curvature geometry. The asymptotic behavior of the expectation value in eq. \eqref{Zgamma5} relates to quantum gravity in four dimensions with cosmological constant. 



\section{Relation with Loop Quantum Gravity (LQG)}\label{sec_relationLQG}

The Chern-Simons expectation value of the $\G_5$-graph operator, eq. \eqref{Zgamma5}, is well-motivated by non-perturbative, covariant LQG in 4-dimensions, where the idea of path integral quantization is adapted to the setting of LQG \cite{Perez2012,Rovelli2011,Rovelli2004}. The quantum dynamics is formulated in terms of boundary state transition amplitudes. These amplitudes naturally extend the notion of transition between initial and final states to the general covariant context \cite{Atiyah1988,Witten1988a}. In quantum gravity these boundary states capture the geometry of the boundary and according to LQG this geometry is encoded in spin networks \cite{Rovelli2004,Thiemann2007,Ashtekar2004,BianchiHaggard2011,BianchiHaggard2012,Haggard2011,HAN2007}. 

In order to calculate such transition amplitudes, in principle one should interpolate using all 4-dimensional histories, or bulk (quantum) geometries, compatible with the given boundary states. At present the theory is defined only in terms of successive truncations,  relying on a specific discretization of the  bulk geometry.\footnote{The question of how to remove the discreteness is a controversial problem beyond the scope of this paper (see e.g. \cite{Oriti2007,Dittrich2014}).} Each of these quantum discrete geometires is a spinfoam. The spinfoam building blocks are usually taken to be 4-simplices, and the total spinfoam amplitude is then constructed as a product of 4-simplex amplitudes  followed by an integration over all the bulk data encoding their geometry. In this section, we briefly review the motivations and construction of the EPRL 4-simplex amplitude \cite{Engle2008}, which is one of the leading candidate models for covariant four-dimensional LQG. We will then explain why and how the Chern-Simons expectation value eq. \eqref{Zgamma5} is a deformation of the EPRL 4-simplex amplitude that includes the cosmological constant.


\subsection{Lorentzian EPRL 4-Simplex Amplitude}\label{sec_EPRL}

First order gravity in tetrad-connection variables can be expressed as a constrained topological $\Slc$ $BF$ theory in 4-dimensions, which is known as the Plebanski formulation of classical gravity \cite{Plebanski1977}. The EPRL 4-simplex amplitude is constructed by imposing constraints on quantum $\Slc$ $BF$ theory. The first-order action of $\Slc$ $BF$ theory on a 4-dimensional manifold $\frak M_4$ is
\be
S_{BF} : = - \frac{1}{2} \int_{\frak M_4} \prec B\wedge \F[\A]\succ\;,
\ee
where $\F[\A]=\rmd\A+\A\wedge\A$ is the curvature of the $\slc$ connection $\A := (A,\bar{A})$, $B$ is an $\sldc$ valued two form, and $\prec\cdot,\cdot\succ$ is one of the two invariant, nondegenerate, bilinear forms of $\sldc$. Specifically, it is the one that couples boosts with rotations, i.e. $\prec X, Y\succ := \frac{1}{2}\epsilon_{IJ}^{\phantom{IJ}KL} X^{IJ}Y_{KL}$ (see \autoref{app_notations} for some information on these bilinear forms). Notice that $B$ is the momentum conjugated to the connection $\A$.

The quantization of $BF$ theory is given by the functional integration
\be
\int\cd \A\cd B\ \E^{-\I S_{BF}}=\int \DD\A \;\delta\left( \F[\A]\right),
\label{Bintegration}
\ee 
where the equality implements the integration over the momentum $B$ and defines the associated second-order theory.
Given a 4-manifold $\frak M_4$ with boundary $\partial \mathfrak{M}_4 =\mathfrak{M}_3$, let $\psi=\psi(\A_\partial)$ be a (gauge invariant) wave function of the connection boundary state, where $\A_\partial$ is the connection $\A$ restricted to $\frak M_3$. Then the BF amplitude of such a state is
\be
\langle BF|\psi\rangle = \int \DD\A \;\delta\left( \F[\A]\right)\;\psi[\A_\partial].
\label{BF_ampl}
\ee
When a gauge invariant state $\psi$ has support only on a graph $\G\subset \frak M_3$, and it depends on the connection $\A$ only through the holonomies $G_\ell[\A]$,  with $\ell$ an edge of the graph $\G$, then we call it a \textit{spin-network state} and write $\psi_\G$,
\begin{align}
\psi_\G[\A_\partial] = \psi_\G\big(\big\{G_{\ell}[\A_\partial]\big\}\big).
\end{align}
As normalizable states, the $\psi_\G$ belong to $\mathcal{L}^2(\Slc^{\otimes L})$, $L$ being the number of edges $\ell$ in $\G$. It is then convenient to use the distributional basis $\{\psi_{s}\}_\G$ such that the wave functions $\psi_s$ contain only one $\Slc$ unitary irrep $(j_\ell,\rho_\ell)$ on each edge $\ell$,  and an $\Slc$ intertwiner $I_n$ at each vertex $n$ of $\G$. In this sense, $s$ can be seen as a collective index labeling these data, $s=(j_\ell,\rho_\ell;I_n)$.

A special case is that of $\mathfrak{M}_4$ a 4-simplex with its associated boundary $\mathfrak{M}_3 \cong S^3 $. In this case it is natural to consider the graph $\Gamma_5$ dual to the 4-simplex boundary. Then a state $\psi_{\G_5}$ depends on ten $\Slc$ holonomies $G_\ell[\A_\partial]$, and is required to be invariant under gauge transormations at each of its five vertices. The graph\footnote{In this section and the following, we commit a slight abuse of notation; we use the symbol $\Gamma_5$, which was already introduced for the Wilson-line operator of eq. \eqref{graphoperator} within Chern-Simons theory, for a particular graph in $S^3$.} $\G_5$ is represented in \autoref{fig_gamma5}, although $BF$ theory is not sensitive to the crossing.

The functional integration of $BF$ theory on a triangulated manifold can be written as the product of 4-simplex amplitudes followed by a summation over all the intermediate (boundary) states of every 4-simplex boundary.

From eq. \eqref{BF_ampl} we see that the amplitude $\langle BF|\psi_{\G_5}\rangle$ is nothing but the integral of $\psi_{\G_5}$ over the space of flat connections. In particular, since it is already gauge invariant, we immediately obtain
\be
\langle BF|\psi_{\G_5}\rangle = \psi_s(\mathbb I),
\label{BF_dyn}
\ee
which is the evaluation of the spin-network function $\psi_{\G_5}$ on trivial holonomies $\mathbb I$.

The classical action of first-order general relativity in the the tetrad $e$ and connection $\A$ variables is\footnote{We work in units where the reduced gravitational constant $\kappa:=8\pi G_N$ is equal to 1.}
\be
S_\text{GR} := - \frac{1}{2}\int_\mathcal{M} \prec(e\wedge e) \wedge \F[\A] \succ .
\ee
This action can be twisted without altering its equation of motion (at least in the absence of fermions) by adding to it the so-called Holst term \cite{Holst1996}, to obtain what is known as the Holst action of general relativity
\begin{align}
S_\text{H}& := - \frac{1}{2}\int_\mathcal{M} \prec(e\wedge e) \wedge \F[\A] \succ + \frac{1}{\gamma} <(e\wedge e) \wedge \F[\A] >. 
\label{Holst}
\end{align}
In the last expression we have introduced the Barbero-Immirzi parameter $\gamma$, which we shall require to be real,\footnote{Generalizations to complex Barbero-Immirzi parameter are of interest. Clearly, a very special role is played by $\gamma=\pm i$.} as well as the second nondegenerate bilinear form $<\cdot,\cdot>$ on $\slc$. This bilinear form is related to the first one by $<\cdot,\cdot> = -\prec \star \cdot,\cdot \succ$, where $\star$ denotes the usual Hodge star operator; in other words $<X,Y> = X^{IJ} Y_{IJ}$ (see \autoref{app_notations} for details). The interest of the Holst formulation is that it twists the phase space of the theory and makes it easier to quantize (see \cite{Rovelli2004,Thiemann2007,Ashtekar2004,HAN2007}).

Both the standard and Holst's first order formulation of general relativity can be put in the form of a constrained $BF$ theory, where the $B$ field is required to take the simple form 
\be
B\stackrel{!}{=} e\wedge e,
\label{B=ee}
\ee
here and below $\stackrel{!}{=}$ indicates equality after the imposition of a constraint. 
In particular, the relevant $BF$ action for the Holst formulation is
\be
S_{\text{H}BF} := - \frac{1}{2}\int_\mathcal{M} \prec \left[\left( 1-\frac{1}{\gamma}\star \right) B \right]\wedge \F \succ .
\ee 
Notice that in this action the $B$ field is no longer the momentum conjugate to the connection $\A$. Indeed, calling the conjugate momentum $\Pi$, one finds\footnote{Notice that the relation between $B$ and $\Pi$ is not invertible if $\g=\mp i$. Indeed, in this case, the (anti-)selfdual part of $B$ is projected out of the theory.}
\be
\Pi = \lt(\star + \gamma^{-1}\rt) B \qquad\text{or equivalently} \qquad B = \frac{\g}{\g^2+1}\lt(1-\gamma\star \rt) \Pi .
\ee
Nonetheless, because $B$ and $\Pi$ are linearly related, integrating over one or the other, as in eq. \eqref{Bintegration}, does not make any difference.

Imposing the simplicity constraints of eq. \eqref{B=ee} modifies topological $BF$ theory, unfreezing local degrees of freedom and yielding general relativity.

The EPRL 4-simplex amplitude is obtained by imposing the simplicity constraint on the $BF$ amplitude $\langle BF|\psi_{\G_5}\rangle$. In this context the simplicity constraint is quantized to a constraint operator that acts on the boundary state $\psi_{\G_5}$. Imposing this quantum constraint reduces the available boundary states $\psi_{\G_5}$ to a proper subspace. Implementation of the simplicity constraint is described briefly in the following, however see e.g. \cite{Engle2008,Freidel2008,Dupuis2010,Livine:2007ya,Dupuis:2011wy,Dupuis2011,Livine2007,Ding2011} for details.

Given a triangulation of the 4-manifold $\frak M_4$, the $\slc$-valued 2-form field $B$ can be understood as a (anti-symmetric) bivector $B^{IJ}$ associated to each triangle. It turns out that for a simplicial decomposition, a more manageable linear version of the simplicity constraints can be employed in the quantization. Given any tetrahedron in the triangulation, all the bivectors associated to the triangles of the tetrahedron are constrained to satisfy
\be
U^I B_{IJ}\stackrel{!}{=}0,\ \ \ \ \text{or}\ \ \ \ 
U^I\left[(1-\gamma\star)\Pi\right]_{IJ} \stackrel{!}{=}0,
\label{simplicity_Pi}
\ee
with $U^I$ a unit time-like normal, which can be fixed to $U^I=(1,0,0,0)^T$ by an $\Slc$ gauge transformation. Upon quantization, $\Pi$ becomes a derivative operator acting on the connection variables. More precisely, the operator associated to $\Pi$ is a right invariant vector field on $\Slc$ that acts as a derivative on the spin-network functions $\psi_\G(\{G_\ell\})$. For this reason, in the quantum theory, $\Pi^{IJ}$ acts on $\psi_\G$ as the $\slc$ Lie algebra generator $\mathcal{J}^{IJ}$. Hence, by decomposing the $\sldc$ generators into boosts and rotations with respect to the frame $U^I=(1,0,0,0)^T$, that is, $K^i=\mathcal{J}^{0i}$ and $J^i=\frac{1}{2} \epsilon^{0i}_{\phantom{0i}jk}\mathcal{J}^ {jk}$, the linear simplicity constraint can be restated at the quantum level as
\be
(K^i-\gamma J^i )_\ell|\psi_\Gamma\rangle\stackrel{!}{=} 0 .
\ee
The constraint operators, however, do not commute among themselves, and thus cannot be imposed strongly on the states. One solution to this issue is to impose them weakly, i.e. in expectation value
\be
\langle \psi_s| (J^i-\gamma^{-1}K^i )_\ell |\psi_s \rangle \stackrel{!}{=} 0 ,
\label{simplicity_JK}
\ee
another is to use the master constraint technique \cite{Thiemann2006,Thiemann2006a,Han2006,Han2010,Han2010a,Engle2008,Ding2011}. 

Implementation of the quantum simplicity constraint reduces in a nontrivial fashion the possible $\SLDC$ unitary irreps in $\psi_\G$, i.e. after the imposition of the constraints only some $s=(j_\ell,\rho_\ell;I_n)$  are allowed. It turns out that the irreps that survive have a constant ratio between $\rho_\ell$ and $j_\ell$
\be
\rho_\ell=\gamma j_\ell,
\ee
where $\g$ is Barbero-Immirzi parameter that appears in the Holst action of eq. \eqref{Holst}. Consequently, the allowed $\Slc$ intertwiners are reduced to a finite-dimensional subspace, specifically, to those contained in the image of the injection $\Y$ of SU(2) intertwiners (see eq. \eqref{Ymap}). The resulting spectrum of $\Slc$ spin-network functions, after the reduction, has the same expression as the class of $\G_5$-graph operators defined previously, and \textit{after} the imposition of the constraints (notated with the exclamation mark) one has
\be
\psi_{\G_5,s}\Big(G_\ell[\A]\Big) \stackrel{!}{=} \Gamma_5\lt( j_\ell, i_n \big| A,\bA\rt),
\ee
where $s=\Big(j_\ell,\rho_\ell=\gamma j_\ell; I_n= I\Big(i^{\{j_\ell\}}_n\Big) \Big)$ and $\A=(A,\bA)$. Notice, however, that the previous equation can be only formal, since its left hand side is defined within $BF$ theory while the right one is defined within Chern-Simons theory, nonetheless we will use it as a notational short-hand.

The leitmotiv of the EPRL construction, then, is the treatment of quantum gravity as a constrained $BF$ theory, with the constraint imposed after quantization. The simplicity constraint imposes geometricity conditions on the boundary state for each 4-simplex, while the $BF$ theory dynamics is retained inside the 4-simplex. This is analogous to the Regge calculus of simplicial general relativity \cite{Regge1961,Feinberg1984,Friedberg1984}, where inside each (small) 4-simplex the geometry is trivially flat, while the full manifold geometry (e.g. metric, curvature) is reflected both in the shape of the flat 4-simplices and in the gluing between them. This perspective and the relation with Regge calculus has been confirmed through large-$j$ asymptotic analysis \cite{Barrett2010,Han2012,Han2013a,Conrady2008,Han2011c,Han2014b,Han2014a,Han2013}. Once again, the question of how to remove this discreteness is controversial and is beyond the scope of this paper.

Thus, the EPRL spinfoam amplitude of a single 4-simplex $\sigma$ is given simply by evaluating the \textit{constrained} spin-network functional $\G_5\lt(\vec j,\vec i \;\big| A,\bA\rt)$ at the trivial connection, with the last requirement following from the imposition of $BF$ dynamics within each single spin-network vertex,
\be
Z_{EPRL}\lt(\sigma\big|\vec j, \vec i\;\rt):= \langle BF |\,\G_5\lt(\vec{j},\vec{i}\;\rt)\rangle = \G_5\lt(\vec{j},\vec{i}\;\big|\, 0\rt).
\ee
This is effectively a transition amplitude of a boundary SU(2) spin-network state. A substantial body of results in LQG shows that three-dimensional quantum geometry is described by SU(2) spin networks \cite{Rovelli2004,Thiemann2007,Ashtekar2004,BianchiHaggard2011,BianchiHaggard2012,Haggard2011,HAN2007}. Thus, the EPRL spinfoam amplitude is understood as a transition amplitude between boundary quantum geometries. The full spinfoam amplitude is given by first multiplying all the amplitudes $Z_{EPRL}(\sig)$ of the bulk 4-simplices, and then summing over the intermediate boundary states labelled by $(\vec{j}, \vec{i})$.

Before moving on, it is important to notice that the way the simplicity constraint are implemented makes use of the time gauge, in which the frame $U^I$ is chosen to have the form $U^I=(1,0,0,0)^T$. This can be done without loss of generality, since covariance will be restored explicitly in the following steps. Nonetheless, it is important to keep this fact in mind, since it will turn out to be crucial for the geometrical interpretation of the Wilson graph operator in the asymptotic limit.


\subsection{Deformation and Cosmological Constant}\label{sec_lEPRL}

The Chern-Simons expectation value of the $\G_5$-graph operator, eq. \eqref{Zgamma5}, can be understood as a deformation of the above EPRL construction of the 4-simplex amplitude, by including a cosmological constant in the $BF$ theory. We propose this expectation value as a new spinfoam 4-simplex amplitude in LQG that properly includes the cosmological constant in the theory. 

In this section we repeat the construction of the last section with an extra cosmological term inserted in the Holst-twisted $BF$ action,
\begin{align}
S_{\text{H}\L BF} 
& =- \frac{1}{2}\int_\mathcal{M} \prec \left[\left(1-\frac{1}{\gamma}\star\right)B\right]\wedge  \F[\A]\succ - \frac{\Lambda}{6}\prec \lt[\left(1-\frac{1}{\gamma}\star\right)  B\rt] \wedge B\succ\;.
\label{HlBF}
\end{align}
It is obvious that when the simplicity constraint $B^{IJ}=e^I\wedge e^J$ is imposed, $S_{\text{H}\L BF}$ reduces to the Holst action of gravity with the proper cosmological constant term proportional to $\L \det(e)$ (see \autoref{app_GRconv} for our conventions). Note, that the term proportional to $\L/\g$ drops out once the simplicity constraints are imposed. However, the extra term is necessary to obtain the expected equations of motion under variations of the $B$ field; indeed, by using the fact that the operator $(1-\g^{-1}\star)$ is invertible, one obtains $\cf[\A]=\frac{\L}{3}B$, which in turn yields for simple $B=e\wedge e$ the result
\be
\cf[\A]=\frac{\L}{3}e\wedge e .
\ee 
In standard $BF$ theory, the solution of the equations of motion corresponds to a flat geometry if the connection $\A$ is viewed geometrically. However, now the solution has been deformed to correspond to constant curvature geometry. Next, we employ a methodology similar to that of the EPRL model to construct the deformed 4-simplex spinfoam amplitude. The bulk dynamics is fixed to that of $\text{H}\L BF$ within the 4-simplex, while the geometrical simplicity constraint is imposed at the quantum level to the boundary state. The construction should relate (in a certain regime) to Regge calculus with \textit{constant curvature} 4-simplices. This expectation is confirmed by the asymptotic analysis in the main body of the paper.

Consider the functional integration of the $\text{H}\L BF$ theory in a single 4-simplex $\sig$. Let $\psi_{\G_5}$ be again the $\Slc$ spin-network function on the dual 4-simplex graph $\Gamma_5$. The $\text{H}\L BF$ 4-simplex amplitude of the boundary state $\psi_{\G_5}$ can then be written in the same way as in eq. \eqref{BF_ampl},
\be
\langle\text{H}\L BF|\psi_{\G_5}\rangle = \int \DD \A\DD\Pi \;\exp\left(-\rmi S_{H\L BF}\right) \psi_{\G_5}[\A_\partial]\;.
\ee
The integration over $\Pi$ is Gaussian and can be performed straightforwardly, omitting irrelevant normalization factors it gives
\be
\langle \text{H}\L BF|\psi_s\rangle = \int \DD \A \;\exp\left\{
\frac{3\I}{4\Lambda}\int_\sig \prec  \F \wedge \left[\left(1-\frac{1}{\gamma}\star\right) \F\right] \succ
 \right\} \psi[\A_\partial]\;.
\label{HlBF_tot}
\ee

Neglecting for one moment the $1/\g$-terms, the resulting action is precisely the evaluation of the second Chern form of $\A$ on $\sig$. To regain this interesting form even in the presence of the Holst contributions, we decompose the curvature into its self-dual and anti-self-dual parts with respect to the $\star$ operator.

The self-dual and anti-self-dual parts of a Lie algebra element $X\in\sldc$ are given by
\be
X_\pm:=\frac{1}{2}(1\mp\I\star) X\;,
\label{sd_asd}
\ee
respectively. Notice that in the previous equation the imaginary unit is necessary because $\star^2=-1$; this is due to the Lorentzian spacetime signature. This means that we have to complexify $\sldc$ before decomposing it into its self- and anti-self-dual parts. The relations $\star X_\pm =\pm \I X_\pm$, which eq. \eqref{sd_asd} implies, mean that $X_\pm$ have three complex independent components each; the same number as two complex $\frak{su}(2)_\mathbb{C}$ algebras. It turns out that this is no coincidence, the self- and anti-self-dual parts of the complexified $(\sldc)_\mathbb{C}$ actually form two commuting complexified $\frak{su}(2)_\mathbb{C}$ algebras:
\be
(\sldc)_\mathbb{C}=\frak{su}(2)_\mathbb{C}^+\oplus{\frak{su}(2)_\mathbb{C}^-}\, ,
\ee
where $\pm$ label the action of $\star$ on the two complex subalgebras. Therefore, the self-dual (or the anti-self-dual) part of $(\sldc)_\mathbb{C}$ must be isomorphic to the real $\sldc$ algebra. The real $\sldc$ we started from can be regained by requiring $X_-{=}\overline{X}_+$, where the overbar stands for complex conjugation.

Two technical ingredients are needed before continuing. First, note that $<X_+,X_->=\prec X_+,X_-\succ=0$, since $P_\pm:=\frac{1}{2}(1\mp\I\star)$ are orthogonal projectors tailored to the action of the Hodge $\star$.  We also define
\be
T_\pm^i :=
\frac{1}{2}\left(J^i\pm \I K^i\right)
\ee
 to be the generators of $\frak{su}(2)^\pm_\mathbb{C}$, respectively, and check that 
\be
\prec T^i_\pm, T^j_\pm \succ = \pm \I\delta^{ij}
\quad\text{and}\quad [T^i_\pm , T^j_\pm ] = 
\epsilon^{ij}_{\phantom{ij}k} T^k_\pm .
\ee
See \autoref{app_notations} for full details on the notation and any necessary clarification. 

With this decomposition in hand, the Lagrangian density appearing in eq. \eqref{HlBF_tot} can be rewritten as
\begin{align}
\prec \F \wedge \left(1-\gamma^{-1}\star\right)\F \succ
&
= \frac{(\gamma-\I)}{\gamma} \prec F\wedge F\succ +
\frac{(\gamma+\I)}{\gamma} \prec \bF\wedge \bF\succ ,
%
\end{align}
where $F=P_+\F$ and $\bF=P_-\F$ are the self-dual and anti-self-dual parts of the curvature $\F$.  Moreover, as is well known, they are also equal to the curvature of the self-dual and anti-self-dual parts of the (real) connection $\A$, namely of $A$ and $\bA$, respectively.

The Lagrangian can be recast in terms of traces in the fundamental $\frak{su}(2)$ representation. We set $T^k_\pm :=  \tau^k$ with $\tau^k:=-\frac{i}{2}\sigma^k$ and $\{\sigma^k\}_{k=1,2,3}$ the Pauli matrices (see \autoref{app_notations}), so that
\be
\prec T^i_\pm,T^j_\pm\succ=\mp2\I\,\Tr(T^i_\pm T^j_\pm ) \equiv \mp2\I\,\Tr\left(\tau^i \tau^j  \right)=\pm\I \delta^{ij}.
\ee
Then, the Lagrangian of eq. \eqref{HlBF_tot} reads
\be
\prec \F \wedge \left(1-\gamma^{-1}\star\right)\F \succ
=-2\I \frac{(\gamma-\I)}{\gamma} \Tr\left(F \wedge F\right)  +2\I \frac{(\gamma+\I)}{\gamma} \Tr\left(\bar F \wedge \bar F\right),
\ee
and we have the second Chern form $\Tr\left(F \wedge F\right)$ appearing explicitly in the action.

According to the Chern-Weil theorem (see, e.g. \cite{Nakahara2003}), the integral of the second Chern form over the interior of a 4-simplex $\mathfrak{M}_4=\sig$ can be evaluated as the integral of the Chern-Simons form on its boundary $\partial \sig \cong S^3$
\begin{align}
\int_\sig \Tr\left( F\wedge F\right)  &=\int_{\partial \sigma\cong S^3} \Tr\left( A\wedge \D A + \frac{2}{3} A\wedge A \wedge A \right).
\end{align}
The complex-conjugated relation holds for the anti-self-dual part of the curvature.

In order to simplify notation and agree with common conventions, we introduce the holomorphic Chern-Simons functional
\be
W[A] := \frac{1}{4\pi}\int_{S^3} \Tr\left( A\wedge \D A + \frac{2}{3} A\wedge A \wedge A \right)\;.
\ee
The normalization is chosen so that $W[A]$ is SU(2)-gauge invariant modulo $2\pi\Z$, and therefore the exponential $\exp\left( \I k\, W[A] \right)$ is gauge invariant provided $k\in\mathbb{Z}$.

Finally, we rewrite the $\text{H}\L BF$ amplitude of the boundary state $\psi_{\G_5}$ on a 4-simplex in terms of $\Slc$ Chern-Simons theory
\be
\langle \text{H}\L BF|\psi_{\G_5}\rangle = \int \cd A \cd \bar{A} \;\exp\left(- \I\frac{h}{2}W[A] -\I\frac{\bar{h}}{2}W[\bA]   \right)\;\psi_{\G_5}[A,\bA]\;.
\label{HlBF_CS}
\ee
This is the Chern-Simons expectation value of a graph operator $\psi_{\G_5}[\A]=\psi_{\G_5}[A,\bar{A}]$. The complex Chern-Simons couplings $h$ and $\bar{h}$ are related to the cosmological constant $\L$ and to the Barbero-Immirzi parameter by
\be 
h=\frac{12\pi}{\L}\lt(\frac{1}{\g}+\I\rt) \qquad \text{and}\qquad \bar{h}= \frac{12\pi}{\L}\lt(\frac{1}{\g}-\I\rt).
\ee
The coupling $\bar{h}$ is the complex conjugate of $h$ provided $\gamma\in\mathbb{R}$, which we will always assume. Note that the action
\be
\text{CS}[A,\bA]=\frac{h}{2}W[A]+\frac{\bar{h}}{2}W[\bA]
\ee
which appears in eq. \eqref{HlBF_CS} is always real and equal to $\Re( h \,W[A] )$. 

A compact gauge group $\Su\times \Su$ version of eq. \eqref{HlBF_CS} was proposed by one of the authors in \cite{Han2011}, it covered the quantum deformation of a Euclidean spinfoam model. A similar proposal to eq. \eqref{HlBF_CS} also appeared in \cite{Wieland2011}, where the Wilson graph operator was different, but the Chern-Simons action had exactly the same complex weight. In that case the Chern-Simons weight was fixed by requiring that $\exp\lt(-\I \text{CS}[A,\bA] \rt)$ formally solved the Hamiltonian constraint of (Holst) general relativity expressed in the complex Ashtekar variables.\footnote{See also the works \cite{Randono2006,Randono2006a}, in turn inspired by \cite{Kodama1988,Smolin1995,Smolin2002}.} As will become clear later, this precise form of the Chern-Simons weight is also necessary for the semiclassical analysis of the amplitude to admit a clear geometrical interpretation, and so eventually explains the requirement $\frac{1+\I\g}{h}\in\R$, imposed in the subsequent analysis, from a geometrical perspective.

It is sometimes useful to split $h$ into its real and imaginary parts:
\be
h = k + \I s\;,\quad \text{where}
\quad k = \frac{12\pi}{\L\g} \;\text{and}\; s=\frac{12\pi}{\L}\,.
\ee
Since the action $\text{CS}[A,\bA]$ is gauge invariant only modulo $2\pi$, for $\exp\lt(i\, CS\rt)$ to be gauge invariant, we need
\be
k \in \mathbb{Z} \quad\text{and} \quad 
s \in \g\mathbb{Z}\;.
\ee
It is now straightforward to deform the EPRL 4-simplex amplitude to one including a cosmological constant. Indeed, following the recipe of the previous section, we only need impose the simplicity constraints {on} the boundary states $\psi_{\G_5}$. The space of possible boundary states is then reduced to the subspace of $\G_5$-graph operators $\G_5\lt(\vec{j},\vec{i}\;\big|A,\bA\rt)$ labeled by SU(2) spin-network data $\lt(\vec{j},\vec{i}\;\rt)$. Therefore, using eq. \eqref{HlBF_CS} to replace eq. \eqref{BF_ampl} and imposing the quantum simplicity constraints, we obtain the deformed EPRL 4-simplex amplitude\footnote{The papers in which this formula first appeared (basically simultaneously) are \cite{Han2011} and \cite{Wieland2011}. }
\begin{align}
Z_{\L \text{EPRL}}\lt(\sigma\big|\vec j, \vec i\;\rt)& := \langle \text{H}\L BF |\, \G_5\lt(\vec{j},\vec{i}\;\rt)\rangle \nonumber\\
& =  \int \DD A \cd\bA\;\exp\left(- \I\frac{h}{2}W[A] -\I\frac{\bar{h}}{2}W[\bA]   \right)\;\G_5\lt(\vec{j},\vec{i}\;\big|A,\bA\rt).
\label{lEPRL_implicit}
\end{align}
Note that in the last expression the connection in the bulk of the 4-simplex has disappeared, and only its values on the boundary, which is isomorphic to $S^3$, play a role (we omit the subscript $\partial$ on the connections $A,\bA$).
But, this is precisely the Chern-Simons evaluation of the $\G_5$-graph operator $Z_\text{CS}(S^3;\G_5)$ introduced in eq. \eqref{Zgamma5}, 
\be
Z_{\L \text{EPRL}}(\sig)=Z_\text{CS}(S^3;\G_5).\label{LQG-CS1}
\ee
Because Chern-Simons theory is sensitive to the crossings appearing in the projection of a graph to the plane, we need to make a choice for the knotting of the graph $\G_5$ used to define $\G_5\lt(\vec{j},\vec{i}\;\big|\,A,\bA\rt)$. We choose such a knotting as in \autoref{fig_gamma5}. This choice turns out to be well motivated geometrically, as will be explained later on. 

Importantly, the above construction and the result eq. \eqref{LQG-CS1} illustrate the relation between $\Slc$ Chern-Simons theory and 4-dimensional covariant LQG with cosmological constant at the level of a 4-simplex.  

Before continuing let us comment briefly on the requirement
\be
\frac{12\pi}{\Lambda \g} \equiv k \in \mathbb Z \,,
\ee
which has indeed a very natural interpretation. To see this let us introduce the cosmological radius of curvature $R_\Lambda := \sqrt{3/|\Lambda|}$. Then, the previous condition reads
\be
4\pi R_\Lambda^2 = \g |k| \in \g\mathbb N,
\ee
which says that the area of the cosmological horizon is quantized in units of the quantum of area. {This is also nicely consistent with the fact that in SU(2) Chern-Simons theory of level $k$ only observables with spins up to $|k|/2$ are allowed. Indeed, even if the Chern-Simons theory we are using is $\Slc$ Chern-Simons with complex level $h=k+\I s$, as we will show later on, it  reduces to an SU(2) Chern-Simons with level k close to the vertices of the graph.} Hence the previous condition just says that the cosmological horizon has (twice) the area associated with the maximal allowed spin.\footnote{Even the factor of 2 can be heuristically explained: a convex spherical triangle has always area less than $2\pi$ and at least two (``degenerate'') triangles are needed to cover the surface of a sphere.}


\section{Semiclassical and Zero-Cosmological-Constant Limits}\label{sec_largej}

Let us reintroduce physical units in the previous formulae, the goal being to distinguish the semiclassical and the zero-cosmological-constant limits. It turns out that the semiclassical limit corresponds to the double-scaling limit of $Z_\text{CS}(S^3;\G_5)$ in eq. \eqref{Zgamma5} (or equivalently of $Z_{\L EPRL}(\sig)$ in eq. \eqref{lEPRL_implicit}), which we recall consists in taking $j_{ab},h\to\infty$ uniformly and keeping the ratios $j_{ab}/h$ fixed. On the other hand, the zero-cosmological-constant limit is taken by only sending $h\to\infty$ while keeping $j_{ab}$ fixed.\footnote{ Of course, $h$ is the Chern-Simons coupling here and is not to be confused with Planck's constant; we will always use the reduced form to distinguish the latter, $\hbar$.}

We start from the $\L\mathrm{BF}$ action. In our diffeomorphism invariant treatment, we consider coordinates to be just labels, and therefore dimensionless. Dimensional units are then carried by the metric $g_{\mu\nu}$, which has units of length squared. This can be read directly from $\D s^2 = g_{\mu\nu} \D x^\mu \D x^\nu$, since $\D s$ is a physically meaningful and measurable quantity (we will keep $c=1$). From this starting point, it is most natural to assign units to the tetrad field via $g_{\mu\nu} = \eta_{IJ} e^I_\mu e^J_\nu$; therefore $e$ has units of length.

The field $B$ will eventually (i.e. after the imposition of the  simplicity constraints) turn out to be equal to $e\wedge e$, so it has the units of an area. Because we have chosen dimensionless coordinates, the connection is also dimensionless. One can see this from the formula for the covariant derivative $\mathrm{D} \cdot= \D \cdot+ [\A,\cdot]$. Consequently, the curvature $\F$ is dimensionless, too.

From this discussion, it follows that, for the $\L BF$ action of eq. \eqref{HlBF} to have the correct units, it must first be divided by a constant with units of an area, and then multiplied by a constant carrying the units of an action. To eventually recover general relativity, as well as the usual path integral prescription where the action is weighted by $\hbar$, these constants must be the squared Planck length $\ell_P^2=8\pi \hbar G_N$ and $\hbar$ respectively
\be
\frac{1}{\hbar}S_{\L BF} = -\frac{1}{2\ell_P^2} \int_{\frak M_4} \prec \lt[\left(1-\gamma^{-1}\star\right) B\rt] \wedge \F\succ - \frac{\Lambda}{6} \prec  \lt[\left(1-\gamma^{-1}\star\right) B\rt] \wedge B\succ\,.
\ee
From this formula it is also clear that $\gamma$ is dimensionless, while $\Lambda$ has units of inverse area. 

A moment of reflection shows that after the integration of the $B$ field,  $\Lambda$ being the only dimensionful quantity left, units can be restored by simply replacing $\Lambda\mapsto \ell_P^2\Lambda$. Therefore, for the dimensionless Chern-Simons coupling, the restoration of physical units gives
\begin{align}
h=\frac{12\pi}{\ell_P^2\L}\lt(\frac{1}{\g}+\I\rt).
\end{align}

The kinematics of LQG \cite{Rovelli2004,Thiemann2007,Ashtekar2004,HAN2007} predicts that $\gamma\sqrt{ j(j+1)}$ are the  eigenvalues of the quantum area operator in Planck units, where $j$ is the SU(2) irrep label entering the knotted graph operator $\G_5({j}_{ab},\xi_{ab}\,|\,A,\bA)$. Therefore $j_{ab}$ is related to the physical (dimensionful) area $\text{a}_{ab}$ by
\be
\sqrt{ j_{ab}(j_{ab}+1)} = \frac{1}{\gamma \ell_P^2}\text{a}_{ab}.
\ee

The semiclassical limit is obtained by sending $\hbar\rightarrow0$, and accordingly $\ell_P^2\rightarrow0$, while keeping the dimensionful quantities $\text{a}_{ab}$ and $\L$ invariant and finite.\footnote{The Barbero-Immirizi parameter $\g$ is fixed to be a finite constant here. See, however, \cite{Rovelli2006,Bianchi:2006uf,Bianchi2009,Bianchi2011,Zhang3poitnfunct,Magliaro2011,MAGLIARO2013,Han2014b,Han2014a,Han2013} where the scaling of $\g$ is also involved in the limit.} This limit corresponds exactly to the double-scaling limit where $j_{ab},h\to\infty$ uniformly with the ratio $j_{ab}/h$ fixed. This double-scaling provides a generalization of the spinfoam large-$j$ limit (see e.g. \cite{Barrett2010,Conrady2008,Han2012,Han2013a,Rovelli2006,Bianchi:2006uf,Bianchi2009,Bianchi2011,Zhang3poitnfunct,Barrett2003,Freidel2003}) to our deformed EPRL spinfoam amplitude including a cosmological constant.

The other physically interesting limit we consider is that of vanishing cosmological constant. It can be obtained by sending $\Lambda\rightarrow0$, holding all other quantities fixed. This limit arises for $h\rightarrow\infty$ while keeping all other quantities finite. Therefore, the zero-cosmological constant limit of this theory can be obtained by projecting the Chern-Simons sector onto its classical solutions on $\partial \mathfrak{M}_4$. But, the Chern-Simons classical equations of motion simply impose flatness of the connection $\A_\partial$ on the entirety of $\partial \mathfrak{M}_4$, much as $BF$ theory would do. Therefore, it turns out, consistently, that the $\Lambda\rightarrow0$ limit of $Z_\text{CS}(S^3;\G_5)$ is the usual EPRL spinfoam amplitude.

There is a final limit that would be interesting to consider in more detail, namely, taking the Barbero-Immirzi paramter to infinity, $\g\to\infty$. On the spinfoam side this limit reduces the EPRL graph operator to the Lorentzian Barrett-Crane one \cite{Barrett1998,barrett2000lorentzian}, while on the $\Slc$ Chern-Simons theory side it approaches the sector of the theory where a quantum-group interpretation is available. Interestingly, a quantum deformed version of the Lorentzian Barrett-Crane model has been introduced and studied in \cite{Noui2003}. It would therefore be intriguing to study the relationship between the latter model and the one we propose here. This is left for future work.


\section{Integral Representation of the Knotted Graph Operator}\label{sec_lEPRL_explicit}

In this section we begin the analysis of the asymptotic behavior of the Chern-Simons expectation value $Z_\text{CS}(S^3;\G_5\,|\,j_{ab},\xi_{ab})$ in the double-scaling limit of eq. \eqref{eq_DoubScal}. We rewrite the knotted graph operator $\G_5({j}_{ab},\xi_{ab}\,|\,A,\bA)$ of eq. \eqref{Zgamma5} or \eqref{lEPRL_implicit} as an explicit integral; this allows us to apply the stationary phase method to the asymptotic analysis. The following derivation is a generalization of the path integral asymptotics for the EPRL spinfoam amplitude \cite{Barrett2010,Conrady2008,Han2012,Han2013a,Han2014c}.

One of the key ingredients is an expression for the $\Slc$ Wigner matrices of an unitary irrep $D^{(j,\gamma j)}(A,\bar{A})$ in the coherent state basis (see \autoref{sec_CSandGraph}). The Hilbert space $\mathcal H^{(j,\rho)}$ of the $(j,\rho)$ unitary irrep can be given \cite{Ruhl1970} in terms of homogeneous functions of two complex variables $(z^0,z^1)$ with degree $(-1+\I \rho + j; -1 +\I\rho - j)$, i.e. such that for any $\o\in\mathbb{C}\setminus\{0\}$
\be
f(\o z^\alpha) = \o^{-1+\I \rho + j} \bar{\o}^{-1 +\I\rho - j} f( z^\alpha),
\label{homog}
\ee
where $z^\alpha$ is a two-component spinor ($\alpha=0,1$). Now, given a general spinor $z^\alpha$, we build the SU(2) matrix 
\be
g(z):=\frac{1}{\sqrt{\langle z,z\rangle}}\left(
\begin{array}{cc}
z^0 & -\bar{z}^1 \\
z^1 & \bar{z}^0
\end{array}
\right) \equiv \frac{1}{\sqrt{\langle z,z\rangle}}(z, Jz),
\label{u(z)}
\ee
where we have introduced $J:(z^0,z^1)^T \mapsto (-\bar{z}^1,\bar{z}^0)^T$. We also recall the expression of the Hermitian inner product on $\mathbb{C}^2$, $\langle z, w\rangle := \delta_{\dot\alpha\alpha} \bar{z}^{\dot\alpha}w^\alpha$. The restriction of the canonical basis $f^{(j,\rho)}_{l,m}(z) = \langle z\,|(j,\rho);l,m\rangle$ of $\mathcal{H}^{(j,\rho)}$ to normalized spinors ($z$ such that $\langle z,z\rangle =1$) is given by
\be
f^{(j,\rho)}_{l,m}(z) = \sqrt{\frac{2l+1}{\pi}} D^l_{m\, j}\left(u(z)\right),
\label{can_basis}
\ee
where $D^l\left(u\right)$ is the usual SU(2) Wigner matrix of $u\in \mathrm{SU(2)}$ in the spin-$l$ irrep. Evaluating $f^{(j,\rho)}_{l,m}(z)$ on an unnormalized spinor by using the homogeneity of eq. \eqref{homog}, we obtain
\be
f^{(\rho,j)}_{l,m}(z) = \sqrt{\frac{2l+1}{\pi}} \langle z,z\rangle^{\I\rho-1-l} D^l_{m\,j}\left(u(z)\right).
\ee
The action of $g \in \SLDC$ on the canonical basis $f^{(\rho,j)}_{l,m}(z)$ is given by
\begin{align}
\left(g\triangleright f^{(\rho,j)}_{l',m'}\right)(z)= f^{(\rho,j)}_{l,m}(g^T z),
\end{align}
where $g^T$ is the transpose of $g$ in the fundamental representation. The inner product of $\mathcal{H}^{(j,\rho)}$ is given by 
\be
\langle (j,\rho);l,m| (j,\rho);l',m'\rangle := \int_{\mathbb{CP}^1} \rmd\mu(z) \; \overline{f^{(\rho,j)}_{l,m}(z)} f^{(\rho,j)}_{l',m'}(z) = \delta_{j,j'} \delta_{m,m'}\;,
\ee
where $\rmd\mu(z):=\frac{\I}{2}(z_0\D z_1-z_1\D z_0)\wedge(\bar{z}_0\D \bar{z}_1-\bar{z}_1\D \bar{z}_0)$.

Recall that the coherent states $|j,\xi\rangle$ are contained in the knotted graph operator $\G_5({j}_{ab},\xi_{ab}\,|\,A,\bA)$. There is an important factorization property of these coherent states; for example, when we compute the SU(2) irrep matrix element in the coherent state basis we find, 
\be
\langle j, \xi | h | j, \xi' \rangle 
= \langle \xi, h \xi' \rangle^{2j}
\label{2j}
\ee
for any $h\in \mathrm{SU(2)}$; here $\xi,\xi'\in \mathbb{C}^2$ are understood to be normalized 2-spinors. Now, recall the injection $\Y$ of eq. \eqref{Ymap}. We would like to represent the coherent state $\Y|j,\xi\rangle\equiv |(j,\g j),j,\xi\rangle\in\ch^{(j,\g j)}$ by an homogeneous function of two complex variables. By eq. \eqref{can_basis}, we can write explicitly the highest weight state $\Y|j,j\rangle\equiv |(j,\g j),j,j\rangle$
\be
f^{j}_j(z)^{(j,\g j)}=\sqrt{\frac{\dim(j)}{\pi}}\lag z,z\rag^{i\g j-1-j}(z^0)^{2j}.
\ee
Therefore, by definition the coherent state $\Y|j,\xi\rangle\equiv |(j,\g j),j,\xi\rangle$ can be represented using
\be
\lt|(j,\g j);j,\xi\rag \equiv f^{j}_\xi(z)^{(j,\g j)}=f^{j}_j\Big(g(\xi)^tz\Big)^{(j,\g j)}=\sqrt{\frac{\dim(j)}{\pi}}\lag z,z\rag^{i\g j-1-j}\lag\bar{z},\xi \rag^{2j}.
\ee

Let us introduce notation to describe the quantities living on the $\Gamma_5$-graph of \autoref{fig_gamma5}. Graph vertices are labeled by $a,b,\dots\in\{1,\dots,5\}$, and hence, edges are labeled by unordered couples of indices $(ab)$. Further, we fix orientation: for all pairs $(a,b)$ with $a<b$ orient the edge from $b$ to $a$. The $\Slc$ holonomy along the edge $(ab)$, from $b$ to $a$, is then denoted $G_{ab}$; this makes it natural to introduce the convention $G_{ab}^{-1}=G_{ba}$. The spins for each edge are $j_{ab}=j_{ba}$. There are two (different) normalized spinors sitting at the two endpoints of edge $(ab)$, $\xi_{ab}$ and $\xi_{ba}$, respectively at vertex $a$ and $b$.\footnote{The idea behind these conventions is that $G_{ab}$ can act on $\xi_{ba}$ because the index $b$ appears in sequence; this guarantees that both quantities are defined in the same reference frame. Similarly $G_{ab}G_{bc}$ is a valid expression, while e.g. $G_{ab}G_{ac}$ or $G_{ab}\xi_{ab}$ are not.}

The knotted graph operator $\G_5({j}_{ab},\xi_{ab}\,|\,A,\bA)$ can then be written (see \cite{Barrett2010} for details)
\be
\G_5({j}_{ab},\xi_{ab}\,|\,A,\bA)=\int_{\Slc}\prod_{a=1}^5\rmd g_a\prod_{a<b}\langle j_{ab}, J \xi_{ab} | Y^\dagger g_a^{-1}G_{ab} g_b Y| j_{ab} , \xi_{ba}\rangle,
\ee 
where $G_{ab}=G_{ab}[A,\bA]$ denotes the holonomy of the Chern-Simons connection $(A,\bA)$, and $\rmd g_a$ is the Haar measure on $\Slc$. Each factor can be conveniently recast  using the above expression for $\Y|j,\xi\rangle$, yielding
\begin{align}
\langle j, J\xi | Y^\dagger g Y| j, \xi '\rangle_j 
& = \int_{\mathbb{CP}^1} \rmd\mu( z) \; \overline{f^{(\gamma j, j)}_{j,j}\left(\xi^T z\right)}f^{(\gamma j, j)}_{j,j}\left((g\xi')^T z\right) \nonumber \\
& = \frac{2j+1}{\pi} \int_{\mathbb{CP}^1} \rmd\mu_{g} (z) \; \exp\left[ S_w(z,g,\xi,\xi',j) \right],
\label{EPRL_wedge}
\end{align}
an expression valid for all $g\in\Slc$. For ease of notation, we have introduced the scale invariant measure on $\mathbb{CP}^1$
\be
\rmd\mu_{g} (z) := \frac{\rmd\mu( z)}{   \langle\bar{z}, \bar z \rangle     \langle g^\dagger \bar{z}, g^\dagger \bar z \rangle}\,,
\ee
and the ``spinfoam wedge action''
\be
S_w(z,g,\xi,\xi',j) := 2j \ln \frac{  \langle J \xi, \bar{z}\rangle   \langle g^\dagger \bar{z},\xi' \rangle }{ \langle \bar{z}, \bar z \rangle^{1/2}   \langle g^\dagger \bar{z}, g^\dagger \bar z \rangle^{1/2}}
+ 2\I\gamma j \ln \frac{\langle g^\dagger \bar{z}, g^\dagger \bar z \rangle^{1/2} }{\langle \bar{z}, \bar z \rangle^{1/2} }.
\label{wedge_action}
\ee
Hence, we now have the following integral representation of the $\G_5$ knotted graph operator,
\begin{align}
\G_5({j}_{ab},\xi_{ab}\,|\,A,\bA)=\prod_{a<b}\left(\frac{2j_{ab}+1}{\pi}\right)\int_\SLDC  \prod_a \rmd g_a \int_{\mathbb{CP}^1}  \prod_{a<b} \rmd\mu_{g_a G_{ab}g_b^{-1}}( z_{ab}) \;\exp(S_{\Gamma_5})
\label{fgamma_explicit}
\end{align}
for which we introduce the $\G_5$-graph spinfoam action
\begin{align}
{S}_{\Gamma_5} & = \sum_{a<b} 2j_{ab} \ln \frac{  \langle J \xi_{ab}, \bar{z}_{ab} \rangle   \langle g_b^\dagger G_{ab}^\dagger (g_a^\dagger)^{-1}\bar{z}_{ab},\xi_{ba} \rangle }{ \langle \bar{z}_{ab}, \bar{z}_{ab} \rangle^{1/2}    \langle g_b^\dagger G_{ab}^\dagger (g_a^\dagger)^{-1} \bar{z}_{ab}, g_b^\dagger G_{ab}^\dagger (g_a^\dagger)^{-1} \bar{z}_{ab} \rangle^{1/2} }+ \notag\\
&\hspace{4cm}+2\I\gamma j_{ab}\ln \frac{ \langle g_b^\dagger G_{ab}^\dagger (g_a^\dagger)^{-1}\bar{z}_{ab}, g_b^\dagger G_{ab}^\dagger (g_a^\dagger)^{-1} \bar{z}_{ab} \rangle^{1/2} }{ \langle  \bar{z}_{ab},  \bar{z}_{ab} \rangle^{1/2} }\;. 
\label{SG5}
\end{align}
Note that $S_{\Gamma_5}$ is \textit{neither} a holomorphic \textit{nor} an anti-holomorphic function of the Chern-Simons connection. This follows from the fact it is derived using the unitary irreps of $\Slc$.

To shorten the formulae, we group all the measure factors in eq. \eqref{fgamma_explicit} into 
\be
\D\Omega_{g,z} = \prod_{a<b}\left(\frac{2j_{ab}+1}{\pi}\right) \prod_a \rmd g_a \prod_{a<b} \rmd\mu_{g_a G_{ab}g_b^{-1}}( z_{ab}).
\ee

Finally, we obtain the Chern-Simons expectation value of the $\G_5$ knotted-graph operator expressed in the path-integral form
\be
Z_\text{CS}(S^3;\G_5\,|\,j_{ab},\xi_{ab}) = \int \DD A\cd\bA \int \D\Omega_{g,z} \;\exp\left( 
I_{\Gamma_5} [j_{ab},A,\bA,g_a,z_{ab},\xi_{ab}]
\right)\;,
\label{lEPRL_VA}
\ee
where the total action $I_{\G_5}$ is
\be
I_{\Gamma_5}  [j_{ab},A,\bA,g_a,z_{ab},\xi_{ab}] := - \I\frac{h}{2} W[A] - \I\frac{\bar{h}}{2} W [\bA] + S_{\Gamma_5}[j_{ab},A,\bA, g_a,z_{ab},\xi_{ab}].
\ee

The action $I_{\G_5}$ is invariant (modulo $2\pi \Z$) under local $\Slc$ gauge transformations of the Chern-Simons theory at any point $x\in S^3$. In particular, whenever $x_a\in S^3$ is the position of the vertex $a$ of $\Gamma_5$, one finds that the $g_a$ and $G_{ab}$ are modified
\be
g_a\mapsto G(x_a) g_a, \qquad \text{and} \qquad G_{ab} \mapsto G(x_a) G_{ab} G(x_b)^{-1}.
\ee 
We can use this gauge freedom at the vertices of the graph to set all the $g_a$ to the identity. Through this gauge fixing, the (infinite) integral $\int\prod_a\rmd g_a$ drops out of $\rmd\O_{g,z}$. We will keep referring to the gauge-fixed spinfoam and total actions with the same letters $S_{\G_5}$ and $I_{\G_5}$,
\be
I_{\G_5}&=& -\I \text{CS}[S^3\,|\, A,\bA]+S_{\Gamma_5}[j_{ab},A,\bA,z_{ab},\xi_{ab}]\nonumber\\
& =& -\I \frac{ h}{2} W[A] -\I \frac{\bar{h}}{2} W[\bA] + \sum_{(ab), a>b} 2 j_{ab} \ln \frac{  \langle J \xi_{ab}, \bar{z}_{ab} \rangle   \langle  G_{ab}^\dagger \bar{z}_{ab},\xi_{ba} \rangle }{ \langle \bar{z}_{ab}, \bar{z}_{ab} \rangle^{1/2}    \langle  G_{ab}^\dagger  \bar{z}_{ab},  G_{ab}^\dagger  \bar{z}_{ab} \rangle^{1/2} }
+2\I\gamma j_{ab}\ln \frac{ \langle  G_{ab}^\dagger \bar{z}_{ab},  G_{ab}^\dagger  \bar{z}_{ab} \rangle^{1/2} }{ \langle  \bar{z}_{ab},  \bar{z}_{ab} \rangle^{1/2} }\;.\notag\\
\label{IG5}
\ee

Still there remains an SU(2) gauge symmetry of $\E^{I_{\G_5}}$ 
\be
G_{ab}\mapsto h_a G_{ab} h_b^{-1},\ \ \ \ \xi_{ab}\mapsto h_a \xi_{ab},\quad \text{and} \quad  z_{ab}\mapsto h_a z_{ab},\ \ \ \ \forall\ h_a\in\mathrm{SU(2)},\label{SU(2)gauge}
\ee
as well as a scaling gauge symmetry of the $z_{ab}$
\be
z_{ab}\mapsto \kappa z_{ab},\ \ \ \ \forall\ \kappa\in\C\setminus\{0\}.\label{scaling gauge}
\ee 
It is practical to use this last symmetry to fix the norm of the $z_{ab}$ to unity.

In the zero-cosmological-constant limit $h\to\infty$, the connections $A,\bA$ become trivial on $S^3$. Then $G_{ab}$ is purely gauge $G_{ab}=g_a^{-1}g_b$, and $I_{\G_5}$ reduces to the usual spinfoam action in \cite{Barrett2010} by a change of variables $\bar{z}_{ab}\mapsto g_a^\dagger z_{ab}$.

Another interesting property of the knotted graph operator we have just described, that is $\int \D \Omega \exp S_{\Gamma_5}$, is that it is essentially invariant under the reversal of any of its edges. Indeed, under this operation the graph operator acquires only a sign $(-1)^{2j_{ab}}$. This is a nontrivial fact in the present formulation. However, it is true by construction in the formulation of \cite{Barrett2010}.

Importantly the total action $I_{\G_5}$ is linear in both the spin $j_{ab}$ and the Chern-Simons couplings $h$ and $\bar{h}$. The double-scaling limit can be conveniently carried out by uniformly rescaling $j_{ab}\to\l j_{ab}$ and $h\to\l h$, and sending $\l\to\infty$.\footnote{Note that $\lambda\to\infty$ here is just a dimensionless way to speak about the $\ell_P^{2}\to0$ limit of the previous section.} The total action scales as $I_{\G_5}\to\l I_{\G_5}$. Thus the asymptotic behavior of the Chern-Simons expectation value $Z_\text{CS}(S^3;\G_5\,|\,j_{ab},\xi_{ab})$ can be studied using stationary phase methods. 

For ease of notation, in what follows we will drop the subscript $\Gamma_5$ from the action functionals.


\section{Stationary Phase Analysis}\label{sec_SPA}

The stationary phase method studies the asymptotic behavior of the following type of integral as $\l\to\infty$ (Theorem 7.7.5 in \cite{Hormander1998})
\be
f(\l)=\int\rmd x\ r(x)\ \E^{\l S(x)}\label{flambda},
\ee
where $S(x)$ and $r(x)$ are smooth, complex valued functions, and $\mathrm{Re}S\leq 0$. For large parameter $\l$ the dominant contributions to the integral come from the \textit{critical points} $x_c$ of $S(x)$ that satisfy $\Re S(x_c)=0$. The asymptotic behavior of the above integral for large $\l$ is then given by
\be
f(\l)=\sum_{x_c}\lt(\frac{2\pi}{\l}\rt)^{\frac{\text{rnk}(x_c)}{2}}\frac{\E^{i\mathrm{Ind}H'(x_c)}}{\sqrt{|\det H'(x_c)|}}r(x_c)\E^{\l S(x_c)}\lt[1+\text{O}\lt(\frac{1}{\l}\rt)\rt]
\ee
for isolated critical points $x_c$. Here $\text{rnk}(x_c)$ is the rank of the Hessian matrix $H_{ij}(x_c)=\partial_i\partial_j S(x_c)$ at the critical point $x_c$ and $H'(x_c)$ is the invertible restriction onto $\mathrm{ker}H(x_c)^\perp$.
 Finally $\mathrm{Ind}H'(x_c)$ is a Maslov index, generalizing the $\pi/4$ one finds in the standard stationary phase analysis of a one variable function (see \cite{Esterlis2014} for a recent discussion of Maslov indices and their computation). If $S(x)$ does not have any critical points, $f(\l)$ decreases faster than any power of $\l^{-1}$. 

In the last section the Chern-Simons expectation value $Z_\text{CS}(S^3;\G_5\,|\,j_{ab},\xi_{ab})$ was put in a form adapted to the stationary phase analysis. The asymptotic behavior in the doubling scaling limit $j,h\to\infty$ and $j/h$ fixed, (or $\l\to\infty$) is obtained, at the leading order, by finding all the critical points of the action and evaluating the integrand $\E^{\l I_{\G_5}}$ at each critical point. (Though, we do not attempt to calculate the scaling of the ``amplitude determinant'' associated to the determinant of the reduced Hessian.) In $Z_\text{CS}(S^3;\G_5\,|\,j_{ab},\xi_{ab})$ there are two types of integration variables, $(A,\bA)$ and $z_{ab}$. The critical equations are given by the variational principle with respect to these variables $\delta_{A,\bA}I=\delta_{z_{ab}}I=0$, and the requirement that the real part of the action $\Re(I)$ is at its maximum (which will shortly be shown to be zero). The data $j_{ab}$ and $\xi_{ab}$ are not involved in the integral, so they are consider fixed, external data, or from the LQG point of view, the boundary data of this (basic) spinfoam amplitude.

\subsection{Real Part of the Action}

The total action $I_{\G_5}$ is generally a complex number. Nonetheless, its Chern-Simons  part is equal to $\I\Re(h W[A])$ and is therefore purely imaginary. The only real contributions come, therefore, from the knotted graph operator $S$. A quick inspection of $S$ in eq. \eqref{SG5} shows that the only real contribution comes from the first term of this equation and is equal to
\be
\Re(I)=\Re(S) = \sum_{a<b} j_{ab} \ln \frac{  |\langle J\xi_{ab}, \bar{z}_{ab} \rangle|^2  | \langle G_{ab}^\dagger \bar{z}_{ab},\xi_{ba} \rangle|^2 }{ \langle \bar{z}_{ab},  \bar{z}_{ab} \rangle   \langle G_{ab}^\dagger\bar{z}_{ab}, G_{ab}^\dagger \bar{z}_{ab} \rangle}\;.
\ee
The Cauchy-Schwarz inequality, together with the normalization of the spinors $\xi_{ab}$, proves that $\Re(I_{\G_5}) \leq 0$. Therefore, the critical equation $\Re(I_{\G_5}) \stackrel{!}{=}0$ gives
\be
J\xi_{ab} \stackrel{!}{=} \alpha^a_b 
 \bar{z}_{ab} 
 \qquad\text{and}\qquad 
 \xi_{ba} \stackrel{!}{=} \alpha^b_a 
 G_{ab}^\dagger \bar{z}_{ab}\;.
\label{xi_propto_gz}
\ee
for some complex numbers $\alpha_{ab}, \alpha_{ba} \in \mathbb C$. The above equations imply relations among the spinors $\xi_{ab}$. Using $JgJ^{-1} = {g^\dagger}^{-1}$ $\forall g\in\SLDC$, along with $J^{-1} = -J$, one finds
\be
 \xi_{ab} \stackrel{!}{=} - \E^{-\psi_{ab}-\I \varphi_{ab}}
  G_{ab}  (J\xi_{ba})\;.
\label{Jxi_Jxi}
\ee
where $ \E^{-\psi_{ab}} := |\alpha^a_b / \alpha^b_a| \in \R^+$ and $\varphi_{ab} := \text{arg}( \alpha^a_b / \alpha^b_a)\in[0,2\pi)$. 

\subsection{Variation of the $\mathbb{CP}^1$ Variables $z_{ab}$}

Next, we consider variations with respect to the $\mathbb{CP}^1$ variables ${z}_{ab}$. For definiteness, we fix the scaling symmetry of eq. \eqref{scaling gauge} by choosing the section of $\mathbb{CP}^1$ given by normalized spinors $\langle z_{ab}, z_{ab} \rangle =1$. A general variation of $z\in\mathbb{C}^2$ is given by $\delta z= \omega z + \epsilon (Jz)$, with $\epsilon,\omega\in\mathbb{C}$.  Since we work at linear order, the $\epsilon$- and $\omega$-variations are independent and do not influence each other. The rescaling of the $z_{ab}$ variables has been gauge-fixed, so the $\omega$-variation is not allowed, and can be discarded. Thus, we need only consider the $\epsilon$-variation. A short calculation, in analogy with \cite{Han2012,Han2013a}, shows that $\delta_{z_{ab}}I_{\G_5}\stackrel{!}{=}0$, on the $\Re(I_{\G_5})=0$ hypersurface, translates into the requirement
\be
 J \xi_{ab}\stackrel{!}{=} \E^{\psi_{ab}+\I \varphi_{ab}} 
  G_{ab} \xi_{ba}\;.
\label{xi_xi}
\ee 
The proportionality constants of eq. \eqref{Jxi_Jxi} and \eqref{xi_xi} are inverses of one another. This fact is a consequence of the orthonormality and completeness of the two bases $\{\xi_{ab},J\xi_{ab}\}$ and $\{\xi_{ba},J\xi_{ba}\}$, as well as the fact that $G_{ab}$ has unit determinant.


\subsection{Variation of Chern-Simons Connection}

It is well known that the variational principle of Chern-Simons theory gives $F(A)\stackrel{!}{=}0\stackrel{!}{=}F(\bA)$, i.e. that $A$ and $\bA$ are flat connections on the 3-manifold $\Fm_3$ on which the theory is defined. In the presence of Wilson-lines, e.g. Wilson-loops and knotted graph operators, the variations with respect to $A$ and $\bA$ give flat connections on the graph complement.\footnote{Given a general, knotted graph embedded in $\Fm^3$ it can be thickened to a region including a tubular neighborhood of the graph; the graph complement 3-manifold is obtained by removing this tubular neighborhood from $\Fm_3$.} The critical equation we obtain here is then
\be
F(A)\stackrel{!}{=}0\stackrel{!}{=}F(\bA)\ \ \ \ \text{on}\ \ \ \ M_3:=S^3\setminus\G_5,\label{F=0}
\ee
where $M_3$ is the $\G_5$-graph complement on $S^3$. 

When Wilson-lines are included in the theory, the on-shell Chern-Simons connection usually gives a singular curvature on the Wilson-lines. As an equivalent description of the same fact, the flat connection $(A,\bA)$ on the graph complement $M_3$ is nontrivial, since $M_3$ has nontrivial topology. There is a nontrivial holonomy along the transverse cycles that go around each tube  surrounding a Wilson-line. This fact can be viewed as an analog of the Aharonov-Bohm effect \cite{Bianchi2014}. The holonomy around each tube can be thought of as the boundary data for the equation of motion eq. \eqref{F=0}, implemented on $\partial M_3$, which is a genus-6 Riemann surface (see \autoref{ribbon}). These boundary data, for the flat connection, are determined by the knotted-graph operator. They are derived from the variational principle of the coupled system $I$ including both the Chern-Simons action and the contribution $S$ of the knotted-graph operator.  

The above argument is implemented concretely in our context by the following derivation. To begin with we calculate the variation of the Chern-Simons part of the action with respect to the connection, yielding\footnote{In \autoref{app_variations} the reader can find the main calculations of this section spelled out in some detail.}
\be
\frac{\delta W[A]}{\delta A^i_\mu(x)} = -\frac{1}{8\pi}\epsilon^{\mu\rho\sigma}F_{\rho\sigma}^i[A](x)\;.
\ee

For what concerns the knotted-graph-operator part of the action $S$, its dependence on the connection $A:=A^i_\mu \tau_i \D x^\mu$ is limited to the holonomies $G_{ab}$,
\be
G_{ab}[A] := \mathbb P\exp \int_{\ell_{ab}} A .
\ee
Hence, we first calculate $\delta G_{ab}[A]/\delta A_\mu^i(x)$ 
\begin{align}
\frac{\delta G_{ab}[A]}{\delta A_\mu^i(x)} = \left( \int_0^1 \delta^{(3)}(x-\ell(s)) \frac{\D \ell_{ab}^\mu}{\D s} \D s\right) G_{a, s_{ab}} \tau_i G_{s_{ab},b}\;,
\end{align}
with $s_{ab}$ now understood to be the (supposedly) unique solution of the condition given by the delta-function argument, i.e. $\ell_{ab}(s_{ab})=x$. We will often write the two-dimensional distribution appearing in this equation symbolically as
\be
\delta_{\ell}^{(2)\;\mu}(x):=  \int_0^1 \delta^{(3)}(x-\ell(s)) \frac{\D \ell^\mu}{\D s} \D s \;.
\ee
For the variation of the hermitian conjugate holonomy $G_\ell^\dagger$ with respect to  $\bA$, we find
\be
\frac{\delta (G_{ab})^\dagger[\bA]}{\delta \bA_\mu^i(x)} = \left( \int_0^1 \delta^{(3)}(x-\ell_{ab}^{-1}(s)) \frac{\D (\ell^{-1}_{ab})^\mu}{\D s} \D s\right) (G_{s_{ab},b})^\dagger \tau_i  (G_{a, s_{ab}})^\dagger\;.
\ee
Note that $\delta G^\dagger/ \delta A = \delta G / \delta \bA=0$, where $A$ and $\bA$ are considered independent.

Using the previous equations, we can compute the variation of the total action $I_{\G_5}$ with respect to $A$
\begin{align}
\left(\frac{\delta I}{\delta A_\mu^i(x)}\right)_{\Re(I)=0} 
&=+\I\frac{h}{16\pi} \epsilon^{\mu\rho\sigma}F^i_{\rho\sigma}(x) +(1+\I\gamma) \sum_{(ab),a>b} j_{ab} \;\langle \xi_{ba} , \left[(G_{s_{ab},b})^{-1} \tau^i G_{s_{ab},b} \right] \xi_{ba}\rangle\;
\delta_{\ell_{ab}}^{(2)\;\mu}(x)\;.
\end{align}
Once again in our notation $G_{s,b}=G_{b,s}^{-1}$. Note that eq. \eqref{xi_propto_gz} has been used to simplify the final expression. Similarly, we find
\begin{align}\displaybreak[0]
\left(\frac{\delta I}{\delta \bA_\mu^i(x)}\right)_{\Re(I)=0} 
&=+\I\frac{\bar h}{16\pi} \epsilon^{\mu\rho\sigma}\bF^i_{\rho\sigma}(x) -(1-\I\gamma) \sum_{(ab),a>b} j_{ab} \;\langle \left[(G_{s_{ab},b})^{-1} \tau^i G_{s_{ab},b} \right] \xi_{ba} , \xi_{ba}\rangle\;
\delta_{\ell_{ab}}^{(2)\;\mu}(x)\;.
\end{align}
Comparison of the two variations, shows that they are minus the complex conjugate of one another, once $\bA$ is taken to be the complex conjugate of $A$:
\be
\left(\frac{\delta I}{\delta \bA_\mu^i(x)}\right)_{\Re(I)=0} = -\;\text{c.c.}{\left(\frac{\delta I}{\delta A_\mu^i(x)}\right)_{\Re(I)=0}}\; .
\ee

Finally we find the following on-shell expression for the curvature $F$ as a distribution on $S^3$ (the critical equation for $\bar{F}$ is obtained by complex conjugation)
\begin{align}
\frac{ \I h}{16\pi} \epsilon^{\mu\rho\sigma}F^i_{\rho\sigma}(x) \stackrel{!}{=} - (1+\I\gamma) \sum_{(ab),a>b} j_{ab} \;\langle \xi_{ba} , \left[(G_{s_{ab},b})^{-1} \tau^i G_{s_{ab},b} \right] \xi_{ba}\rangle\;\delta_{\ell_{ab}}^{(2)\;\mu}(x)
 \label{transv_curv}
\end{align}
which is singular on the $\G_5$-graph and vanishes on the graph complement $M_3=S^3\setminus\G_5$. This distributional curvature results in nontrivial holonomies $H_{ab},\bar{H}_{ab}$ along the non-contractible cycles transverse to each edge $\ell_{ab}$ of $\G_5$. Their definition and calculation is part of the next section.


\section{Flat Connections on Graph Complements}\label{sec_flatgraphcomplement}

In this section, we want to recast eq. \eqref{transv_curv} expressing its information content in terms of holonomies. To do this, we introduce the graph complement $M_3 = S^3\setminus \Gamma_5$. The complement is obtained by removing an (infinitesimally) \textit{thickened} graph $\G_5$ from $S^3$. Geometrically this corresponds to removing a solid cylinder for each edge and a 3-ball for each vertex of $\G_5$; this leaves a set of hollow tubes and spheres in $S^3$ that make up an inner boundary of $M_3$.

Within the graph complement $M_3$ there are two different types of holonomies: those transverse to one of the tubes, call them $H_{ab}$, and the longitudinal holonomies along the tubes, call them $G_{ab}$. Note that the transverse holonomies $H_{ab}$ are non-contractible loops of $M_3$, and therefore can take nontrivial values. This is the case here, since according to eq. \eqref{transv_curv}, they do acquire a non-zero contribution from the presence of distributional curvature along $\G_5$ itself. 

Before delving into explicit calculations, we need to define the transverse and longitudinal holonomies precisely, that is, we need to fix the set of paths along which they are calculated. This is crucial because the paths defining the longitudinal holonomies cannot run naively along the graph defect, but need to be (infinitesimally) displaced from it and a precise prescription is needed. We shall see that the longitudinal holonomies must satisfy constraints that heavily depend on the specific graph one studies, and in particular on the number and nature of its crossings.\footnote{The graph crossings are a property of the projection of the graph onto the plane, and  not a direct property of the graph itself. However, we will see that such a projection naturally fixes a set of paths for the longitudinal holonomies, and---more intrinsically---it is these paths that are sensitive to the topology of the graph and that satisfy relations intuitively associated with the
 crossing. 
 \label{fnote_xing}} For this reason, we first briefly go one step back and justify our particular choice of 
 graph.

\begin{figure}[t]
\centering
\includegraphics[height=4.5cm]{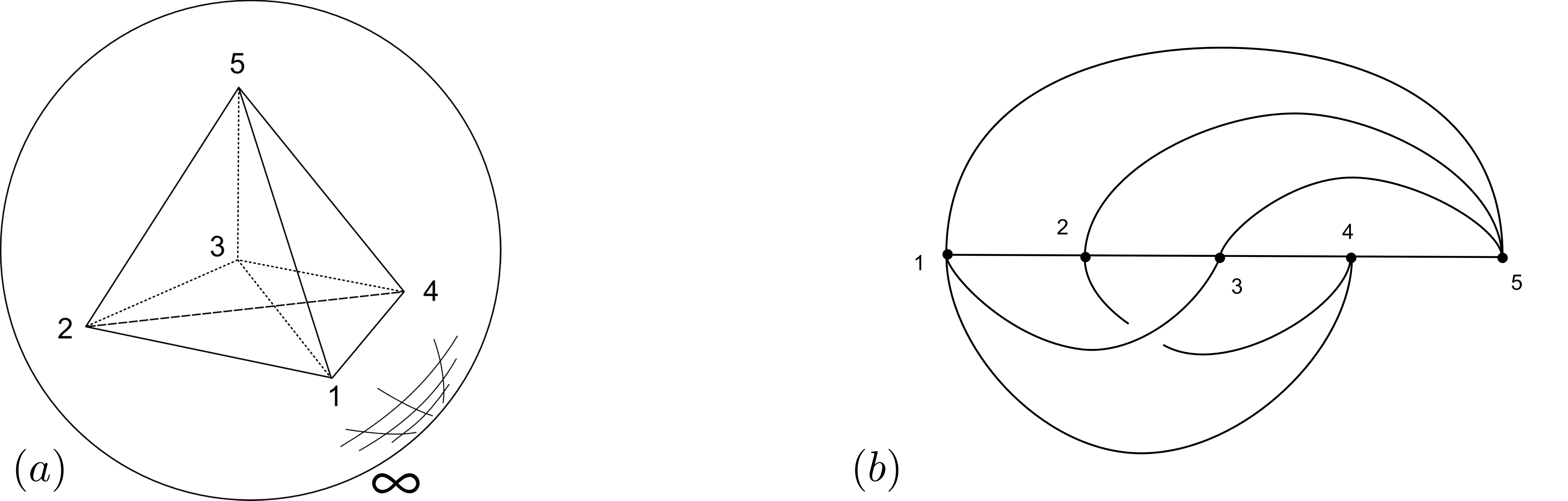}
\caption{(a) The stereographic projection of $\Gamma^5 \subset S^3$ to $\R^3$. The point from which the projection has been performed is mapped onto the 2-sphere at infinity. Point 3 is visually in the interior of the tetrahedron (1245). However, this picture should be more precisely thought of as a triangulation of the whole $\R^3\cup\{\infty\}\simeq S^3$; therefore, the interior of tetrahedron (1245) is actually what appears to be its exterior in the picture. Because the stereographic projection has been performed from the interior of this tetrahedron, it is consequently ``blown up'' to infinity. (b) The graph $\G_5$.}
\label{fig_4simplexproj}
\end{figure}

We have chosen the specific prescription for $\G_5$ that corresponds to the two-dimensional projection of the 1-skeleton of the dual to the boundary of a 4-simplex triangulating a 3-sphere. This is because, as will be clear shortly, we eventually identify the vertices of the graph with the five tetrahedra on the boundary of the 4-simplex, and the ten edges with its ten triangular faces. To understand the graph structure, first note that the dual of a 4-simplex is again a 4-simplex. Thus, to understand how $\G_5$ comes about we just need to see how the 1-skeleton of a 4-simplex projects into the plane $\R^2$. The first step is to project the 3-sphere containing the 1-skeleton of the 4-simplex onto $\mathbb R^3$. This is easily done by stereographic projection from a general point of the 3-sphere (i.e. not one which belongs to the 4-simplex skeleton). The result of this is depicted in \autoref{fig_4simplexproj}~(a), which represents a tetrahedron with an extra vertex on the inside, this vertex is then connected to the four other vertices.\footnote{To visualize that this is the result of the stereographic projection, one can proceed as follows. Qualitatively, the stereographic projection can be understood as the identification of a point of the 3-sphere with the two sphere at infinity in $\mathbb R^3$. In this sense the vertex on the inside of the tetrahedron of \autoref{fig_4simplexproj}~(a) is just a regular vertex of the 4-simplex connected to the other four vertices, and similarly the four tetrahedra sharing this vertex correspond to the four tetrahedra on the boundary of the original 4-simplex. The last tetrahedron is actually the \textit{exterior} of the outer tetrahedron of \autoref{fig_4simplexproj}~(a). That is, the point from which the stereographic projection proceeded was in the interior of this tetrahedron.} From this picture it is not hard to see that there is a way to project onto the plane such that this is only one crossing---this last crossing is impossible to eliminate. The result is the graph $\G_5$ shown again in \autoref{fig_4simplexproj}~(b).

\begin{figure}[t]
\begin{center}
\includegraphics[width=.9\textwidth]{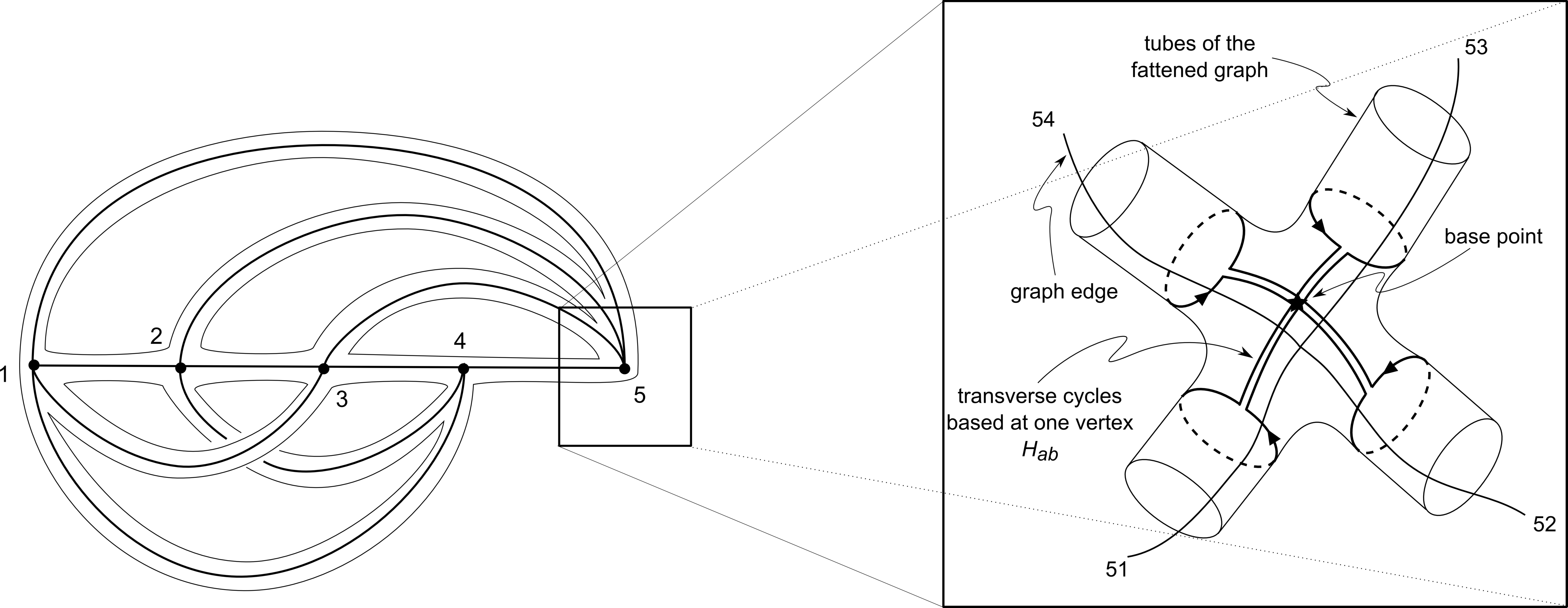} 
\caption{A top view of the fattened $\G_5$-graph. The two-dimensional boundary $\partial M_3$ of the graph complement $M_3$ is a genus-6 Riemann surface. The left panel depicts the longitudinal holonomies running along the tops of the tubes. The right panel is a zoomed in inset of vertex 5 and shows the structure of the transverse paths.} 
\label{ribbon}
\end{center}
\end{figure}

The next step consists in slightly thickening the graph, and choosing a set of paths running along the exterior of the tubes along which one will eventually calculate the transverse ($H_{ab}$) and longitudinal ($G_{ab}$) holonomies. This amounts to a choice of graph framing and is essential to the definition of the longitudinal holonomies.\footnote{See e.g. \cite{Baez1994} for an introduction to these concepts.} The simplest choice is the blackboard framing, where the paths are picked to run along the top of the tubes, as in the left panel of \autoref{ribbon}. The transverse path used to calculated  the transverse holonomy at the vertex $a$ around the edge $(ab)$, named $H_{ab}(a)$, is constructed as follows. First of all, it is based at vertex $a$, the point where the longitudinal paths meet; then, it follows the longitudinal path towards vertex $b$ an infinitesimal amount and winds once around the tube $(ab)$ in a right-handed sense with respect to the outgoing direction from the vertex; finally, it goes back to the base point again along a piece of the longitudinal path. We are now ready to express the equations of motion of the connection (eq. \eqref{transv_curv}) in terms of the holonomies $G_{ab}$ and $H_{ab}(a)$.

\paragraph*{Longitudinal holonomies} Equation \eqref{transv_curv} states that the curvature of $(A,\bA)$ vanishes on the graph complement manifold $M_3 = S^3 \setminus \G_5$. This means that all the holonomies defined along paths contractible within $M_3$, must be trivial. The longitudinal paths are generated by the following six independent cycles: (125), (235), (345), (124), (123), and (234). All these cycles, except (234), are homotopically trivial in the $\mathbb R^3$ complement of the fattened graph.\footnote{Imagine pulling these paths away from the graph from above; then they can be continuously deformed into trivial loops.} This means that they are trivial also in $M_3$. Hence, introducing the natural notation $G_{ba} := G_{ab}^{-1}$, we have
\be
G_{ab}G_{bc}G_{ca} = \mathbb{I} \qquad\text{for} \quad a,b,c \;\text{such that } \{a,b,c\}\neq\{2,3,4\} \;\text{and}\; a\neq b \neq c\,.
\label{A2_BIS_1}
\ee
The path associated to the cycle (234), however, cannot be contracted without winding around some of the tubes. In particular, within $M_3$ this path can be shrunk around the tube $(13)$. Taking vertex 3 as the base point for this cycle, we then immediately obtain the relation
\be
G_{34}G_{42} G_{23} = H_{13}(3)\,.
\label{A2_BIS_2}
\ee

The previous two equations can equivalently be expressed as the requirement that there exists a set of $\{g_a\}_{a=1,\dots,5}\in\Slc$ such that the longitudinal holonomies $G_{ab}$ are
\be
G_{ab}=g_a^{-1} g_b\,, \qquad\text{except}\qquad G_{42} = g_4^{-1}\lt[ g_3 H_{13}(3) g_3^{-1}\rt] g_2\,.
\label{Gab_BIS}
\ee

Note that these equations pick out a preferred couple of edges (the pair that crosses), from a configuration originally symmetric in all the edges. In a sense, this is just a matter of representation and eventually we will see that it is due to our choice of framing (see \autoref{fnote_xing}). 

\paragraph*{Transverse holonomies} The paths transverse to the tubes enclose the curvature singularity located on the graph edge. Thus, they acquire a nontrivial value:
\begin{align}
H_{ab}(b) &= \mathbb P\exp \int_{{\frak D}_{ab}} F(a) \notag\\
& = \exp \left[-\frac{8\pi}{\I h}\left( \frac{1}{\g} + \I \right) \g j_{ab} \lt\langle \xi_{ba}, \tau^j\xi_{ba} \right\rangle \tau_j \right] \notag\\
&= \exp \left[+\frac{4\pi}{h}\left( \frac{1}{\g} + \I \right) \g j_{ab} \lt\langle \xi_{ab}, \sigma^j \xi_{ab} \right\rangle \tau_j \right] \notag\\
&= \exp \left[+\frac{4\pi}{h}\left( \frac{1}{\g} + \I \right) \g j_{ab} \hat n_{ba}^j \tau_j \right],
\label{eq_7.4}
\end{align}
where the first equality is a statement of the non-Abelian Stokes theorem \cite{Fishbane1981,Broda2004}, and the transverse surface ${\frak D}_{ab}$, bounded by the loop around the edge $(ab)$, is taken infinitesimally close to the node; the second equality follows from eq. \eqref{transv_curv}, which says that the only non-trivial contribution to $H_{ab}(b)$ comes from the singularity at the graph itself (we evaluate this expression at $s=0$); 
and the last equality is just a consequence of the definition of the unit vector
\be
\hat n_{ba} := \lt\langle \xi_{ba}, \vec \sigma \xi_{ba} \right\rangle \,.
\label{eq_unitvector}
\ee
As a consequence of the equations of motion eq. \eqref{Jxi_Jxi} and eq. \eqref{xi_xi}, which state how the spinors $(\xi_{ba}, J\xi_{ba})$ are parallel transported by the longitudinal holonomies $G_{ab}$, one finds that
\begin{align}
G_{ab} H_{ab}(b) G_{ab}^{-1} & = \exp \left[+\frac{4\pi}{ h}\left( \frac{1}{\g} + \I \right) \g j_{ab} \lt\langle \xi_{ba}, \sigma^j \xi_{ba} \right\rangle G_{ab}\tau_j G_{ab}^{-1}\right] \notag\\
& = \exp \left[+\frac{4\pi}{ h}\left( \frac{1}{\g} + \I \right) \g j_{ab} \lt\langle \xi_{ba}, G_{ab}^{-1}\sigma^j G_{ab} \xi_{ba} \right\rangle \tau_j \right] \notag\\
& = \exp \left[+\frac{4\pi}{ h}\left( \frac{1}{\g} + \I \right) \g j_{ab} \lt\langle J\xi_{ab}, \sigma^j J \xi_{ab} \right\rangle \tau_j \right] \notag\\
& = \exp \left[-\frac{4\pi}{ h}\left( \frac{1}{\g} + \I \right) \g j_{ab} \hat n_{ab}^j \tau_j \right] \,.
\end{align}
Here we used the mathematical identities $(G^\dagger)^{-1} = -J G J $ for all $G\in\mathrm{SL}(2,\mathbb C)$, $\langle J\xi, \vec\sigma J\xi \rangle = -\langle \xi,\vec\sigma \xi\rangle$, and $\sum_{i=1}^3 \sigma^i_{\alpha\beta} \sigma^i_{\alpha'\beta'} = 2 \delta_{\alpha\beta'} \delta_{\alpha'\beta}-\delta_{\alpha\beta}\delta_{\alpha'\beta'}$, as well as the definition of eq. \eqref{eq_unitvector}. Now, note that the holonomy $ G_{ab}H_{ab}(b)G_{ab}^{-1}H_{ab}(a)$ is associated to a contractible cycle within $M_3$, and must therefore be trivial. Thus, 
\be
H_{ab}(a)= \exp\left[+\frac{4\pi}{h}\left( \frac{1}{\g} + \I \right) \g j_{ab} \hat n_{ab}^j \tau_j \right] \,,
\ee
which is perfectly consistent with the previous definition of $H_{ab}(b)$. 

Henceforth, we will use the following lighter notation
\be
G_{ba}:=G_{ab}^{-1}, \quad H_{ab} := H_{ab}(a), \quad \text{and}\quad H_{ba} := H_{ab}(b).
\ee
In this way, the previous results can be rewritten as
\begin{align}
& H_{ab}= \exp\left[-\frac{4\pi}{h}\left( \frac{1}{\g} + \I \right) \g j_{ab} \hat n_{ab} .\vec\tau \right], \\
& \text{and} \quad G_{ab} H_{ba} G_{ba} = H_{ab}^{-1},
\end{align}
for any $a$ and $b$, such that $a\neq b$. The last equation could alternatively have been deduced from the fact that the particular composition of paths defining the holonomies $G_{ab}, \; H_{ab}$, and $H_{ba}$ that it uses is contractible in the graph complement.

Since most of the $G_{ab}$ factorize into $g_a^{-1} g_b$, it is also useful to introduce another set of variables,
\be
\tilde H_{ab} := g_a H_{ab} g_a^{-1},
\ee
in terms of which the parallel transport equations take the form
\be
\left\{
\begin{array}{ll}
\tilde H_{ba}  = \tilde H_{ab}^{-1}  &\quad\text{for } (ab)\neq(24) \\
{}&{}\\
\tilde H_{42} = \tilde H_{13}^{-1} \tilde H_{24}^{-1} \tilde H_{13}   &
\end{array}\right.  ,
\ee
where the different r\^ole played by the edge $(24)$ is inherited from eq. \eqref{Gab_BIS}. We will sometimes refer to the latter equation as the crossing relation, since it is due to the crossing in $\G_5$.

There is one last set of equations that can be deduced from the vanishing of the curvature in the graph complement $M_3$. At each vertex the transverse loops can be composed to form a contractible path; this path must be associated to a trivial holonomy. Therefore one obtains the five constraints
\be
H_{ab_4} H_{ab_3} H_{ab_2} H_{ab_1} = \mathbb{I} \qquad \forall a,
\ee
where the indices $b_i$ are all different and range in $\{1,\dots,5\}\setminus a$. Crucially, the order in which these holonomies appear is completely determined by the choice of framing of the graph. This equation is also permutation invariant in the $b_i$ (this will be of some importance later). With our choice of framing the five constraints that we obtain are\footnote{These equations can be read off by circling counterclockwise from edge to edge around every vertex in the graph $\G_5$, as drawn in \autoref{ribbon} (the composition of the holonomies in our notation reds from right to left).}
\begin{align}
\left\{\begin{array}{l}
H_{12} H_{13} H_{14} H_{15}= \mathbb{I}\\
H_{23} H_{24} H_{21} H_{25}= \mathbb{I}\\
H_{34} H_{31} H_{32} H_{35}= \mathbb{I}\\
H_{41} H_{42} H_{43} H_{45}= \mathbb{I}\\
H_{54} H_{53} H_{52} H_{51}= \mathbb{I}\\
\end{array}\right.
\label{eq_5vertices}
\end{align}
In what follows, we will refer to these equations as closure equations, closure conditions, or simply closures.

Before proceeding further, a couple of remarks. First, note that, when $(1+\rmi\g)h^{-1}\in\R$ as we propose, the transverse holonomies become SU(2) holonomies if they are evaluated close to the vertices, where they take the form given in eq. \eqref{eq_7.4}.\footnote{Though, if they are parallel transported between the vertices, they are genuine $\Slc$ holonomies.} This fact, which will play an important r\^ole in the geometric reconstruction of the following sections (see also \autoref{sec_parity}), is a consequence of the specific Wilson-graph operator we are using, and in particular of the properties of the map $\Y$ involved in its definition. More precisely, it is a consequence of the fact that in the construction of the graph operator, we solved the simplicity constraints in time gauge (see \autoref{sec_EPRL}). Finally, note also that in the calculation of $H_{ab}$ we ignored any contribution of the parallel transport from the graph (where the divergent curvature is located) to the base point of the holonomy. This can be heuristically interpreted as a gauge fixing of the holonomy from the vertex of the graph to the base point of the transverse loops. However, one should be more careful in devising appropriate regularization procedures if one wanted to be mathematically precise. 



\section{Summary of Critical Equations}\label{sec_equations}

We briefly summarize the results of \autoref{sec_SPA} and \autoref{sec_flatgraphcomplement}. These results specify the stationary phase points dominating the asymptotics of $Z_\text{CS}(S^3;\G_5\,|\,j_{ab},\xi_{ab})$, eq. \eqref{lEPRL_VA}. 

\begin{description}

\item[Parallel Transports:] From $\delta_{z_{ab}} S=0$ and $\Re(S)=0$, we have obtained the following parallel transport equations   relating the spinors $\xi_{ab}$ and $\xi_{ba}$ at opposite ends of the edge $\ell_{ab}$
\be
\xi_{ab}=-\E^{-\psi_{ab}-\I\varphi_{ab}}G_{ab} J\xi_{ba},\quad \text{and} \quad
J\xi_{ab}=\E^{\psi_{ab}+\I\varphi_{ab}}G_{ab} \xi_{ba}.\label{gluing}
\ee

\item[Transverse Holonomies:] The variation with respect to the Chern-Simons connection $(A,\bA)$ yields a distributional curvature on $S^3$ with support on the graph $\G_5$
\be
\frac{ \I h}{16\pi} \epsilon^{\mu\rho\sigma}F^i_{\rho\sigma}(x) \stackrel{!}{=} - (1+\I\gamma) \sum_{(ab),a>b} j_{ab} \;\langle \xi_{ba} , \left[(G_{s_{ab},b})^{-1} \tau^i G_{s_{ab},b} \right] \xi_{ba}\rangle\;\delta_{\ell_{ab}}^{(2)\;\mu}(x)
\label{distributionalcurvature}
\ee
where the complex conjugate equation holds for the curvature of $\bA$, $\bF$.
From this one deduces that the holonomy transverse to each edge $\ell_{ab}$ is nontrivial and has the form
\be
H_{ab}= \exp\lt[+{\frac{4\pi}{ h}\lt(\frac{1}{\g}+\I\rt) \g j_{ab}\hat{n}_{ab}\cdot\vec{\t}} \rt] \in \text{SU(2)} \subset \Slc.
\label{monodromiesresume}
\ee
Note that $j_{ab}=j_{ba}$ but $\hat n_{ab} \neq \hat n_{ba}$, since these unit vectors are defined by $\hat n_{ab} := \lag \xi_{ab},\vec \sigma \xi_{ab}\rag\in \mathbb R^3$ and generally $\xi_{ab}\neq \xi_{ba}$.

From the parallel transport equations it follows that
\be
G_{ab} H_{ba} G_{ba}  = H_{ab}^{-1}, \quad \forall a, b \quad \text{with} \quad a \neq b. 
\label{gluingH}
\ee
These parallel transport conditions are weaker than those for the spinors $\{\xi_{ab},J\xi_{ab}\}$. 

Again, the complex conjugate equations hold for the holonomies of the connection $\bA$.

\item[Flat Connection on $M_3$:]Another consequence of eq. \eqref{distributionalcurvature} is that the Chern-Simon connection must be flat on the graph complement $M_3$. This forces constraints on the transverse and longitudinal holonomies. In particular, the transverse holonomies must satisfy the closure equations
\be
H_{ab_4}H_{ab_3}H_{ab_2}H_{ab_1}=\mathbb{I},\label{Hclosure_summary}
\ee 
where the convention is that the $b_i$ can be read off $\G_5$ by circling counterclockwise around each vertex. The longitudinal holonomies can all be trivialized, except for $G_{13}$,
\be
G_{ab}=g_a^{-1}g_b,\ \ \ \ \text{except}\ \ \ \ G_{42}=g_4^{-1}\lt[g_3 H_{31} g_3^{-1}\rt] g_2,  \label{Gab}
\ee
for some $g_a\in\Slc$. These equations will henceforth be called the crossing conditions.

\end{description}

When expressed in terms of the $\tilde H_{ab} := g_a H_{ab} g_a^{-1} \in g_a\text{SU(2)} g_a^{-1} \subset \Slc$, the parallel transport equations, and the crossing condition read
\be
\tilde H_{ba} = \tilde H_{ab}^{-1} \ \ \ \ \text{except}\ \ \ \ \tilde H_{42}=\tilde H_{13}^{-1}  \tilde H_{24}^{-1} \tilde H_{13}\,,
\ee
while the closure equations maintain their form
\be
\tilde H_{ab_4}\tilde H_{ab_3}\tilde H_{ab_2}\tilde H_{ab_1}=\mathbb{I}.
\ee
We stress once more the importance of the non-trivial fact that the $\{\tilde H_{ab}\}_{b, b\neq a}$ at each vertex are in specific SU(2) subgroups of $\Slc$. This will be crucial for the geometrical interpretation of the critical equations.


\section{Geometry from Critical Point Equations I: The Idea}\label{sec_idea}

In this section, we start the core analysis of the paper, which will continue in the following two sections where we will deal with all the technical aspects. These three sections are dedicated to the correspondence between the (non-degenerate) solutions of the critical point equations just discussed and an essentially unique 4-dimensional 4-simplex geometry of constant curvature $\L$. This result is crucial, since it paves the way to showing that, in the semiclassical limit, the expectation value of the graph operator associated to a given simplicial complex is peaked around configurations corresponding to meaningful 4-dimensional simplicial geometries. This is an important step towards a definition of a path integral for (discrete) quantum gravity with a cosmological constant.\footnote{Once again, in this paper we do not try to tackle the difficult question of the continuous limit.} Moreover, the missing step---at least at the level of a single 4-simplex---of showing that the oscillatory weight of the given simplicial complex reproduces in the proper limit (a discretized version of) the Einstein-Hilbert action is the subject of the last sections of the paper.

Before delving into all the details, we will give a brief summary of the correspondence between the spinfoam critical data and the simplicial geometry using purely qualitative arguments. We hope this will be a useful guide to follow the technical constructions of the next sections.

Define the \textit{spinfoam critical data} to be a set of $\big\{j_{ab}, \xi_{ab}, G_{ab}, H_{ab}(a)\big\}$ that fufills the critical point equations summarized in the last section.  What is the geometrical content of these variables?

\begin{description}
\item[Spins $\mathbf{j_{ab}}$] As you might expect from the loop quantization, these variables correspond to the areas of the triangular faces of the 4-simplex. To be more precise, the physical area of one such triangle is $\mathrm{a}_{ab}/\ell_P^2=\gamma \sqrt{j_{ab}(j_{ab}+1)}$, which is approximately $\gamma j_{ab}$ in the semiclassical limit considered here (see \autoref{sec_largej}). This happens because spins are associated to the eigenvalues of the angular momentum operator $J^i$, which in turn relates to the quantization of the discretized area density field $B=e\wedge e$ (see \autoref{sec_relationLQG}).
\item[Spinors $\mathbf{\xi_{ab}}$] First, recall that the normalized spinors $\xi_{ab}$ have a (conventionally) fixed phase, and therefore have just two, rather than three, real parameters. Indeed, they map to the space of 3-dimensional unit vectors, via
\be
\hat n: = \lag \xi, \vec\sig \xi\rag.
\ee
 To see what these vectors geometrically correspond to, let us start by taking the vanishing-cosmological-constant limit (flat limit) of one of the closure equations, say at vertex 5: 
\be
H_{51} H_{52} H_{53} H_{54} = \mathbb{I} \xrightarrow{\L\to 0} +\frac{\L}{3}\sum_{b=1}^4 \gamma j_{5b} \hat n_{5b} \cdot\vec\tau = O(\L^2)\,.
\ee
With the previous interpretation for the spins, this gives at lowest order in $\Lambda$: $\sum_{b=1}^4 \mathrm{a}_{5b} \hat n_{4b} = \vec 0$, which can be interpreted via Minkowski's theorem \cite{Minkowski1989} as the equation defining the unique geometric tetrahedron (up to rotations and parity inversion) having face areas $\mathrm{a}_{5b}$ and outgoing face normals $\hat n_{ab}$. For brevity we will call $\vec{\mathrm{a}}_{ab}$ the \textit{area vector} of the face $(ab)$. Now, we claim that the $\hat n_{ab}$'s are still interpretable as (spacial) normals to the triangular faces also in the curved case. This is made possible by the natural condition that all the subsimplices of the curved 4-simplex are flatly embedded (have vanishing extrinsic curvature).
\end{description}

The interpretative framework for the spins and the spinors just discussed is part of the rigorous geometric quantization of the space of shapes of polyhedra (see e.g. \cite{Barbieri1997,Baez1999,Conrady2009,Bianchi2011PRD,BianchiHaggard2011,BianchiHaggard2012,Haggard2013,Haggard2011} for the quantization of flat polyhedra). 

\begin{description}
\item[Holonomies $\mathbf{H_{ab}}$] In light of the previous discussion and eq. \eqref{monodromiesresume} it is clear that we want to assign to the holonomies $H_{ab}$ the r\^ole played by the area vectors in the flat case. This is indeed possible, and in both the following section and a companion paper \cite{companionMink} we discuss these results in detail. However, the basic idea is very simple: in a curved geometry one can get non-trivial information about a surface just by going around its boundary. This is a consequence of the (non-Abelian) Stokes theorem \cite{Fishbane1981,Broda2004}, which enormously simplifies in the case in which (\textit{i}) the space has constant curvature $\l$,\footnote{It will become clear later why we have not used the symbol $\L$.} and (\textit{ii}) one considers only flatly embedded surfaces. Under conditions (\textit{i}) and (\textit{ii}) it is not hard to show that---in three dimensions---the (torsionless) parallel transport $U_{\partial\fs}$ around a surface $\fs$ is given by
\be
O_{\partial\fs} = \exp\lt(+ \frac{\l}{3}\mathrm{a}_\fs \hat{\mathfrak{n}}_{\fs} \cdot \vec J \, \rt),
\ee
where $\mathrm{a}_\fs$ is the area of $\fs$, $\hat{\mathfrak{n}}_\fs$ is its spacial normal\footnote{The sign of $\hat{\mathfrak{n}}_\fs$ is related to the direction in which the boundary of the surface is circulated via a right-handed convention.} parallel transported to the base point of $O_{\partial\fs}$, and the $\{J^{i}\}$ are the generators of SO(3). This is exactly the form the $H_{ab}$ have in their vector representation $\OO_{ab}$.\footnote{We could have considered the holonomy of the spin connection around $\fs$ to obtain an expression analogous to that of the $H_{ab}$. We have avoided this for technical reasons that will be clarified later.} We see that it is important that the holonomies $\{H_{ab}\}_{b, b\neq a}$ at one vertex are all in the same SU(2) subgroup of $\Slc$: this means that the surfaces they are associated with all have a common timelike normal, and hence define a spacelike frame. 

\item[Holonomies $\mathbf{G_{ab}}$] Finally, the holonomies $G_{ab}$ are to be interpreted as the parallel transport holonomies between different reference frames. Specifically, the holonomy $G_{ab}$ allows one to parallel transport any geometrical quantity from the frame of tetrahedron $b$ to that of tetrahedron $a$. This is manifest from the parallel transport equations for the spinors $\xi_{ab},\,\xi_{ba}$, eq. \eqref{gluing}, which also imply those for the $H_{ab},\,H_{ba}$, eq. \eqref{gluingH}. Notice that the gauge-invariant information carried by the $G_{ab}$ must then have the geometrical interpretation of hyper-dihedral angles between the two boundary tetrahedra $a$ and $b$. Notice also the meaning of the crossing equation (eq. \eqref{Gab}): if all the $G_{ab}$ could have been trivialized into products of the type $g_a g_b^{-1}$ it would have meant that the 4-simplex was flat, since the composition of changes of reference frame would be trivial, i.e. independent of the path chosen. This is not the case, since a non trivial cycle exists (going around vertices 3, 2, 4, and back), and therefore the reconstructed geometry must be curved.
\end{description}
For a relationship (in the flat 4-simplex context) between the continuous-geometry picture and discrete data of the type above, see e.g \cite{Freidel2013}.


\subsection{Framing of $\G_5$ and its Dual Geometry}\label{sec_framing}

In this section we show how to translate the paths defining the holonomies $H_{ab}$ and $G_{ab}$ in $M_3=S^3\setminus\G_5$ into paths on the one-skeleton of the 4-simplex. This is an essential preliminary step for the reconstruction theorem. In particular, we will show that a choice of graph framing, like ours, completely fixes these paths in a consistent way.

The strategy is well illustrated by the transverse holonomies $H_{ab}$. We would like to interpret these as the holonomies resulting from parallel transport along the edges bounding a face of a curved tetrahedron. But, if this interpretation is to hold, what is the appropriate order to circuit all four tetrahedral faces such that the closure eq. \eqref{Hclosure_summary} holds, with the appropriate ordering of the $H_{ab}$? Surprisingly, these path orderings are completely encoded in a refined understanding of the $\G_5$ graph. This section explains the relevant structure of $\G_5$ in detail, proceeding from the four-dimensional down to the three-dimensional geometry. 

\begin{figure}[t]
\begin{center}
\includegraphics[width=.4\textwidth]{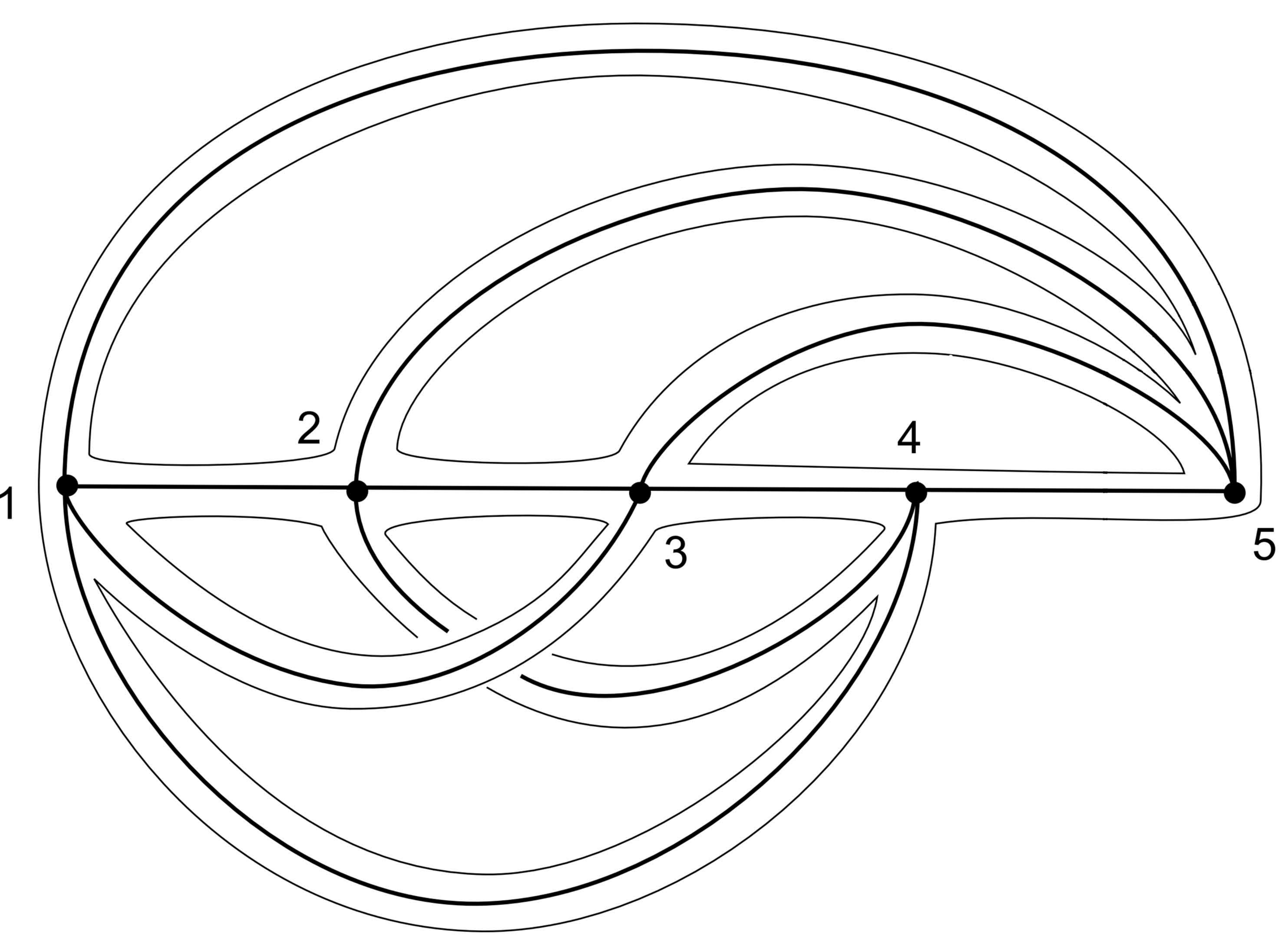} 
\caption{Top view of the fattened $\G_5$-graph. The faces of the dual 4-simplex are constructed by appropriately ``filling in the holes'' bounded by triples of edges.} 
\label{fig_soloribbon}
\end{center}
\end{figure}

In order to understand which path one is supposed to follow on the 4-simplex 1-skeleton, one has to recall the origin of the $\G_5$ graph itself: this graph is the \textit{dual} of the 4-simplex boundary, which means that its vertices are dual to tetrahedra and its edges to triangular faces. In principle the dual of the 4-simplex sides are given by the ten faces of the dual 4-simplex of which $\G_5$ is the 1-skeleton. These are given by the ten 2-surfaces bounded by three-edge-long closed sequences of edges in $\G_5$. In spite of the fact that the graph \textit{per se} does not contain any explicit information as to what these surfaces concretely are, we shall show that, once a graph framing (of the type discussed above) has been picked, there exists a natural prescription that uniquely fixes these 2-surfaces. The prescription states that the 2-surfaces should not intersect each other, except along their boundary edges. It will be useful to refer to \autoref{fig_soloribbon} as the construction proceeds. 

To see how this prescription fixes the 2-surfaces, let us start by fixing without loss of generality the faces (125), (235), and (345) to lie in the 2-plane in which their boundary edges are drawn, henceforth called the blackboard plane. This can be visualized as simply ``filling in the holes'' these triples of edges form. Consider now face (145), and think of it as also lying in the blackboard plane, but now extending out to infinity.\footnote{Recall that the three-space in which the graph is embedded is actually compactified by the identification of the sphere at infinity with a point. Therefore this face is forming a dome above---or, equivalently, below---the graph.} In order not to intersect any of the faces we have already fixed, faces (135) and (245) must be contained on either the upper or the lower half-space with respect to the blackboard plane. Obviously they cannot lie in the same half-space, and the framing fixes face (135) to lie in the upper part, and face (245) in the lower one. The same reasoning applies consistently to faces (134) and (124). Finally, faces (123) and (234) are fixed to be transverse to the blackboard plane, each on one side of it. Clearly, what we have just described should be understood up to smooth deformations. 

\begin{figure}[t]
\centering
\includegraphics[width=.8\textwidth]{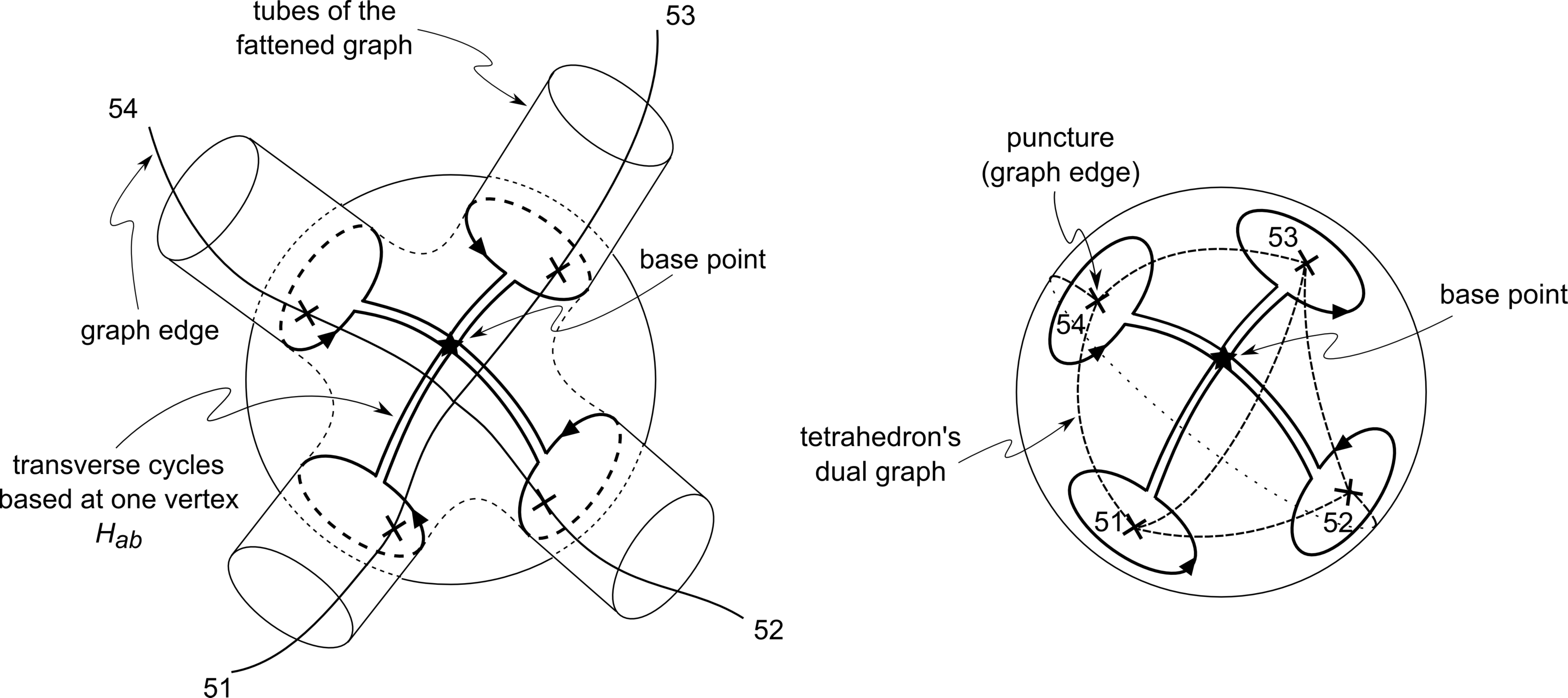}
\caption{A close-up of the region close to the fifth vertex of the thickened $\G_5$ graph. The paths along which the transverse holonomies $H_{ab}$ are calculated are represented with thick solid lines. All of them follow a right-handed outward-pointing path around the edges of the graph. We have also depicted a virtual sphere around the vertex of the graph in both panels. The sphere is pierced by the graph edges, these punctures are represented by $\times$'s. The right panel shows the intersections of the faces of the graph with the sphere around the vertex (dashed lines), as deduced from our choice of framing for $\G_5$. The line connecting punctures $(52)$ and $(54)$ traverses the hidden back side of the sphere. The intersection pattern of these lines with the paths defining the transverse holonomies allows the reconstruction of the full path structure on the tetrahedron, shown in the next figure.}
\label{fig_tetdual}
\phantom{space}
\includegraphics[width=.8\textwidth]{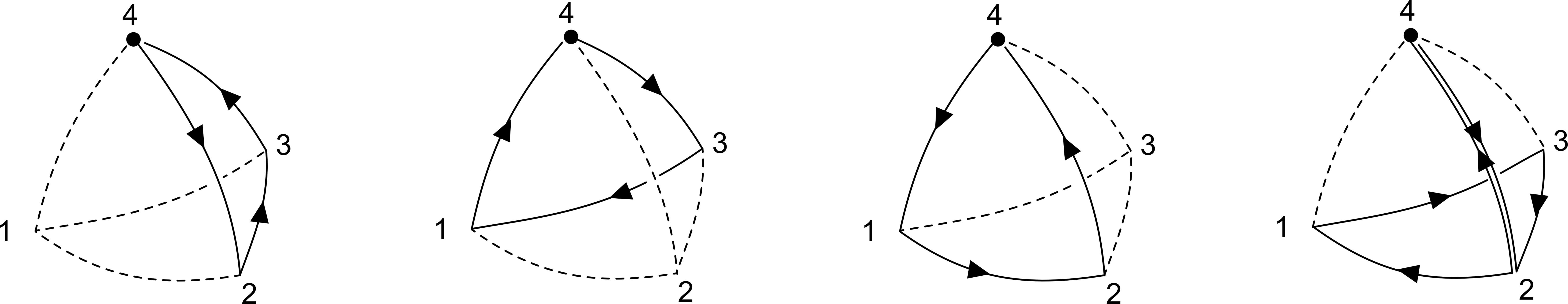}
\caption{The ``simple'' path on tetrahedron 5, dual to vertex 5 (see the previous figure). The images display the path around faces 1 to 4, as reconstructed from the framing of $\G_5$. Notice, that the path around the fourth face is necessarily different from the preceding ones.}
\label{fig_simplepath}
\end{figure}

The paths defining the transverse holonomies are also picked out by the choice of framing. Let us focus on a single vertex, and consider a small sphere around it (see \autoref{fig_tetdual}, where vertex 5 is used as an example). This sphere is pierced by the edges of $\G_5$, which therefore identify four punctures on its surface. The tubes around each edge cut out circles on the sphere around each puncture. The paths of the transverse holonomies can be chosen to live on the surface of the sphere, and simply go around the punctures along the aforementioned circles. 

The final missing ingredients are the duals to the tetrahedron's sides; since they are non-intersecting surfaces joining at the graph vertex, and in its neighborhood bounded by couples of graph edges, their intersections with the sphere are given by non-intersecting lines connecting couples of punctures (dashed lines in the right panel of the figure). We have just seen that these lines are also uniquely determined by the choice of graph framing. 

Now, the lines connecting the punctures on the sphere form a tetrahedron dual to the one we want to associate to the vertex: its vertices, the punctures, should correspond to the tetrahedron's faces, and conversely, its faces correspond to the tetrahedron's vertices, and finally the sides correspond to the tetrahedron's sides themeselves (though in an ``orthogonal'' sense\footnote{For example, at tetrahedron 5, the line joining punctures 1 and 3 is dual to the side in the 4-simplex shared by the triangular faces (51) and (53), and therefore is dual to the 4-simplex side connecting the 4-simplex vertices $\bar{2}$ and $\bar{4}$.}). Therefore, we see that each path going around a puncture corresponds to a path on the tetrahedron which starts at one vertex (the same for all of them) and visits some other vertices, in a precise order, before coming back to the original one. A moment of reflection shows that this set of paths corresponds to what we have called a ``simple path'' on the 1-skeleton of the tetrahedron in \cite{companionMink}; for clarity and completeness, a simple path is illustrated in \autoref{fig_simplepath}, which should be self-explanatory. 

Observe that in the simple path of \autoref{fig_simplepath} a special role is played by the ``special side'' $\overline{24}$ of the tetrahedron 5, and that this statement is independent of the position of the base point: had we moved it somewhere else on the sphere, this would amount to adding extra segments at the beginning and at the end of \textit{all} four face paths, which would correspond to a \textit{global} gauge transformation (i.e. to conjugation by an SU(2) or $\Slc$ element of all the $H_{5b}$). Such a global transformation has no effect on the reconstruction we are going to perform.

Notice that we have drawn right-handed paths on the tetrahedron. In this manner all the faces are circulated counterclockwise, and using a right-hand rule one can assign outgoing normals to each face. Had we used a left-handed tetrahedron, and at the same time stayed with a right-hand rule for assigning the normals, we would have obtained ingoing normals. We work with the right-handed convention. It will be important to keep this arbitrary choice in mind for the last part of the paper. 

With a little bit of work, e.g. by choosing all the tetrahedra to be oriented subsimplices of a right-handed 4-simplex $[\bar{1}\bar{2}\bar{3}\bar{4}\bar{5}]$, one finds that all of the faces of all the tetrahedra are circulated in a right-handed sense, and that the ``special sides'' of the five tetrahedra are respectively:
\be
\overline{24} & \text{ for tetrahedra } 1,3,5, \notag\\
\overline{45} & \text{ for tetrahedron } 2, \\
\text{and} \quad \overline{25} & \text{ for tetrahedron } 4. \notag
\ee
It then becomes clear that every face is circulated in opposite directions when considered from each of the two tetrahedra sharing it. This is in agreement with the parallel transport equation
\be
H_{ab} = G_{ab}H_{ba}^{-1} G_{ba}.
\ee

However, since we want to understand the paths on the 4-simplex, and not just within each tetrahedron, we have to be slightly more careful than this. First, notice that all the $\{H_{ab}\}_{b,b\neq a}$ share the same base point, and so do $H_{ab}$ and $G_{ab}H_{ba}^{-1} G_{ba}$ (otherwise the closure and parallel transport equations would not be gauge covariant). It is then immediate that the $\{\tilde H_{ab} := g_a \tilde  H_{ab} g_a^{-1}\} $ must \textit{all} share the same base point. Now the parallel transport equations together with the crossing relations give
\be
\tilde H_{ba} = \tilde H_{ab}^{-1} \ \ \ \ \text{except}\ \ \ \ \tilde H_{42}=\tilde H_{13}^{-1}  \tilde H_{24}^{-1} \tilde H_{13}\,.
\label{eq_except}
\ee
This equation says that the paths associated to all the triangular faces in the 4-simplex, except those of face (24), do not depend on the tetrahedron they are associated to. This is a quite natural consequence of the pattern of special sides discussed here.\footnote{This is not a \textit{logical} consequence, instead it relies on an hypothesis of simplicity.} Indeed, we see that the 4-simplex vertices $\bar{2}$ and $\bar{4}$ play a completely symmetric r\^ole, and that either of the two vertices is part of all the special sides but one. Let us pick the 4-simplex vertex $\bar{4}$ to be the base point of all the paths on the 4-simplex, and suppose that the ``special paths'' of tetrahedra 1, 3, 5, and 2 all start there (as in \autoref{fig_simplepath}). We immediately see that eq. \eqref{eq_except} is automatically verified for all faces $\{(ab)\}_{a,b\neq4}$ not involving tetrahedron 4. Now, consider tetrahedron 4; it cannot be naturally based at vertex $\bar{4}$ like the others, and by looking at its special side, we see that its ``internal'' base-point should be either vertex $\bar{2}$ or $\bar{5}$. Vertex $\bar 2$ is the choice which ensures consistency with the crossing relation. Indeed, in this way every face path associated with tetrahedron 4, must start at vertex $\bar{4}$, go through the side $\overline{42}$ and then go along the usual sequence of sides, to finally come back to $\bar 4$ along side $\overline{24}$. In this way the side $\overline{42}$ plays in the 4-simplex a r\^ole similar to that of a tetrahedral special side, and allows eq. \eqref{eq_except} to be valid for faces $(4b)$ with $b\neq2$. 

\begin{figure}[t!]
\centering
\includegraphics[height=4cm]{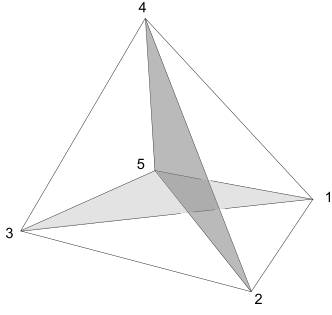}
\caption{A depiction of the oriented 4-simplex. All tetrahedra can be interpreted as having outward pointing normals, except the external tetrahedron (i.e. tetrahedron 4). Indeed, the interior of tetrahedron 4 is actually the region of $\mathbb R^3$ extending to infinity. This was also discussed when justifying the particular form of the graph $\G_5$.}
\label{fig_4simplexpersp}
\end{figure}

Finally, since every path has now been fixed, one just has to check that  $\tilde H_{42}=\tilde H_{13}^{-1}  \tilde H_{24}^{-1} \tilde H_{13}$ (see \autoref{fig_4simplexpersp}). %
The sequence of sides defining $\tilde H_{42}$ is (read from right to left):\footnote{We have tried to highlight the r\^ole played by the special sides with the bracket notation.}
\be
\overline{42} \lt\{ \overline{25} \lt[\overline{53}\;\overline{31}\;\overline{15} \rt] \overline{52} \rt\} \overline{24}\,,
\ee
while that defining $\tilde H_{24}$ is
\be
 \overline{45}\lt[\overline{51}\;\overline{13}\;\overline{35}\rt] \overline{54}.
\ee
It is clear that to equate the inverse of the first to the second, one has to conjugate the latter by 
\be
\overline{45}\;\overline{52}\;\overline{24}\,,
\ee
which is nothing but the path defining $\tilde H_{13}$.

This set of paths allows us to geometrically interpret the holonomies $H_{ab}$, and especially the $\tilde H_{ab}$, in terms of parallel transports along the 1-skeleton of the 4-simplex.


\section{Geometry from Critical Point Equations II: Curved Tetrahedra}\label{sec_curvedtets}

In this and the following sections we show how, using the critical point equations, together with the interpretation of the holonomies $H_{ab}$  just sketched (and soon to be made more rigorous), one can reconstruct a curved-tetrahedral geometry at every vertex of the graph. In this paper, we will just show how to recover the tetrahedral geometry from the holonomies in a constructive way, and skip the rigorous proof of the consistency of this reconstruction.\footnote{In particular, one needs to show that the tetrahedron reconstructed with the procedure presented here has areas compatible with those appearing in the holonomies. This is \textit{a priori} nontrivial.} Anyway, this result follows from the reconstruction theorem for the 4-simplex geometry. We present, in some detail, the tetrahedral result because it is more intuitive, it proceeds constructively, and it offers most of the features of the 4-simplex reconstruction.

The key equation is the closure condition (since for the moment we work at a fixed vertex, we simplify the labeling of our variables)
\be
H_4 H_3 H_2 H_1= \mathbb{I},
\ee
however, in the following we will focus on the derived equation
\be
\OO_4 \OO_3 \OO_2 \OO_1 = \mathbb{I}.
\label{OOOO}
\ee
where $\OO_b\in\text{SO}(3)$ is the vectorial (spin 1) representation of $H_b\in \mathrm{SU}(2)$. We leave the discussion of the relation between the SU(2) group elements $H_{ab}$ and their SO(3) counterparts for the last part of the paper. As we explained in the previous section the $\OO_b$ are interpreted as parallel transports along specific, simple paths on the tetrahedron 1-skeleton. The ordered composition of all the paths associated to a tetrahedron is equivalent to the trivial path, hence the identity on the right-hand side of the closure equation.


\subsection{Flatly Embedded Surfaces}\label{sec_flatlyembeddedsurf}

First we discuss some of the claims of \autoref{sec_idea}. Consider a 4-dimensional spacetime $(\Fm_4,g_{\alpha\beta})$, with no torsion, constant curvature $\l$, and tetrad 
$e^I_\alpha$. In this spacetime, consider a bounded 2-surface $\fs$ that is flatly embedded in $\Fm_4$, i.e. such that the wedge product of its space- and time-like normal fields, $\mathfrak{n^\alpha}$ and $\mathfrak{u}^\alpha$ respectively, is preserved by parallel transport on the surface $\fs$.\footnote{This requirement is equivalent to asking that the extrinsic curvature of $\fs$ vanish, and generalizes the concept of planes in $\R^3$ to curved spaces. Further examples of flatly embedded surfaces are equatorial (great) spheres in $S^3$ and great hyperboloids in $\mathbb{H}^3$. {It is not too difficult to show that the vanishing of the extrinsic curvature on $\fs$ is equivalent to asking that $\fs$ be totally geodesic, see e.g. \cite{Aminov2001}. }} Then, the holonomy around $\fs$ of the torsionless spin connection $\omega^{IJ}_\alpha := \E^{I\beta}\nabla_\alpha e^J_\beta$, is given in the spinor representation by\footnote{
This formula follows from the non-Abelian Stokes theorem $\mathcal P \exp \oint_{\partial\fs} \omega = \mathcal P \exp \int_{\fs} \mathcal R(\omega)$, where $\mathcal R$ is the curvature 2-form of $\omega$, and the fact that on constantly curved manifolds and flatly embedded surfaces, the integrand appearing in the previous formula is actually a constant. To see this last point, notice that for a constantly curved manifold one has
 $$
\mathcal{R}^{IJ}[\omega]= \frac{\lambda}{3}e_\alpha^{[I} e_\beta^{J]} \D x^\alpha\wedge\D x^\beta 
$$
which, when restricted to $\fs$ and after some manipulation, becomes 
\begin{align*}
\mathcal{R}^{IJ}[\omega]\big|_{\fs} 
&= \frac{\l}{3} e_\alpha^{[I} e_\beta^{J]}  \frac{\partial x^{\alpha}}{\partial s^1}\frac{\partial x^{\beta}}{\partial s^2}\D s^1\wedge\D s^2 
= \frac{\l}{3} \Big[ \star (\underline{\frak u}\wedge\underline{\frak n} )\Big]^{IJ} \sqrt{\det g_{(2)} } \D s^1 \wedge \D s^2
\end{align*}
where $(s^1,s^2)$ are two coordinates parametrizing the surface, and $g_{(2)}$ is the metric restricted to it. Now, using the fact that the surface is flatly embedded one sees that under the parallel transport needed for the $\mathcal P\exp$ of the non-Abelian Stokes theorem, the term in parentheses is transported onto itself within the surface. Hence, the result follows just by integrating the area of $\fs$, introducing the notation $ \underline{\mathfrak{u}}^{[I} \underline{\mathfrak{n}}^{J]}= \frac{1}{2} \lt(\underline{\mathfrak{u}}\wedge\underline{\mathfrak{n}}\rt)^{IJ}$, and finally taking the self dual part in the internal indices.
 }
\be
U_{\partial \fs}[\omega_+]=\exp\lt[ +\frac{\l}{3} \mathrm{a}_\fs \lt( \star\,\underline{ \mathfrak{u}}\wedge\underline{\mathfrak{n}} \, \rt)_+^k \tau^k\rt],
\ee
where the subscript $+$ indicates the self-dual part of an object,  $\mathrm{a}_\fs$ is the area of $\fs$, and we have defined  $\underline{\mathfrak{n}}^I :=e^I_\alpha \frak{n}^\alpha$ and $\underline{\mathfrak{u}}^I := e^I_\alpha \frak{u}^\alpha$ to be the internal spacelike and timelike normals to the surface, respectively. Notice that the latter are understood to be evaluated at the base point $O\in\partial \fs$ of the holonomy $U_{\partial\fs}$. In the \textit{(future pointing) time gauge} $\underline{\mathfrak{u}}^I =  \delta^I_0$ and $\underline{\mathfrak{n}}^I = \delta^I_k \hat{\underline{\frak{n}}}^k$, and therefore{ (see also \autoref{app_notations} for details on the notation)}
\begin{align}
\lt(\star\,\underline{ \mathfrak{u}}\wedge\underline{\mathfrak{n}} \,\rt)_+^k  
& = 
\I \,\epsilon^{k}_{\phantom{k}ij}\underline{\mathfrak{u}}^{[i}\underline{\mathfrak{n}}^{j]} + 2 \underline{\mathfrak{u}}^{[0}\underline{\mathfrak{n}}^{k]} 
  =
 \underline{\mathfrak{u}}^{0}\underline{\mathfrak{n}}^{k} 
 = 
\underline{\hat{\mathfrak{n}}}^{k}\,,
\end{align}
where the last two equalities hold in time-gauge (more specifically, the last one holds for a future pointing time-gauge. See \autoref{app_Lorentziangluing}).

Finally, by henceforth dropping the underscore, $\underline{\hat{\frak n}} \mapsto \hat{\frak n}$, we obtain (in future pointing time gauge)
\be
U_{\partial{\fs}}[\omega_+] =
\exp \lt(+ \frac{\l}{3} \mathrm{a}_{\fs} \hat{\frak n} \cdot \vec \tau\rt),
\ee
and hence in the vectorial representation
\be
O_{\partial{\fs}}[\omega_+] =
\exp \lt(+ \frac{\l}{3} \mathrm{a}_{\fs} \hat{\frak n} \cdot \vec J\rt).
\ee

We clearly see that the functional dependence of the holonomies $O_{ab}$ on $\lt(\hat n, \g j, -\frac{12\pi}{h}(\g^{-1}+\I ) \rt)$ is completely analogous to that of the parallel transports $O_{\partial \fs}$ on $\lt(\hat{\frak n} , \mathrm{a}_{\fs}, \l\rt)$. However, all these variables are mixed with one another, and it is possible to distinguish them only up to some ambiguities. As we have already emphasized, a physically well-motivated candidate for a spinfoam analogue of the area exists, and is $\mathrm{a}_{\fs} \leftrightarrow \gamma j$. This identification also fixes the magnitude of $\lambda$ to $\frac{12\pi}{h}\lt(\frac{1}{\g} + \I \rt)$, since the vectors $\hat n$ and $\hat{\mathfrak n}$ are both normalized. However, things are slightly  more complicated than this, since the signs of the cosmological constant and of the unit vector cannot be distinguished \textit{a priori}. Moreover, there is also a sign ambiguity arising from the area, since there is no way of telling apart $\frac{\lambda}{3}\mathrm{a}_{\fs}\in(0,2\pi)$ and $\lt(2\pi-\frac{\lambda}{3}\mathrm{a}_{\fs}\rt)\in(0,2\pi)$ at the outset.

Beyond the (partly ambiguous) properties of area, curvature, and orientation, the shape of $\fs$ is not better defined for the moment. The best we can do to define its shape with the limited data at our disposal, is to further constrain its geometric degrees of freedom; for example, by requiring each vertex of the graph to be identified with the simplest curved geometrical object with four faces, a homogeneously curved tetrahedron. The viability of this requirement is a consequence of a theorem reviewed in the following section (details can be found in the companion paper \cite{companionMink}). At this stage, the fact that all the previous ambiguities are \textit{consistently} solved throughout the whole 4-simplex is very surprising, nonetheless it will be a consequence of the equations of motion.

Before continuing, notice that in particular the sign of the cosmological constant (or equivalenty the curvature) is a priori totally free at each face. The aforementioned theorems cure this problem, at the level of each tetrahedron and also at the level of the whole 4-simplex. An important consequence of this discussion is that our model cannot be considered a quantization of gravity with a fixed-sign cosmological constant: it is rather a quantization of gravity with \textit{a} cosmological constant, the sign of which is determined \textit{dynamically}, and \textit{only semiclassically}, by the imposed boundary conditions (in this case the external $j_{ab}$ and $\xi_{ab}$). 



\subsection{Constant Curvature Tetrahedra}\label{sec_constcurvtets}

According to our requirements, the faces of the curved tetrahedron are spherical or hyperbolic triangles, with a radius of curvature equal to $R=\sqrt{3/|\L|}$. This means that their areas must lie in the interval $[0,6\pi/|\L|]$, or $[0,3\pi/|\L|]$, respectively.\footnote{{On the sphere we always consider the triangle to be the smaller of the two portions in which the sphere is subdivided by three points connected by geodesic arcs. We also require this region to be convex.}} The spherical case is no problem, since SU(2) group elements have the right periodicity in their argument. Even more compellingly, by looking at the deformed SU$(2)_q$ representations 
 with $q=\exp \frac{4\pi \I}{k+2}$, and $k=\Re(h)=\frac{12\pi}{\g\Lambda}$, one only finds spins up to $|k|/2$, which translates into $\g j \leq 6\pi /| \L|$.\footnote{Still, the need to quantize gravity \textit{and} Chern-Simons simultaneously, which leads to a deformation of the gauge group and hence of its representations, raises the interesting question of whether the construction of the EPRL amplitude is still justified in this setting or if it needs to be modified. An hint at an answer comes from the modified phase space of LQG  suggested by the new closure condition. The classical phase space is investigated in the companion paper \cite{companionMink}.} The hyperbolic case, on the other hand, is more subtle. We do not try to deal with all the subtleties here, since a thorough discussion can be found in \cite{companionMink}. Nonetheless, we anticipate the fact that these subtleties can give rise---in certain cases determined by the choice of the spins---to non-standard geometries that extend across the two sheets of the two-sheeted hyperboloid. 

In any event, care is needed in identifying $\g j$ with the face area even in the spherical case, since the ambiguity $ \g j \leadsto \lt(2\pi R^2 - \g j \rt)$ may arise. We will come back to this point later on.
\color{black}

From now on, we will consider the reconstruction at the vertex 5 of $\G_5$. As previously shown, the closure equation in the vector representation is
\be
\OO_4 \OO_3 \OO_2 \OO_1 = \mathbb I
\ee
and the special side is $\overline{24}$. We will take the base point to be vertex $\bar 4$. Because all the holonomies are based at vertex $\bar 4$, all the $\hat{\frak n}_{b}$ are defined there, which we notate $\hat{\frak n}_b(4)$. However, recall that the property of being flatly embedded means having vanishing extrinsic curvature, and so this makes the normal to a face well-defined at any of its points. The faces 1, 2 and 3 (faces are labeled using the opposite vertex) contain vertex $\bar 4$, and this means that $\hat{\frak n}_1(4), \hat{\frak n}_2(4), \hat{\frak n}_3(4)$ can be directly interpreted as normals to their respective faces, while $\hat{ n}_4(4)$ is the vector obtained after parallel-transporting $\hat{\frak n}_4$ from its face to vertex $\bar 4$, via the edge $\overline{24}$ (see \autoref{fig_simplepath}). That is,
\be
\hat{\frak n}_4(4) = \oo_{42} \hat{\frak n}_4(2) ,
\ee
where $\oo_{cb}$ is the vector representation of the holonomy from vertex $\bar b$ to vertex $\bar c$, along the side $\overline{cb}$. Notice that this is not part of our critical data and is highly gauge dependent. As we will see, this will not pose any problem to the reconstruction, it is just an intermediate step towards a well-defined expression.

From this understanding of the normals we can find the (cosine of the external) dihedral angles along the edges $\overline{41}$, $\overline{42}$, and $\overline{43}$. They are 
\be
\cos \phi_{bc} = \hat{\fn}_b(4) \cdot \hat{\fn}_c(4) , \quad \text{for } (b,c)\neq(2,4).
\ee

For the dihedral angle along the edge $(24)$, more care is needed. Certainly, we have
\be
\cos \phi_{24} = \hat{\fn}_2(1) \cdot \hat{\fn}_4(1) = \hat{\fn}_2(3) \cdot \hat{\fn}_4(3)\,.
\label{eq_dihedral}
\ee
The problem is that $\hat{\fn}_4$ when based at vertices 1 or 3 does not relate to the critical data in a gauge-independent way. However, we can relate, e.g.,  $\hat\fn_2(1)$ and $\hat\fn_4(1)$ with their values at vertex $4$,\footnote{Note that, since the surface is flatly embedded $\OO_2 \hat\fn_2= \hat \fn_2$, and therefore $\oo_{13}\oo_{34}\hat\fn_2 = \oo_{14}\hat\fn_2$.}
\begin{align}
\cos \phi_{24} &= \hat{\fn}_2(1) \cdot \hat{\fn}_4(1) = \lt[\oo_{14}  \hat{\fn}_2(4) \rt] \cdot\lt[\oo_{12}\oo_{24} \hat\fn_4(4)\rt],
\end{align}
Bringing $\oo_{14}$ across the dot product  and using the group orthogonality property we obtain, 
\be
\cos \phi_{24} =\hat \fn_2(4) \cdot \lt[\OO_3^{-1}\hat\fn_4(4) \rt].
\ee
Similarly, starting at vertex 3, one finds
\be
\cos \phi_{24} = \hat \fn_2(4) \cdot \lt[\OO_1\hat\fn_4(4) \rt].
\ee
The fact that these two equations are consistent with one another follows immediately from the closure condition and the fact that the surfaces are flatly embedded, which gives $\OO_b \hat\fn_b=\hat\fn_b$.

It is also possible to give an expression of the cosines of the dihedral angles directly in terms of the holonomies, the data actually specified by the critical point equations,
\begin{subequations}
\begin{align}
\cos \phi_{bc} & = \pm_b\pm_c\frac{\frac{1}{2}\Tr\lt( \OO_b \OO_c\rt) - \frac{1}{4}\Tr\lt(\OO_b\rt)\Tr\lt(\OO_c\rt)  }{  \sqrt{1-\frac{1}{4}\Tr^2\lt(\OO_b\rt)}\sqrt{1-\frac{1}{4}\Tr^2\lt(\OO_c\rt)}  } , \quad \text{for } (b,c)\neq(2,4)\\
\cos \phi_{24} & = \pm_2\pm_4\frac{\frac{1}{2}\Tr\lt( \OO_2 \OO_3^{-1}\OO_4 \OO_3\rt) - \frac{1}{4}\Tr\lt(\OO_2\rt)\Tr\lt(\OO_4\rt)  }{  \sqrt{1-\frac{1}{4}\Tr^2\lt(\OO_2\rt)}\sqrt{1-\frac{1}{4}\Tr^2\lt(\OO_4\rt)}  }\,,
\end{align}
\end{subequations}
which are a sort of normalized, connected two-point functions of the holonomies.  The signs $\pm_b$ are exactly those appearing in the relation between the geometrical and spinfoam data: $\hat{\frak n}_b = \pm_b \hat n_b$. When the tetrahedron is known, the signs can be fixed by knowing the signs of $\sin\lt(\frac{\l}{3} \mathrm{a}_b\rt)$, which fixes the branch of the square root. However, when the tetrahedron is not known {a priori}, as when one is given only the spinfoam data, this ambiguity is due to the fact that given an holonomy we are unable to distinguish between a rotation of $\theta$ around an axis and a rotation of $2\pi-\theta$ around the opposite axis. This means that we cannot initially decide whether $\g j_b$ or $\lt(2\pi R_\lambda^2 - \g j_b\rt)$ will be the geometrical area of the face. Nonetheless, this ambiguity can be fixed by requiring that the tetrahedron be convex; this is easily accomplished by imposing positivity on the triple product $\hat\fn_1(4)\cdot \lt[\hat\fn_2(4) \times \hat\fn_3(4)\rt]$, and its like. Indeed, these triple products can be expressed as normalized connected three-point functions of the holonomies, similar to the two-point functions above. These expressions present  analogous sign ambiguities that can therefore be fixed by the convexity requirement (again, see \cite{companionMink} for details). In what follows we will largely proceed as if this ambiguity was not present, however we will come back to it toward the end of the paper, in \autoref{sec_2pi-area}.

Once we have unambiguously fixed the cosines of the dihedral angles, these can be used to construct the Gram matrix of the tetrahedron
\be
\big({}^3\mathrm{Gram}\big)_{bc} = - \cos \phi_{bc}.
\ee
The determinant of ${}^3 \mathrm{Gram}$ determines whether the tetrahedron is hyperbolic or spherical \cite{Alekseevskij1988}, therefore providing the crucial information
\be
\sgn\,\det\big({}^3 \mathrm{Gram} \big)  = \sgn \,\l .
\ee
This fixes the sign of the cosmological constant at a given vertex. Consequently, there is no freedom, within a vertex, to change this sign, and a unique correspondence between the spinfoam and geometric data can finally be established. Note that flipping the sign of the cosmological constant does not change the Gram matrix, since it corresponds to flipping \textit{all} the $\pm_b$. This fact is crucial, since it means that $\sgn\lt(\lambda\rt)$ can actually be calculated. 

Finally, from the Gram matrix one can fully reconstruct the curved tetrahedron. In practice this amounts to repeatedly applying the spherical (and/or hyperbolic) law of cosines to first calculate the face angles of the tetrahedron and then its side lengths. The fact that this algorithm leads to a tetrahedron actually consistent with the initial data is non-trivial. This fact is proved for a single tetrahedron in the companion paper \cite{companionMink}, but can also be seen as a consequence of the more general reconstruction theorem for the whole 4-simplex proved later in this paper.

\section{Geometry from Critical Point Equations III: Curved 4-Simplex}\label{sec_curved4simplex}

As in the three-dimensional case, the first step is to clarify the topology of the paths on the four simplex given the equations of motion for the holonomies $H_{ab}$ and $G_{ab}$ calculated in \textit{a certain graph framing}. This was done at the end of \autoref{sec_framing}. We recall here the results of that discussion.

The equations of motions can be written in terms of the variables $\tilde H_{ab}$, which all share the same base-point in the 4 simplex. They take the following form:
\be
\left\{
\begin{array}{l}
\tilde H_{ba} = \tilde H_{ab}^{-1} \quad \text{except}\quad \tilde H_{42}=\tilde H_{13}^{-1}  \tilde H_{24}^{-1} \tilde H_{13}\\
\tilde H_{ab_4}\tilde H_{ab_3}\tilde H_{ab_2}\tilde H_{ab_1}= \mathbb{I}
\label{eq_tildedcritical}
\end{array}
\right. .
\ee
Modulo gauge transformations (that is, global parallel transports), the base point of the $\tilde H_{ab}$ can be understood to be vertex 4. The first of these equations tell us that all faces $(ab)\neq(24)$ are traversed along the same path (though in opposite directions) when considered as part of tetrahedron $a$ or tetrahedron $b$. The relation between the holonomies $\tilde H_{24}$ and $\tilde H_{42}$ tells us that the net difference in their paths, is given by face $(13)$. This difference is due to two facts about face $(24)$: (\textit{i})  from the perspective of tetrahedron $2$, this is the face opposite to the base point $\overline{4}$, and must be reached via a special edge, here this is edge $\overline{45}$; (\textit{ii}) from the perspective of tetrahedron $4$ it is also the face opposite to the tetrahedron's base point, vertex $\overline{2}$. Moreover, tetrahedron 4 is itself opposite to the global base point $\overline{4}$, and must be reached via the ``4-simplex special edge'', that is, edge $\overline{42}$. See \autoref{fig_4simplexpersp}.

Now that the paths have been clarified, we begin to analyze the geometry in detail. The strategy we adopt is as follows: First we show that one can associate to a solution of eq. \eqref{eq_tildedcritical} a set of five 4-dimensional hyperplanes in $\R^5$ equipped with the metric $\stackrel{5}{\eta}(\epsilon)_{\alpha\beta}:=\text{diag}(-1,1,1,1,\epsilon)$, where the sign $\epsilon=\pm1$ is determined by the specific values of the holonomies. To given hyperplanes and signature of $\stackrel{5}{\eta}(\epsilon)$ one can easily associate a set of $2^5$ different curved 4-simplices, with the same sign of the curvature as $\epsilon$. Antipodal 4-simplices are related by a flip in orientation, which might be a parity-inversion (for de Sitter) or time-reversal (for anti-de Sitter). Nonetheless, as discussed in \autoref{sec_parity}, these two transformations are degenerate with respect to our bivectorial description. To each of these 4-simplices one can then associate the holonomies along the paths described above, and the ``reconstructed'' holonomies necessarily satisfy eq. \eqref{eq_tildedcritical}. 

Thus, the question is whether among the reconstructed 4-simplices there exists one (and only one) that produces the same holonomies with which we began. We show this fact indirectly. That is, we show that there are at most $2^5$ sets of holonomies that reproduce the same set of 4-dimensional hyperplanes in $\R^5$ (up to rotations and boosts). We also show that within each such set, holonomies come identified in pairs related by an orientation changing transformation. 
%
%
 In the following we will show one by one the previous claims in order to complete the proof.

\subsubsection*{Determination of the hyperplanes and of the sign of the curvature} The closure relations for the tetrahedra can be expressed as SU(2) closure equations. This means that at each vertex the four $\{H_{ab}\}_{b,b\neq a}$, when seen as elements of SO(1,3), stabilize the unit timelike normal $u=(1,0,0,0)^T$, and can therefore be thought of as defining a spacelike frame in which the tetrahedron lives. Notice that if the tetrahedron is not degenerate, as we shall suppose, this is the only 4-vector which is invariant under the action of all of them. As a consequence, the parallel transported closure equations, i.e. those involving the $\{ \tilde H_{ab} \}_{b,b\neq a}$, uniquely identify the future-directed, unit timelike normals $N_a:= \bLambda_a u$, where $\bLambda_a$ represents the $g_a\in\Slc$ action on $\R^{1,3}$. In summary, there exists a unique  $N_a \in \R^{1,3}$ such that: it has norm equal to $-1$, it is future pointing, and it is stabilized by \textit{all} $\{ \tilde \OO_{ab}\}_{b, \, b\neq a}$, where $\tilde\OO_{ab}\in \text{SO}(1,3)^+_\uparrow$ is the representation of $\tilde H_{ab} \in \Slc$ acting on $\R^{1,3}$. This condition on the $\tilde H_{ab}$ implements the non-trivial facts that the $H_{ab}$ are in the SU(2) subgroup of $\Slc$ (existence), and define a non-degenerate tetrahedron (uniqueness). 

The 4-vector $N_a$ is then interpreted as the timelike normal to the $a$-th tetrahedron when expressed in the common frame in which \textit{all} the $\tilde H_{ab}$ are defined. Thus, if this interpretation is valid, it follows that the hyper-dihedral angle $\Theta_{ab}$ between the $a$-th and $b$-th tetrahedra on the boundary of the 4-simplex must be given by
\begin{subequations}
\be
-\cosh \Theta_{ab} := \eta_{IJ} N_a^I N_b^J \equiv  \eta_{IJ} (\bLambda_b^{-1}\bLambda_a u)^I u^J  \,, \quad \text{for } (ab)\neq(24),
\label{eq_coshab}
\ee
where $\eta_{IJ}:=\text{diag}(-1,1,1,1)$ is the Minkowski metric. We have excluded the case $(ab)=(24)$ for similar reasons as in the three-dimensional case of eq. \eqref{eq_dihedral}; the two normals experience incompatible parallel transports before arriving at their common point of definition, as observed at the beginning of this section. Another, maybe more transparent, way to state this fact is by observing that $G_{42}$, which is the parallel transport between the frames of tetrahedra 2 and 4, does not factorize, i.e. it is not simply given by $g_4^{-1} g_2$. In the 4-vector representation, this means---with obvious notation---that $\bLambda_{42}\neq \bLambda_4^{-1}\bLambda_2$.  In fact, the correct formula is
\be
- \cosh \Theta_{24} :=  \eta_{IJ} (\bLambda_{42} u)^I u^J \equiv  \eta_{IJ} (\tilde\OO_{31} N_2)^I N_4^J.
\label{eq_cosh24}
\ee
\end{subequations}
Importantly, this is exactly the same equation one would have written by just looking at the ``simple'' paths on the 4-simplex described in the previous section, and is a perfect analogue of eq. \eqref{eq_dihedral}.\footnote{Explicitly, to compare $N_2$ and $N_4$ one has to parallel transport them to their common face via the appropriate paths contained in tetrahedra $2$ and $4$; if the comparison is made at vertex 5 (any other choice would give the same result), one has to take the inner product between the two time normals only after parallel transporting them respectively along the 4-simplex sides (42) and then (25) in the case of $N_2$, and side (45) in the case of $N_4$. Eventually, this amounts to taking $N_2$ around face (31) before comparing it to $N_4$.} Also, notice that the first equality in the previous formula holds for the dihedral angle at any face, not just for the one at face $(24)$.

As in the tetrahedral case this set of data is enough to reconstruct the curved 4-simplex, including the sign of its curvature. Let us show this by embedding the problem in one dimension more, in such a way that the homogeneous space in which the 4-simplex is defined becomes the ``unit sphere'' of the embedding space. That is, if the 4-simplex is of positive curvature we consider a de Sitter (dS) space embedded in $\R^{1,4}$, conversely if it is of negative curvature we consider an Anti-deSettir (AdS) space embedded in $\R^{2,3}$. Now, the three dimensional subspaces\footnote{These are either 3-spheres $S^3$ or 3-hyperboloids $H^3$, depending on whether one is dealing with dS or AdS spaces.} in which the tetrahedra on the boundary of the 4-simplex live pick out a 4-dimensional hyperplane through the origin of the embedding space. Observe that the converse is also true; given five such 4-dimensional hyperplanes one can easily reconstruct the 4-simplex. More precisely, the latter statement is true up to a $2^5$-fold ambiguity originating from the five binary choices needed to fix which side of the hyperplanes the 4-simplex lies on.\footnote{For more details, and some subtleties arising in the hyperbolic (AdS) case, see the companion paper \cite{companionMink}. We just mention here that one has to consider a two-sheeted AdS space in order to be deal with all possible solutions to the critical point equations. By this, we mean that the spacelike hyperboloid foliating AdS are two-sheeted. Nevertheless, by limiting the boundary data considered, one can always restrict to the more familiar case.} This discussion shows that once we are given the hyperdihedral angles we can determine the curved 4-simplex (and the sign of its curvature) up to rotations (and boosts) and a discrete number of ambiguities. 

To see how this works concretely, consider $\R^5$ with the metric $\stackrel{5}{\eta}(\epsilon)_{\alpha\beta}:=\text{diag}(-1,1,1,1,\epsilon)$, and the sign $\epsilon=\pm1$ to be determined by the specific values of the holonomies, and consider the following four 5-vectors
\be
\mathcal N_a^{\alpha}  := (N_a^I,0), \quad \text{for } a\neq4.
\ee
Each vector has norm $-1$ with respect to the metric $\stackrel{5}{\eta}(\epsilon)$, irrespective of the sign of $\epsilon$. So, the fifth 5-vector $\mathcal N_4$, as well as the sign of $\epsilon$, are determined by the requirements
\begin{subequations}
\begin{align}
\stackrel{5}{\eta}(\epsilon)_{\alpha\beta}\mathcal N_4^{\alpha}\mathcal N_a^{\beta} & = - \cosh\Theta_{4a},\\
\stackrel{5}{\eta}(\epsilon)_{\alpha\beta}\mathcal N_4^{\alpha}\mathcal N_4^{\beta} & = -1,\\
\mathcal N_5^5 &> 0.
\end{align}
\end{subequations}
(The last equation is only needed to obtain a \textit{unique} solution for $\mathcal N_4$, and is otherwise irrelevant.) The 5-vectors $\mathcal N_a$ are then precisely the normals to the five 4-dimensional hyperplanes discussed a moment ago. The second condition can be fulfilled for one sign of $\epsilon$ only, and therefore fixes whether the 4-simplex is positively or negatively curved. Note that with these choices for the $\mathcal N_a$, vertex 4 lies at the ``North (or South) pole'' of dS (or AdS, respectively), i.e. at $(\pm1,0,0,0,0)$.

\subsubsection*{Counting the sets of holonomies satisfying the critical equations}

At this point we need to count how many sets of holonomies satisfy the critical equations. To do so, the strategy is to show that the eqs. \eqref{eq_tildedcritical}, \eqref{eq_coshab} and \eqref{eq_cosh24}, along with the uniqueness property of the $N_a$ discussed in the first paragraph of the last subsection, have at most $2^5$ solutions, with parity transformations relating them in pairs.\footnote{Observe that in the tetrahedral case we could make use of the triple products to fix exactly which solution we were interested in. In the present case this is not possible. The reason is that all the normals to the tetrahedra are taken as future pointing, and therefore some of them will be inward- and others outward-pointing with respect to the 4-simplex. Clearly this problem is related to the fact that it is not possible for a vector to cross the light-cone by means of $\Slc$ transformations.}

{
First, consider eq. \eqref{eq_coshab}. These equations have exactly 4 solutions in terms of the $N_a$, up to a global SO$(1,3)^+_\uparrow$ symmetry. To see this, let us consider the two square submatrices of the 4-simplex Gram matrix
\be
\big({}^4\text{Gram}\big)_{ab} = -\cosh \Theta_{ab},
\ee
given by $a,b\in\{1,2,3,5\}$ and $a,b\in\{1,3,4,5\}$. Call them ${}^4\mathrm{Gram}_{\hat 4}$ and ${}^4\mathrm{Gram}_{\hat 2}$, respectively. These matrices do not contain the ``twisted'' entry associated to the hyperdihedral angle $\Theta_{24}$ (see eq. \eqref{eq_cosh24}). All their entries are given by Lorentzian scalar products as in eq. \eqref{eq_coshab}. Hence, we are trying to solve the equations
\begin{subequations}
\begin{align}
{}^4\mathrm{Gram}_{\hat 4} &= \mathbf N_{\hat{4}}^T \eta  \mathbf N_{\hat{4}} \\
{}^4\mathrm{Gram}_{\hat 2} &=  \mathbf N_{\hat{2}}^T \eta \mathbf N_{\hat{2}}
\end{align}
\label{eq_G=NgN}
\end{subequations}
for the five \textit{future-pointing} 4-vectors $N_a\in\R^4$. The notation is as follows: $ \mathbf N_{\hat{4}}:=\lt(N_1, N_2, N_3, N_5 \rt)$, and $ \mathbf N_{\hat{2}}:=\lt(N_1, N_3, N_4, N_5 \rt)$ are  $4\times4$ matrix whose columns are given by 4 out of 5 of the 4-vectors $N_a$, and $\eta=\text{diag}(-1,1,1,1)$ is the 4-dimensional Minkowski metric. To solve each of the previous equations, start by observing that ${}^4\mathrm{Gram}_{\hat a}$ is symmetric, and can therefore be put in a diagonal form by conjugation via an orthogonal matrix $O_{\hat a}$: ${}^4\mathrm{Gram}_{\hat a}=O_{\hat a}^T D_{\hat a} O_{\hat a}$. At this point, by Sylvester's theorem one has that $D_{\hat a}=\eta E_{\hat a}^2$, where $E_{\hat a}$ can be taken to be a positive diagonal matrix. Thus, we find that $\mathbf{N}_{\hat{a}}$ must be of the form $V_{\hat a} E_{\hat a} O_{\hat a}$, where any $V_{\hat a}\in$O(1,3) is equally viable. Now, time reversal symmetry is broken by the requirement that the $N_a$'s are future pointing. Therefore, considering $\mathbf N_{\hat a}$ to be defined only up to an SO$(1,3)^+_\uparrow$ gauge, one is left with two solutions  related by a parity inversion symmetry for each entry of eq. \eqref{eq_G=NgN}.\footnote{A more pedestrian proof goes as follows. Clearly, we are interested in solutions up to gauge, that is up to the action of an element of SO$(1,3)^+_\uparrow$, acting diagonally on all the $N_a$. We can use this freedom to gauge fix $N_1$ to be $(1,0,0,0)$ and the plane $N_3\wedge N_5$ to be the same as (and have the same orientation of) the plane $\star \hat x\wedge\hat y$, with the spacial part of $N_3$ being parallel to $\hat x$, that is $N_3^I \hat y_I=N_3^I \hat y_I=0$ and $N_3^I \hat x_I>0$. Here, $\hat x:=(0,1,0,0)^T$, and so on. After such a gauge fixing, a moment of reflection shows that the solution to each one of the previous two equations is unique up to reflection with respect to the $N_3\wedge N_5$ plane. In fact, the equation is characterized by a full O$(1,3)$ symmetry, rather than just by an SO$(1,3)^+_\uparrow$ one. } Because we are not yet imposing any relation between $N_2$ and $N_4$, we see that there are at this level exactly 4 different solutions for the five 4-vectors $N_a$.
}

We now move on, to consider the missing information provided by $\big({}^4\text{Gram}\big)_{24}=-\cosh \Theta_{24}$ (eq. \eqref{eq_cosh24}). This clearly imposes a relation between $N_2$ and $N_4$. Nevertheless, this relation is mediated by a new variable, namely $\tilde\OO_{13}$. Therefore the extra information provided by the knowledge of $\cosh \Theta_{24}$ will be used to put constraints on $\tilde\OO_{13}$ (alone). Observe that $\tilde\OO_{13}$ stabilizes $N_1$ and $N_3$, and therefore by choosing a gauge in which $N_1=u=(1,0,0,0)^T$, $\tilde \OO_{13}$ is a pure rotation around the axis $\vec N_3$. Hence, for each solution of eqs. \eqref{eq_coshab} and \eqref{eq_cosh24} one has at most two solutions in terms of $\tilde \OO_{13}$.

We can obtain a similar result for $\tilde \OO_{53}$. Indeed, by using eq. \eqref{eq_5vertices} relative to vertex 3, one gets the identity
\be
\tilde \OO_{31} = \tilde \OO_{43} \tilde \OO_{53} \tilde \OO_{23} \,,
\ee
and by using the fact that $\tilde \OO_{ab}$ stabilizes $N_a$ and $N_b$, one obtains
\be
- \cosh \Theta_{24} =  \eta_{IJ} (\tilde\OO_{31} N_2)^I N_4^J =\eta_{IJ} (\tilde\OO_{53} N_2)^I N_4^J .
\ee
Therefore, for any solution of the eqs. \eqref{eq_coshab} and \eqref{eq_cosh24} thus far considered, there are two additional solutions for $\tilde\OO_{53}$.

We can follow the same reasoning using the closure at vertex 1, obtaining two more solutions for $\tilde \OO_{15}$.

Now, consider the closure equation for vertex 3. 
\be
\tilde \OO_{31} \tilde \OO_{32} \tilde \OO_{35} \tilde \OO_{34} = \mathbb{I}.
\label{eq_temporarylabel}
\ee
At this point we have fixed both $\tilde \OO_{31}$ and $\tilde \OO_{35}$. We claim that this equation has a unique solution compatible with the fact that $\tilde \OO_{32}$ and $\tilde \OO_{34}$ must respectively stabilize  $N_2$ and $N_4$, as well as $N_3$. Indeed, since all the $\tilde \OO_{3a}$ stabilize $N_3$, we can reduce the problem to SU(2). In a gauge where $N_3=u$, one has that $\tilde\OO_{3a}$ is a rotation around $\vec N_a$; for added clarity rewrite eq. \eqref{eq_temporarylabel} as
\be
\tilde \OO_{31} \mathbf{R}_2 = \mathbf{R}_4 \tilde \OO_{53},
\ee
 where $ \mathbf{R}_2$ and $ \mathbf{R}_4$ are rotations of unknown angles around $\vec N_2$ and $\vec N_4$, respectively. Contract this equation with $\vec N_2$ and the result is
 \be
 \mathbf{R}_4 \vec v = \vec w,
 \ee
 where $\vec v$ and $\vec w$ are known vectors (recall that $\mathbf{R}_2\equiv\tilde\OO_{32}$ stabilizes $N_2$). Some thought shows that this equation has at most one solution. Therefore, one can in this way completely fix $\tilde \OO_{32}$ and $\tilde \OO_{34}$.
 
Using analogous argument at vertex 1, it is possible to fix uniquely $\tilde \OO_{12}$ and $\tilde \OO_{14}$. Hence, there are only three holonomies left to be fixed: $\tilde \OO_{25}$, $\tilde \OO_{54}$, and $\tilde \OO_{24}$. This is readily done in a unique way by applying the above arguments at vertices 2 and 5. 
 
{
We are finally left with a total of at most $4\times2\times2\times2=2^5$ solutions, where the factor of 4 comes from the fixing of the $N_a$, and the factors of 2 come from the fixing of $\tilde \OO_{13}$, $\tilde \OO_{53}$, and $\tilde \OO_{15}$.  Notice that one can determine the $N_a$ only up to a reflection in their spacelike components, that is, up to parity.\footnote{This symmetry was taken into account in the preceding counting.} (The fact that the time direction is preserved is due to our hypothesis that the time normals are always future pointing, irregardless of whether they are inward- or outward-pointing with respect to the 4-simplex.) These new solutions can be easily seen to be related to the transformation $A \mapsto \bA$, i.e. $\tilde H_{ab} \mapsto (H_{ba}^\dagger)^{-1}$ (see also \autoref{sec_parity}). This comes as no surprise, since also the five hyperplanes in $\R^5$ subdivide dS, or AdS respectively, in $2^5$ sectors, with a two-fold ``redundancy'' due to parity. We can finally conclude, that any solution to the critical-point equations corresponds to one of the $2^5$ 4-simplices obtained in the way described above.\\%
} %

\subsubsection*{Concluding remark}
One of the main consequences of what we have just shown, is that, in order to solve the critical-point equations, all the tetrahedra must be characterized by the \textit{same} sign of the cosmological constant, hence fixing all the $\nu^a$ at each vertex to be equal.\footnote{Contrary to the last section, we here indicate with $\nu^a$, the sign of $\lambda$ at the vertex $a$.} Hence, the sign of the cosmological constant is a \textit{dynamically} determined quantity that takes---on shell---a consistent value throughout the whole 4-simplex.

%
%
%
%
%


\section{Critical Value of the Action}\label{sec_criticalaction}
 
 In the previous section we showed how the critical points of the $\G_5$ Wilson graph operator obtained in the double scaling limit are related to a curved 4-dimensional simplicial geoemtry. In this and the following section we turn our attention to the evaluation of the spinfoam and Chern-Simon actions at the critical point. We will find that the first corresponds to the Regge action associated to the curved 4-simplex (see \autoref{app_GRconv}), while the second corresponds to the relevant cosmological term. \\

 
 \subsection{The Wilson Graph Operator}\label{sec_criticalwilsongraph}
 
A straightforward calculation shows that by inserting the critical point equations of \autoref{sec_SPA} into the spinfoam action $S_{\G_5}$ of eq. \eqref{IG5}, one obtains 
 \be
S \big|_0 = \I \sum_{a<b} -2j_{ab} \varphi_{ab} - 2\gamma j_{ab} \psi_{ab}.
\label{eq_s0}
 \ee
 
{ Equation \eqref{gluing} clearly shows that the phases $\varphi_{ab}$ appearing in this equation are related to the choice of phase for the spinors $\xi_{ab}$. These phases can be interpreted in the context of ``framed polyhedra'' \cite{Hnybida2014}, where they represent the direction of a unit vector lying \textit{on} the face $(ab)$ of the tetrahedron $a$. 
In our context, these frames are part of the boundary data, and might be chosen so that the first term in the previous expression is zero (the boundary states with this property are called ``Regge states'' in \cite{Barrett2010}). This choice depends on the geometry of the four simplex, see \autoref{sec_NRterms}. In order to keep track of all the dependencies on the geometry, we do not make this choice here, and work with the most general formulae. Anyway, we postpone further discussion of these phases to \autoref{sec_NRterms}..} 

Here, we focus on the second term of eq. \eqref{eq_s0}, the one involving the face areas $\gamma j_{ab}$ and the variables $\psi_{ab}$. We turn to giving a geometrical meaning to the latter variables. Let us start from the parallel transport equations (\ref{gluing}) recast in the more compact form
\be
 (J \xi_{ab}, - \xi_{ab}) = 
G_{ab} ( \xi_{ba}, J \xi_{ba})
\lt(
\begin{array}{cc}
\E^{\psi_{ab}+\I\varphi_{ab}} & 0 \\
0 & \E^{-\psi_{ab}-\I\varphi_{ab}}
\end{array}
\rt),  
\ee
 where $(\xi,J\xi)$ is a $2\times2$ matrix with the spinor $\xi$ as the first column and spinor $J\xi$ as the second. This matrix is the SU(2) rotation mapping the spinor $(1,0)^T$ to $\xi$. In vectorial language, it maps the $\hat z$-axis to $\hat n(\xi)$. Introducing the shorthand $D(\xi):=(\xi,J\xi) \in \text{SU}(2)$, the previous equation can be written 

   \be
 G_{ab} = D(J \xi_{ab})\lt(
\begin{array}{cc}
 \E^{\psi_{ab}+\I\varphi_{ab}} & 0 \\
0 &\E^{-\psi_{ab}-\I\varphi_{ab}}
\end{array}
\rt)^{-1} D(\xi_{ba})^{-1}     \equiv D(J \xi_{ab}) \E^{-(\psi_{ab} + \I \varphi_{ab} ) \sigma_z} D(\xi_{ba})^{-1}  ,
\label{eq_G=DexpD}
 \ee
where we used the fact that $J^2=-1$. In this form the interpretation of the $G_{ab}$ is transparent: in vectorial language, it first rotates the frame of  face $(ab)$, contained in tetrahedron $b$ and defined by its normal $-\hat n_{ba}$, onto the $z$-axis. Then it boosts this frame in the $\hat z$-direction  (orthogonal to the face) and rotates it around $\hat z$ (along the plane of the face), and finally it takes the $\hat z$-axis into the unit normal $\hat n_{ab}$ of the face $(ab)$, contained in tetrahedron $a$. Importantly, we can both recognize the significance of the phase $\varphi_{ab}$ as a rotation angle, and characterize the boost parameter between the frames of the two tetrahedra $a$ and $b$, as being $2\psi_{ab}$. To make this claim completely explicit, it is enough to write eq. \eqref{eq_G=DexpD} in the vectorial representation, yielding
 \be
 \bLambda_{ab} = \bR(J\xi_{ab}) \E^{2 \psi_{ab} \bK_z +2 \varphi_{ab} \bJ_z} \bR(\xi_{ba})^{-1}  ,
 \label{eq_RBR}
 \ee
 where $\bR(\xi)\in \text{SO}(3)\subset \text{SO}(1,3)$ is the vectorial representation of $D(\xi)\in \text{SU}(2)\subset\Slc$. Note that $\bR(\xi)$ stabilizes the four vector $U^I=(1,0,0,0)^T$. Finally, by contracting this equation with $U$ itself, one finds
\be
-\cosh \Theta_{ab} = \eta_{IJ} (\Lambda_{ab} U^I ) U^J = -\cosh 2\psi_{ab}\,,
\ee
and hence
\be
|\Theta_{ab}| = 2\,|\psi_{ab} |.
\ee
 Therefore, $2\,|\psi_{ab} |$ is the magnitude of the hyper-dihedral angle hinged at the face $(ab)$.  
 
 Fixing, in geometrical terms, the sign $\psi_{ab}$ is more subtle.\footnote{We could just take the result of \cite{Barrett2010}, obtained in the context of the flat EPRL model, corresponding to an infinite Chern-Simons coupling (i.e. to a vanishing cosmological constant). Indeed, their result must be valid also in our case: since one can imagine to take the flat limit continuously without letting the dihedral angle ever vanish, by continuity the sign calculated in the flat case must coincide with the one in the curved case. Nonetheless, in order to keep our presentation completely self-contained, we provide an alternative argument.}  From the previous discussion, it should be clear that $\bLambda_{ab}$ is a Lorentz transformation sending the timelike direction of tetrahedron $b$ into the timelike direction of tetrahedron $a$. Moreover, because $\bLambda_{ab}$ is an element of the vectorial representation of $\Slc$, $\Lambda_{ab}\in\text{SO}(1,3)_\uparrow^+$ (the proper orthochronus Lorentz group), it sends future-pointing vectors into future-pointing vectors. All of this said, generically it remains the case that the oriented geometric time-normals of the tetrahedra on the boundary of the 4-simplex are not all future pointing.
 
 So, consider the transformation $\mathbf{L}_{ab}\in\text{SO}(1,3)^+_\uparrow$, defined as a function of the future directed 4-vectors $
N_a$ and $N_b$
 \be
 \mathbf{L}_{ab} := \exp \lt( |\Theta_{ab}| \frac{\varsigma( N_b \wedge N_a}{|N_b\wedge N_a|}  \rt),
 \ee
 where $\varsigma(N_b \wedge N_a ):= (N_b)_{[I} (N_a)_{J]} \mathbf J^{IJ} \in \mathfrak{so}(1,3)$. It is easy to check that $ \mathbf{L}_{ab} N_b = N_a$. Let $\tilde N_b$, and $\tilde N_a$ be the oriented geometric normals to the tetrahedra $b$ and $a$, respectively. One then has, $\tilde N_b = \pm N_b$, with the sign being determined by whether the tetrahedron is future or past pointing. Similar considerations hold for tetrahedron $b$. Define also 
 \be
 \sgn \;\Theta_{ab} = -\sgn \;\eta_{IJ} \tilde N_b^I \tilde N_a^J \;.
 \ee
 This definition fits well with the geometrical requirements of a discrete Lorentzian geometry \cite{BarrettFoxon,Barrett2010}. Then, 
  \be
 \mathbf{L}_{ab} = \exp \lt( \Theta_{ab} \frac{\varsigma(\tilde N_b \wedge \tilde N_a)}{|\tilde N_b\wedge \tilde N_a|}  \rt).
 \ee
Geometrically, the wedge product $\tilde N_b \wedge \tilde N_a$ is orthogonal to the triangle shared by tetrahedra $a$ and $b$, and must therefore be proportional, when calculated in the frame of tetrahedron $b$, to $-u\wedge n_{ba}$, where $u=(1,0,0,0)^T$ and $n_{ba} = (0, \hat n_{ba})^T$. In \autoref{app_Lorentziangluing} it is shown that $(\tilde N_b \wedge \tilde N_a)(b) = - u\wedge n_{ba}$ for all orientations of the normals $\tilde N_b$ and $\tilde N_a$. Hence,
 \be
 \mathbf{L}_{ab}(b) := \exp \lt( -\Theta_{ab} \frac{\varsigma(  u\wedge n_{ba}}{| u\wedge n_{ba}|}  \rt) = \exp\lt( - \Theta_{ab} \hat n_{ba}\vec{\mathbf K} \rt)\,.
 \ee
 Now, observe that both $\mathbf{L}_{ab}(b)$ and $\mathbf{\Lambda}_{ab}$ have the property of sending $u$ into the future pointing normal $N_a(b)$. This means that, when written in the form of a rotation in the source space times a pure boost, their boost parts must agree. The group element $\mathbf{L}_{ab}(b)$ has already been written in the form of a pure boost, while $\mathbf{\Lambda}_{ab}$ can be expressed in the form
 \begin{align}
 \mathbf{\Lambda}_{ab} & =  \lt[\bR(J\xi_{ab}) \E^{2 \psi_{ab} \bK_z } \bR(J\xi_{ab})^{-1} \rt] \lt[\bR(J\xi_{ab}) \E^{2 \varphi_{ab} \bJ_z} \bR(\xi_{ba})^{-1} \rt] \notag\\
 & =   \E^{ 2 \psi_{ab} (- \hat n_{ba}) \vec \bK} \bR',
 \end{align}
 where we used the fact that $\bR(J\xi_{ab}) \hat z = -\hat n_{ba} $, and that $K_z$ commutes with $J_z$. Finally, $\bR'$ is given by the second bracket in the first line and is a pure rotation in the source space. From this decomposition, we can immediately conclude that
 \be
 \exp\lt( - \Theta_{ab} \hat n_{ba}\vec{\mathbf K} \rt) = \exp\lt(- 2 \psi_{ab} \hat n_{ba}  \vec \bK\rt)
 \ee 
 and hence
 \be
 2\psi_{ab} = \Theta_{ab}.
 \ee
 which fixes the sign of $\psi_{ab}$.
 
 This allows us to give the following expression for the spinfoam action at the critical point 
\be
 S\big|_0 = -\frac{\I}{\ell_{P}^2} \sum_{a<b}  \mathrm{a}_{ab} \Theta_{ab} -2\gamma^{-1} \mathrm{a}_{ab}  \varphi_{ab}.
\ee
We will comment later on the r\^ole of the $\varphi_{ab}$ variables.


 \subsection{The Chern-Simons Functional}\label{sec_criticalCS}
 
In this section we evaluate the Chern-Simons functional at the critical point. The calculation is analogous to those performed in the case of knot complements, see e.g. \cite{Kirk1993}. The strategy consists of evaluating the \textit{variation} of the Chern-Simons functional due to a small change in the boundary geometry (encoded in the spins and coherent-state spinors, 
see \autoref{sec_curvedtets}), and showing that this change is the same one would obtain by varying the volume functional. The Schl\"afli identities (see e.g. \cite{eva}, and for a recent symplectic proof \cite{Haggard2014}) state how the volume of a curved 4-simplex responds to any variation $\delta$ of the geometry of the simplex\footnote{{Analogous Schl\"afli identities hold in all dimensions.}}
\be
\lambda \delta V_4 = \sum_{t\subset \sigma} \text{a}_t \delta \Theta_t\,,
\ee
where $\lambda$ is the value of the \textit{geometrical} cosmological constant, and $V_4$ is the 4-volume of the 4-simplex $\sigma$, $\text{a}_t$ is the area of the triangle $t$ and $\Theta_t$ is the dihedral angle hinged by the triangle $t$. Note that the sign of the cosmological constant on the left-hand side is simply given by the character of the 4-simplex, that is, it is positive (or negative) provided the simplex is embedded in dS (or AdS). 
 
 Therefore, let us start by considering the solution $(A,\bA)$ of the critical point equations with boundary data given by $( j_{ab}, \xi_{ab})$ and its small variation $(A+\delta A,\bA + \delta \bA)$ due to the change $( j_{ab}+\delta j_{ab}, \xi_{ab}+\delta\xi_{ab})$ in the boundary data. The Chern-Simons functional depends explicitly only on the connection $(A,\bA)$, thus
 \be
 \delta W[A] := W[A+\delta A] - W[A] = \int_{S_3} \D^3 x \frac{\delta W[A]}{\delta A(x)} \delta A(x) + O(\delta A^2).
 \ee
 Henceforth every equality between small variations is meant up to an $O(\delta A^2)$ unless otherwise stated. 
 
 Using
 \be
 \frac{\delta W[A]}{\delta A_\mu^i(x)} = -\frac{1}{8\pi} \epsilon^{\mu\nu\rho} F^i_{\nu\rho}[A](x)
 \ee
 and the equation of motion for the curvature, eq. \eqref{distributionalcurvature}
 \be
 \epsilon^{\mu\rho\sigma}F^i_{\rho\sigma}(x) \stackrel{!}{=} - \frac{16\pi}{\I h}(1+\I\gamma) \sum_{(ab),a>b} j_{ab} \;\langle \xi_{ba} , \left[(G_{s_{ab},b})^{-1} \tau^i G_{s_{ab},b} \right] \xi_{ba}\rangle\;\delta_{\ell_{ab}}^{(2)\;\mu}(x),
\ee
 we obtain the following expression for $\delta W[A]$
 \begin{align}
 \delta W[A] &=  +\frac{1}{8\pi}  \int_{S_3} \D^3 x \frac{16\pi}{\I h}(1+\I\g)\sum_{(ab),a>b}j_{ab}   \langle \xi_{ba} , \left[(G_{s_{ab},b})^{-1} \tau^i G_{s_{ab},b} \right] \xi_{ba}\rangle\;\delta_{\ell_{ab}}^{(2)\;\mu}(x) \delta A^i_\mu(x)   \notag\\
 & = -\frac{2\I}{h}\lt(\I + \frac{1}{\g} \rt) \sum_{a<b} \g j_{ab}  \int_{\ell_{ab}} \langle \xi_{ba} , \left[(G_{s_{ab},b})^{-1} \delta A(\ell_{ab}(s_{ab})) G_{s_{ab},b} \right] \xi_{ba}\rangle  .
 \end{align}
 
 It is easy to see that $\int_\ell G_{s b}^{-1} \delta A(s) G_{sb} = G_\ell^{-1} \delta G_\ell$, where $\delta G := G[A+\delta A] - G[A]$ signifies a difference between $\Slc$ matrices. Indeed, by first writing $G_\ell$ as 
 \be
G_\ell=\lim_{n\to\infty}\lt[\mathbb I+\int^1_{s_n}A(s)\rt]\cdots\lt[\mathbb I+\int^{s_1}_{0}A(s)\rt]\,,
\ee
where $0<s_1<\cdots<s_n<1$ is a partition of the interval $(0,1)$,  one can directly compute the variation $ G_\ell^{-1} \delta G_\ell$:
\be
G_\ell^{-1}\delta G_\ell&=&\lim_{n\to\infty}\sum_{k=1}^n\lt[\mathbb I+\int^{s_1}_{0}A(s)\rt]^{-1}\cdots
\lt[\mathbb I+\int^{s_{k+1}}_{s_k}A(s)\rt]^{-1}\int_{s_k}^{s_{k+1}}\delta A(s)\lt[\mathbb I+\int^{s_k}_{s_{k-1}}A(s)\rt]\cdots\lt[\mathbb I+\int^{s_1}_{0}A(s)\rt]\nonumber\\
&=&\lim_{n\to\infty}\sum_{k=1}^n\int_{s_k}^{s_{k+1}} G_{s_{k+1}b}^{-1}\delta A(s)G_{s_{k}b}
=\int_\ell G_{sb}^{-1}\delta A(s)G_{sb}.
\ee
In these last equations $b$ denotes the source of the curve $\ell$. We therefore find that 
\be
\delta W[A] = -\frac{2\I}{h}\lt(\I + \frac{1}{\g} \rt) \sum_{a<b} \g j_{ab}  \lag \xi_{ba},G_{ab}^{-1} \delta G_{ab}\xi_{ba}\rag  .
\ee
 
 The parallel-transport equations can be recast as in eq. \eqref{eq_G=DexpD}, where $\psi_{ab}$ and $\varphi_{ab}$ must be understood as a function of the boundary data $(j_{ab}, \xi_{ab})$. Therefore the variation $\delta G_{ab}$ can be expressed as
 \begin{align}
 \delta G_{ab} = \delta D(J\xi_{ab}) D(J\xi_{ab})^{-1} G_{ab}
 + G_{ab} D( \xi_{ba}) \delta D( \xi_{ba})^{-1}
 - D(J\xi_{ab}) \delta(\psi_{ab} + \I \varphi_{ab})\sigma_z \E^{-(\psi_{ab} + \I \varphi_{ab})\sigma_z} D(\xi_{ba})^{-1}. \notag\\
 \label{eq_deltaG}
 \end{align}
 Recall that $D(\xi) := (\xi, J\xi)$, and therefore $D(\xi+\delta\xi) = D(\xi) + D(\delta\xi) $, from which one immediately deduces that $ D(\xi+\delta\xi)^{-1} = D(\xi)^{-1} -D(\xi)^{-1} D(\delta \xi)D(\xi)^{-1}  $ at first order. On the other hand, because $\xi$ is normalized, its variations can only be orthogonal to the spinor itself, that is $\delta \xi = \epsilon J\xi + \I \delta\phi \xi$, for some small $\epsilon\in\mathbb C$ and $\delta \phi\in\mathbb R$. Now, consider the contribution of the second term in the previous equation to the variation $\delta W[A]$, it reads
 \begin{align}
 \lt\langle \xi_{ba}, G_{ab}^{-1} \lt[ G_{ab} D(\xi_{ba}) \delta D(\xi_{ba})^{-1}\rt] \xi_{ba}   \rt\rangle 
& =  - \lt\langle \xi_{ba},   D(\delta \xi_{ba}) D(\xi_{ba})^{-1}\xi_{ba}   \rt\rangle \notag\\
 & =  - \lt\langle \xi_{ba},  D(\delta \xi_{ba}) +_z   \rt\rangle  \notag\\
 & =  - \lt\langle\xi_{ba},   D(\epsilon_{ba} J \xi_{ba}) +_z   \rt\rangle  - \lt\langle \xi_{ba}, D(\I\delta\phi_{ba}  \xi_{ba})  +_z   \rt\rangle \notag\\
 & =  \epsilon_{ba} \lt\langle \xi_{ba}, J\xi_{ba}   \rt\rangle  - \I \delta\phi_{ba}  \lt\langle \xi_{ba},\xi_{ba}  \rt\rangle  \notag\\
 &= -\I \delta\phi_{ba},
 \end{align}
 where $|+_z\rangle := (1, 0)^T \in \mathbb C^2$, and $|-_z\rangle := J|+_z\rangle \equiv (0,1)^T \in \mathbb C^2$. To make the final part of the calculation more explicit notice that in bra-ket notation $D(\xi) \equiv | \xi \rangle\langle +_z | + | J \xi \rangle\langle -_z | $, from which one finds $D(\xi)^{-1} \equiv D(\xi)^\dagger = | +_z \rangle\langle  \xi | + | -_z \rangle\langle J \xi | $, and also $D(J\xi) =| J\xi \rangle\langle +_z | - |  \xi \rangle\langle -_z | $.
 
 With the help of the parallel transport equations and the properties of the complex structure $J$, and in a manner analogous to that just used, one can show that the first term of eq. \eqref{eq_deltaG} contributes to $\delta W[A]$ a quantity proportional to 
 \begin{align}
 \lag \xi_{ba},G_{ab}^{-1} \lt[\delta D(J\xi_{ab}) D(J\xi_{ab})^{-1} G_{ab}\rt]\xi_{ba}\rag  
 &=  \lag r_{ab}^{-1} \E^{-\I\varphi_{ab}} J\xi_{ab}, \delta D(J\xi_{ab}) D(J\xi_{ab})^{-1} \lt(r_{ab}\E^{-\I\varphi_{ab}} J\xi_{ba}\rt)\rag \notag\\
 &= \lag J\xi_{ab},  D(\delta J\xi_{ab}) +_z\rag \notag\\
 &= \lag J\xi_{ab}, (-\bar\epsilon_{ab}\xi_{ab} - \I\delta\phi_{ab} J\xi_{ab}\rag \notag\\
 &=  - \I\delta\phi_{ab}.
 \end{align}
  Therefore, the first order variation of $\delta W[A]$ is
 \be
 \delta W[A] =  -\frac{2\I}{h}\lt(\I + \frac{1}{\g} \rt) \sum_{a<b} \g j_{ab} \Big[ -\I\delta\phi_{ab} - \I\delta \phi_{ba} -  \delta(\psi_{ab} + \I \varphi_{ab}) \Big].
 \ee
 

Now, define
\be
\Phi_{ab} := 2(\varphi_{ab} + \phi_{ab} + \phi_{ba}).
\ee
Later, we will show that this is a function of the spins, independent of the (arbitrary) choice of the phases $\phi_{ab}$ associated to the spinors $\xi_{ab}$.

 Hence, the variation of the Chern-Simons part of the action is
 \begin{align}
 \delta\text{CS}[A,\bA] &:=  \Re\Big( h \,\delta W[A] \Big) = - \sum_{a<b}  \g j_{ab} ( 2\delta\psi_{ab} + \delta \Phi_{ab} ) .
 \end{align}
At this point we can use the results of the 4-simplex reconstruction performed in the previous sections, stating that $\text{a}_{ab} = \ell_P^2 \gamma j_{ab}$ and $2\psi_{ab} = \Theta_{ab}$, to find how the Chern-Simon functional evaluated at the critical point responds to a change in the geometry (encoded in the boundary data $(j_{ab},\xi_{ab})\mapsto(j_{ab}+\delta j_{ab},\xi_{ab}+\delta\xi_{ab})$). According to the Schl\"afli identities it varies just like the volume functional does,
\begin{align} 
\delta \text{CS}[A,\bA]\big|_0 & = - \frac{1}{\ell_P^2}\sum_{a<b} \text{a}_{ab} \delta\Theta_{ab} + \text{a}_{ab} \delta\Phi_{ab}  \notag\\
& =  -\frac{1}{\ell_P^2} \lambda \delta V_4  - \frac{1}{\ell_P^2}\sum_{a<b} \text{a}_{ab} \delta\Phi_{ab}.
 \end{align}
In the last expression we have reintroduced a subscript zero on the left-hand side to emphasize that the variation is taken on-shell. Notice that the sign of the cosmological constant in the previous equation is the one determined by the reconstruction theorem (while its magnitude is equal to $|\Lambda|$). Unfortunately, for the moment we do not have enough control on the phases $\Phi_{ab}$ to be able to give a geometrical meaning to the variation appearing in the second term. We introduce for purely notational purposes the real function $C_\Phi$ such that $\delta C_\Phi=\sum_{a<b} \text{a}_{ab} \delta\Phi_{ab} $. This is possible in principle since, as we shall argue later on, the $\Phi_{ab}$ are functions of the geometry only.

At this point it is enough to integrate the variations to find that
\be
\label{eq_CSCint}
 \text{CS}[A,\bA]\big|_0 =  -\frac{1}{\ell_P^2} \lambda  V_4   - \frac{1}{\ell_P^2}  (C_{\Phi} + C_\text{int}),
\ee
where $C_\text{int}\in\mathbb{R}$ is some integration constant of a topological nature, that is, it is independent of the geometry of the solution (i.e. of the boundary data $(j_{ab},\xi_{ab})$). {In particular, it is expected to depend on the gauge (recall that the Chern-Simons functional $W[A]$ is gauge invariant only modulo $2\pi$) and also on the choice of framing for the graph (for example on whether one decides to replace the path defining $G_{ab}$ with a path winding once more around the edge $\ell_{ab}$, i.e. on whether one decides to perform a Dehn twist on the tube $(ab)$). A more thorough  characterization of this term is left for future work.}
 
In \autoref{app_perturbative}, we perform a perturbative evaluation of the Chern-Simons invariant around a flat solution. This helps to determine what kinds of terms, other than the 4-volume, appear in the evaluation.

 
 \subsection{The total action}\label{sec_criticaltotaction}
 
 Putting together the results from the last two sections, we can finally state the main result of this paper: the total action for the Chern-Simons plus Wilson-graph operator, evaluated at the critical point $(A,\bA,j_{ab},\xi_{ab})$ of the double scaling limit $j\rightarrow \infty, \; h\rightarrow \infty,\; j/h \sim const$, is given by the Regge action of the curved 4-simplex augmented by a cosmological term:
 \be
 I\big|_0 \equiv S\big|_0 -\I  \text{CS}\big|_0 = -\frac{\I}{\ell_P^2} \lt[  \sum_{a<b} \text{a}_{ab} \Theta_{ab} - \lambda V_4 \rt] -\frac{ 2\I}{\g \ell_P^2} \sum_{a<b} \text{a}_{ab} \varphi_{ab} + \frac{\I}{\ell_P^2}  ( C_\Phi + C_\text{int})\,.
 \label{eq_result}
 \ee


\section{Parity-Reversal Symmetry}\label{sec_parity}

Because of the symmetry between the equations involving the selfdual and the anti-selfdual parts of the $\Slc$ connection, it is not hard to realize that the transformation 
\be
\mathcal P : \quad \A=(A,\bA) \quad \mapsto \quad \bar{\A}=(\bA,A),
\ee
is actually a symmetry of the equations of motion. The main consequence of this transformation is that 
\be
\mathcal P : \quad G_\ell[A]:=\mathbb P\exp \int_\ell A \quad\mapsto  \quad G_\ell[\bA] = \mathbb P\exp \int_\ell \bA \equiv (G[A]^\dagger)^{-1},
\ee
where $G_\ell[A]$ stands for an arbitrary holonomy along a path $\ell$, not necessarily a longitudinal one. In terms of the parameters $\psi_{ab}$, this translates into:
\be
\mathcal P : \quad \psi_{ab} \quad \mapsto \quad -\psi_{ab}\,.
\ee
Therefore, $\mathcal P$ changes the sign of the Regge part of the action evaluated at the critical point:
\be
\mathcal P :  \quad  \lt[  \sum_{a<b} \text{a}_{ab} \Theta_{ab} - \lambda V_4 \rt] \quad \mapsto \quad -\lt[  \sum_{a<b} \text{a}_{ab} \Theta_{ab} - \lambda V_4 \rt]\,,
\ee
while the rest of critical action is left unchanged. This transformation can therefore easily be interpreted as a change in the orientation in the reconstructed spacetime. This is a common feature of all the flat spinfoam models starting from the Ponzano-Regge one (see \cite{Barrett2010,Han2012,Han2013a,Engle2013}), but also of the minisuperspace cosmological models where both the wave functions of the expanding and contracting universe appear as solutions of the dynamical equations \cite{Kiefer:2013jqa}.

What might seem puzzling at this point is the fact that in the previous sections we obtained a clear-cut result on the relation between the sign of $\psi_{ab}$ and that of $\Theta_{ab}$. The point is that in the derivation of that result an implicit hypothesis was used; the hypothesis that the reconstructed geometry had the same orientation as that induced by the choice of paths defining the transverse holonomies. We mean that the ``simple'' paths were supposed to circulate around the faces of the tetrahedra in a right-handed, outward-pointing fashion. This was a natural assumption, but not quite a necessary one. Indeed, by changing the parity of the tetrahedra while keeping the same closure equations, one would have found a consistent description by interpreting the $\hat n_{ab}$ as right-handed, inward-pointing normals. This would correspond to taking the mirror image of \autoref{fig_simplepath}. This new construction alters the results of \autoref{app_Lorentziangluing} by a sign, and consequently the same sign change would appear in the equation relating $\psi_{ab}$ and $\Theta_{ab}$. 

From the face-bivector perspective, flipping the spacial normal to the face is the same as flipping the timelike one. Hence, a roughly equivalent description of this change in parity, can be obtained by thinking of it as a change in the time orientation. These two descriptions of the orientation change are equivalent, or maybe it is better to say ``degenerate'', since no actual gauge (Lorentz) transformation connects these two indistinguishable geometries. 



\section{Parity Invariant non-Regge Terms}\label{sec_NRterms}

In this section we want to comment briefly on the extra terms appearing in the asymptotic formula of eq. \eqref{eq_result}. We need to understand the phases $\varphi_{ab}$, and in particular how they can be calculated, and what geometrical meaning they bear. 

The first observation one can make is that the transformation $\mathcal P$ leaves the $\varphi_{ab}$ invariant. At this stage, it is easy to convince oneself that the contribution to the critical action from the non-Regge terms
\be
\text{NR}[\xi_{ab},j_{ab}] := -\frac{ 2\I}{\ell_P^2} \sum_{a<b} j_{ab} \varphi_{ab} + \frac{\I}{\ell_P^2}  ( C_\Phi + C_\text{topo})
\ee
must be $\mathcal P$-invariant. (To this end, note that the variations $\delta \phi_{ab}$ depend only on the choice of the boundary state and are therefore independent of the connection $(A,\bA)$). This tells us that these terms have different symmetry properties, compared to the Regge contribution to the critical action, and can therefore be isolated.

The formulae that fix the values of the $\varphi_{ab}$ (and of the $\psi_{ab}$) are essentially the crossing relations, in one of their forms these are as in eq. \eqref{A2_BIS_1}. Let us consider for example the equation for the cycle (125)
\be
G_{52} G_{21} G_{15} = \mathbb I.
\ee
By using eq. \eqref{eq_G=DexpD}, %
this equation can be given the form
\be
D(\xi_{51})^{-1} 
 D(J \xi_{52}) \E^{-(\psi_{52} + \I \varphi_{52} ) \sigma_z} D(\xi_{25})^{-1} 
  D(J \xi_{21}) \E^{-(\psi_{21} + \I \varphi_{21} ) \sigma_z} D(\xi_{12})^{-1} 
   D(J \xi_{15}) \E^{-(\psi_{15} + \I \varphi_{15} ) \sigma_z}     
      =\mathbb I.
\ee
To simplify this expression one can then transform the spinors
\be
\xi_{ab} \mapsto \xi'_{ab} = \E^{\I\phi_{ab}^c} \xi_{ab}
\ee
in such a way that 
\be
D(\xi_{ab})^{-1}  D(J \xi_{bc}) \mapsto \exp\lt( 2\theta_b^{ac} \tau_y \rt).
\ee
This transformation requires a change in the phases $\varphi_{ab}$
\be
\varphi_{ab} \mapsto \varphi_{ab}' = \varphi_{ab} + \phi_{ab}^c + \phi_{ba}^c.
\ee
This transformation is always possible for a given cycle, but it is not possible to put all the eqs. \eqref{A2_BIS_1} into this form at the same time. We shall come back to this limitation momentarily. 
Now, the equation for the cycle (125) reads
\be
\E^{ -\I \theta_5^{12} \sigma_y}\E^{-(\psi_{52} + \I \varphi'_{52} )\sigma_z}
\E^{ -\I \theta_2^{52} \sigma_y}\E^{-(\psi_{21} + \I \varphi'_{21} )\sigma_z}
\E^{ -\I \theta_1^{25} \sigma_y}\E^{-(\psi_{15} + \I \varphi'_{15} )\sigma_z}
=\mathbb I.
\ee
This equation can be solved explicitly for the phases $\psi_{ab}+\I\phi_{ab}$ as functions of the $\theta_{ab}$ (see \cite{Riello2013}). The result is that $2\psi_{ab}$ is again exactly related to the hyper-dihedral angles (to see this, one has to realize that the $\theta_c^{ab}$ are the dihedral angles between the faces $(ca)$ and $(cb)$ within tetrahedron $c$), and the $\varphi'_{ab}\in\{0,\pi\}$. Note the prime on the variable. Hence, the final result is
\be
\varphi_{ab} + \phi_{ab}^c + \phi_{ba}^c \in\{0,\pi\}.\label{eq_phiprime}
\ee
The phases $\phi_{ab}^c$ depend on the geometry of the 4-simplex. Explicitly one finds 
\begin{align}
\phi_{ab}^c &= \frac{1}{2}\Big( \text{arg}\langle \xi_{ab},J\xi_{ac}  \rangle + \text{arg}(\langle \xi_{ab},\xi_{ac}  \rangle \Big) \notag\\
&= \frac{1}{2} \text{arg}\Big(\langle \xi_{ab},J\xi_{ac}  \rangle \langle \xi_{ab},\xi_{ac}  \rangle \Big). 
\end{align}

Notice, also, that the expressions $\varphi_{ab} + \phi_{ab}^c + \phi_{ba}^c$ are independent of the initial (arbitrary) choice of phase of the $\xi_{ab}$ and therefore eq. \eqref{eq_phiprime} is meaningful.

Finally, one can, at least in principle, deal similarly with the equation for the cycle (234), where $H_{31}$ makes its appearance.

As a last remark, the phase appearing in the Chern-Simons functional evaluation is $\Phi_{ab} := 2(\varphi_{ab} + \phi_{ab} + \phi_{ba})$, and this is also independent from the arbitrary phase of the spinors $\xi_{ab}$. This feature comes as no surprise, since the connection curvature is sourced by a function of the spinors that is completely independent from their overall phases.

We leave a complete treatment of these parity-invariant terms for future work. But, before doing this, we want to point out that they are related to what has been called in \cite{Hnybida2014} the ``framing'' of the triangles. This name comes from the fact that the phase of the spinors can be seen as an arrow in the plane of the triangles, which the holonomies should consistently parallel transport from one triangle and one tetrahedron to the next. If the 4-simplex is flat, after coming back to the starting point, the framing should not change, while a precession could be present if some curvature is present. This is the meaning of the ``crossing'' equation 
\be
G_{34}G_{42} G_{23} = H_{31}.
\ee

{%
To conclude, we draw the interested reader's attention to \autoref{app_perturbative}, where by performing a perturbative evaluation of the Chern-Simons functional, we shed some light on the nature of these non-Regge terms. In particular, these results show that the whole non-Regge term, including the contribution from $C_{\text{topo}}$, is parity invariant up to an integer multiple of $ 2\pi \ell_{P}^2$.%
}



\section{Areas and Spins: a Subtlety}\label{sec_2pi-area}

Because the reconstruction of the geometry is performed using the vectorial representation, it treats holonomies associated to triangles of area $\text{a}$ and $\lt(2\pi R_\lambda^2 - \text{a}\rt)$ in the same way
\be
O = \exp\lt[ \frac{\lambda}{3}\text{a} \hat n \vec J\, \rt]  =  \exp \lt[ \frac{\lambda}{3}\lt(2\pi R_\lambda^2 - \text{a}\rt) \lt(- \hat n \rt)\vec J \,  \rt]\,;
\ee
this comes at the price of simultaneously reversing our interpretation of $\hat n$. This is due to the fact that the trivial SO(3) holonomy and that  along  the equator of a sphere are both associated to the identity. Therefore, the holonomy around a triangle $T'$ obtained by taking the equatorial complement of one side of a given triangle $T$, is the same as the holomy around $T$ itself. Clearly, one of the two is a non-convex triangle and cannot belong to the reconstructed 4-simplex.\footnote{This non-convex triangle $T'$ looks like a a cake missing a slice.} As was made explicit in the discussion of 3-dimensional reconstruction theorem, ``the sign'' of $\hat n$ is fixed by the convexity requirement. This choice, then, has consequences for the interpretation of the area of the triangle, too.

This might appear to be an issue for the calculation of the action at the critical point. Indeed, the quantities that directly appear in the calculation are the spins $ j_{ab}$, which might or might not encode the physical area of the triangle, as a consequence of the ambiguity just described. However, an intriguing coincidence saves the result.

Let us start by studying the spinfoam action associated to a given face, where not $\ell_P^2 \gamma j$ but $\lt(2\pi R_\Lambda^2 -\ell_P^2 \gamma j\rt) $ is the physical area of the corresponding triangle. Then, the spinfoam action of this face becomes at the critical point
\be
-2\I \gamma j \psi = - \frac{\I}{\ell_P^2}\lt(2\pi R_\Lambda^2 - \text{a}\rt) (-\Theta)  =  - \frac{\I}{\ell_P^2} \text{a}\Theta + \frac{6\pi \I}{|\Lambda|}\Theta\,,
\ee
where we have also used the fact that the relation between $2\psi$ and $\Theta$ must be corrected by a minus sign, because the interpretation of the direction $\hat n$ is modified whenever the interpretation of the physical area in terms of the spin is.

However, the evaluation of the Chern-Simons functional also gets modified, and the contribution of this face to the variation $-\I\delta \text{CS}\big|_0$ is
\be
2\I\gamma j \delta \psi =  \frac{\I}{\ell_P^2}\lt(2\pi R_\Lambda^2 - \text{a}\rt) \delta(-\Theta) =  \frac{\I}{\ell_P^2} \text{a}\delta \Theta -\I \frac{6\pi}{|\Lambda|}\delta\Theta\,,
\ee
where we neglected the contribution coming from the phase $\varphi$. Note that the only difference with the ``standard'' case, is the last term, which can easily be integrated without interfering with the Schl{\"a}fli identities; this is because the area of the triangle does not appear in this term at all. Hence, when combining the spinfoam and the Chern-Simons contribution as in eq. \eqref{eq_result}, we obtain absolutely no modification of the Regge part of the action when it is expressed in terms of geometrical quantities. 

{
This result, which appears here as a coincidence, deserves in our opinion further investigation, since it might point toward a more unified and geometrical treatment of the spinfoam and Chern-Simons parts of the total action. In this direction, we already have some hints related to a WKB analysis of the holomorphic 3d block structure hidden behind our construction. We are actively working to make this construction precise \cite{3dblocks}. 
}


\section{Conclusions and Outlook}\label{sec_conclusions}

\subsubsection*{Summary and Result}

In this paper we have studied the expectation value of the non-planar graph operator $\G_5\big(\vec j, \vec i \big| A,\bA\big)$ in $\Slc$ Chern-Simons theory with complex level $h = 12\pi \Lambda^{-1}(\gamma^{-1} + \I)$,
\be
Z_\text{CS}\lt(S^3;\G_5\big|\,\vec{j},\vec{i}\rt)=\int\cd A\cd\bar{A}\ \E^{\I \text{CS}\lt[S^3\,|\,A,\bar{A}\rt]}\ \G_5\lt(\vec{j},\vec{i}\,\big|A,\bar{A}\rt), 
\ee
where
\be
\text{CS}[S^3\,|\,A,\bar{A}]=\frac{h}{8\pi}\int_{S^3}\tr\lt(A\wedge \rmd A+\frac{2}{3}A\wedge A\wedge A\rt)+\frac{\bar{h}}{8\pi}\int_{S^3}\tr\lt(\bA\wedge \rmd \bA+\frac{2}{3}\bA\wedge \bA\wedge \bA\rt).
\ee
This study has been performed in the double scaling limit
\be
j \, , \;|h |\rightarrow \infty \qquad \text{ with } \; j/|h| \sim const \;\text{and} \; \text{arg}(h) = const, 
\ee
corresponding to the semiclassical ($\hbar \rightarrow 0$) limit of the quantum amplitude of a 4-simplex in quantum gravity including a cosmological constant. We showed that the critical point equations obtained in this limit can be interpreted as describing the geometry of a constant curvature 4-simplex flatly embedded in $(3+1)$-dimensional de Sitter or anti-de Sitter spacetime, depending on the details of the spins $j$ and intertwiners $i$. To obtain this result, we showed that there exists a precise correspondence (when neglecting some ``degenerate'' configurations) between (\textit{i}) the moduli space of SO(3) flat connections on the 4-punctured sphere and a homogeneously curved Euclidean tetrahedron, and (\textit{ii}) a particular subclass of the moduli space of SO$(1,3)^+_\uparrow$ flat connections on the graph complement manifold $M_3 = S^3\setminus \G_5$ and a homogeneously curved Lorentzian 4-simplex. More precisely this holds up to an orientation-flipping transformation of the geometry. Notice that to state the second correspondence, we needed to specify a particular subclass of the moduli space of the SO$(1,3)^+_\uparrow$ flat connections on $M_3$. This subclass, is exactly specified by the boundary condition on $\partial M_3$ induced by the specific form of the graph $\G_5$. The most important characteristic of these boundary conditions is certainly the fact that at each vertex of $\G_5$ the four holonomies transverse to the edges meeting at that vertex are contained in the same SO(3) subgroup of SO$(1,3)^+_\uparrow$. Geomtrically, this specifies a three-dimensional spacelike frame for the boundary tetrahedra. Algebraically, this condition can be understood as constraining $\text{arg}(h) = \arctan \gamma$. Once these correspondences have been drawn, we calculated the asymptotic behavior of the amplitude in the double scaling limit (d.s.l.), obtaining 
\be
Z_\text{CS}\lt(S^3;\G_5\big|\,\vec{j},\vec{i}\,\rt) \xrightarrow{\text{d.s.l.}} \E^{\text{NR}[\vec i,\vec j]}\lt[ \mathscr{N_+}\E^{ \I \lt(\sum_{t} \text{a}_t \Theta_t -  \lambda V_4  \rt)} + \mathscr{N_-}\E^{ - \I \lt( \sum_{t} \text{a}_t \Theta_t -  \lambda V_4  \rt)} \rt]\lt[1 + \text{O}(j^{-1}, h^{-1}) \rt].
\ee
{
To write this formula we have taken into account the fact that whenever the connection $\A = (A,\bA)$ is a solution of the equation of motion, also $\bar{\A} = (\bA,A)$ is, hence the two branches. Each branch contributes the Regge action for a curved 4-simplex including the cosmological constant (the two branches coming with opposite orientations),}\footnote{In units where $c = \hbar = \ell_P^2 = 1$.}
\be
S_\text{R} = \sum_{t=1}^{10} \text{a}_t \Theta_t - \lambda V_4\,,
\ee
where $\text{a}_t$, $\Theta_t$ are the areas and dihedral angles associated to the ten triangles of the 4-simplex,  and $V_4$ is its curved 4-volume. Both the areas and the 4-volume are considered to be positive independently of the 4-simplex orientation. 
This result is closely analogous to what happens in the Ponzano-Regge and Turaev-Viro state sum models of three-dimensional gravity, where the tetrahedron amplitude, considered in the appropriate limit, gives the two branches of the 3-dimensional Regge action. However, the above formula also contains a new overall phase that we have called the ``non-Regge'' phase $\text{NR}[\vec i,\vec j]$. We will discuss this phase briefly in the next subsection.

{
At this point we want to stress one feature of this result relating to the origin of the 4-volume term in the asymptotic formula above. This term stems from the evaluation of the purely 3-dimensional Chern-Simons functional on a connection solving the equation of motion discussed above. Unfortunately, this evaluation also produces extra terms, which we were not able to fully interpret geometrically. However, to get rid of these extra terms, one can use the fact that the solutions always come in pairs characterized, after having interpreted them as 4-dimensional simplicial geometries, by opposite spacetime orientations. In fact, the extra terms happen to be orientation invariant. Hence,
\begin{align}
2\lambda V_4 & = \text{CS}[A,\bA] \big|_0 - \text{CS}[\bA,A] \big|_0 \equiv \Re \big(h W[A]\big)\big|_0 - \Re\big(h W[\bA]\big)\big|_0\;,
\end{align}
where the subscript ``0''  simply emphasizes the on-shell evaluation of the functionals.}

{
Before moving to the issues this work has left open, we want to put forward an important feature of the model that is strongly suggested by the results we have just summarized. The space of \textit{dynamical} vacua of spinfoam loop quantum gravity, that is, the space of solutions of its equations of motion, is a subspace of the moduli space of a class of $\Slc$ flat connections on some graph complement 3-manifold. This type of vacuum is very different from the Ashtekar-Lewandowski kinematical vacuum (and also from the Kodama state proposal). In this approach the graph encodes departures from flatness, i.e. departures from the topological phase of gravity, which is essentially given by $(\Lambda)BF$ theory. This is much closer in spirit to the family of alternative dual\footnote{Dual here refers to the fact that it is the connection variable, instead of the flux one, to acquire a definite---though trivial---value.} vacua recently revived by Dittrich and Geiller \cite{Dittrich2014a,Dittrich2014b}, and also bares strong similarities with the framework proposed in the nineties by Crane.
}

\subsubsection*{Open Issues and Outlook}

The present paper can be read as a preliminary test for the proposal of using Chern-Simons theory as a fundamental tool in spinfoams. Having shown that a geometrical analysis of the would-be 4-simplex amplitude is not only possible but also rich and insightful, elevates both our interests and expectations in this line of research. As with any preliminary test, this paper has left many questions unanswered.

{
One issue that definitely needs to be addressed is how to rigorously define our starting point, that is the path integral formula for $Z_\text{CS}\lt( S^3; \G_5 \big| \vec j, \vec i\,\rt)$. We think that this is not a hopeless task, since a lot of recent progress has been made in understanding $\Slc$ Chern-Simons theory. In particular, following the work of Witten, Dimofte, Gukov, Andersen and others \cite{Witten2010,Dimofte2010,Dimofte2011a,Dimofte2011,Gukov2012,Andersen2014}, we intend to construct a path integral using holomorphic 3d blocks and a state integral model. This is in principle possible for the manifold we are interested in, that is $M_3=S^3\setminus\G_5$, since it can be triangulated by ideal tetrahedra.}

{
A rigorous definition of the path integral should also eventually clarify the r\^ole quantum groups play in our construction. To this end, another point that needs clarification is how the phase-space structure of gravity reduces to that of Chern-Simons theory, and in particular how the standard $\Slc$ spin-connection is substituted by the non-commutative Poisson connection of Chern-Simons theory. Answering this question should help in understanding the r\^ole of the new closure relation and of the new deformed spin networks that can be defined starting from the model (as explained in \cite{companionMink} the deformed closure in terms of the transverse holonomies is the classical analogue of $q$-deformed intertwiners).}

From the perspective of the dynamics of quantum gravity, we have already emphasized how our result bares similarities with the asymptotic behavior of the Ponzano-Regge-Turaev-Viro state sum model of 3-dimensional gravity. However, it is of the utmost importance to bare in mind that three-dimensional gravity is topological, and thus the Ponzano-Regge and Turaev-Viro models are---completely consistently---triangulation invariant. In order to claim that we actually have a full-fledged model of 4-dimensional quantum gravity, with its propagating degrees of freedom, there is still a long way to go, and the most critical issue is definitely how to take the continuum limit of our discrete model. A problem that in 3 dimensions is simply not present. A very preliminary step in this direction, is to consider a manifold triangulated by more than a single 4-simplex. This is a step we plan to take soon in a subsequent publication. This problem is also intimately related to the question of what r\^ole graphs other than $\G_5$ might play in the construction. For the moment we just observe that dealing in full generality with closures among more than four holonomies, which would correspond to polyhedra with more than four faces, though attractive, is \textit{a priori} a lot more complicated than the case considered here (see \cite{Haggard2013} for progress on the flat pentahedron). 

Another issue that we leave for future investigations is that of the non-Regge phase
\be
\text{NR}[\xi_{ab},j_{ab}] := -\frac{ 2\I}{\ell_P^2} \sum_{a<b} j_{ab} \varphi_{ab} + \frac{\I}{\ell_P^2}  ( C_\Phi + C_\text{topo}).
\ee
This object depends on the phase convention of the boundary state, which is physically harmless, and also on some of the geometric data. The latter is potentially a dangerous feature for the model, since these data might interfere uncontrollably with the dynamics of the model once one considers more complicated complexes involving bulk triangles whose geometrical data are integrated over (here we refer to the sum over the spins appearing in the spinfoam models). Therefore, the effects of these phases must be studied in the context of an extended triangulation. The hope is that in this context the phases associated to a given triangle add up essentially to zero, in the relevant asymptotic regime. The above discussion applies to all the terms in the previous equation except the very last one, $C_\text{topo}$. This term contains what can be referred to as the topological ambiguities (or choices) that one must make to define the amplitude. As an example we cite the choice of a specific graph framing, which raises in particular the following questions: how would the amplitude change under a change of the graph framing? Can this change be made irrelevant once exponentiated by, for example, quantizing $\gamma$? These are also fundamental questions we will need to address in the future.  

{
To conclude, we add a final aspect we think would be interesting to investigate in the future. We would like to explore the physical properties of the new vacuum suggested by our result, and in particular the nature of its perturbations. See \cite{Han2014b} for a first step in this direction in the case of the zero-cosmological constant EPRL spinfoam model.
}


\section*{Acknowledgements}
HMH, MH, and AR are indebted to the Quantum Gravity group at the Centre de Physique Th\'eorique in Marseille, where they were working when this collaboration started. They would also like to thank, S. Nawata, C. Rovelli, and R. Van der Veen for insightful discussions. MH would like to thank S. Gukov, J. Mour{\~a}o, D. Pei, and Z. Sun for fruitful discussions at various stages of this work, as well as F. Vidotto for her invitation and hospitality at the Radboud Universiteit in Nijmegen, and for her comments on this work. AR warmly thanks A. Baratin, B. Dittrich, L. Freidel, F. Girelli, D. Pranzetti, L. Smolin, and W. Wieland for enlightening discussions on different aspects of this project.\\

\noindent HMH acknowledges support from the National Science Foundation (NSF) International Research Fellowship Program (IRFP) under Grant No. OISE-1159218. MH acknowledges funding received from the People Programme (Marie Curie Actions) of the European Union's Seventh Framework Programme (FP7/2007-2013) under REA grant agreement No. 298786. MH also acknowledges the funding received from Alexander von Humboldt Foundation. HMH and AR were supported by Perimeter Institute for Theoretical Physics. Research at Perimeter Institute is supported by the Government of Canada through Industry Canada and by the Province of Ontario through the Ministry of Research and Innovation.


\appendix

%
%


\section{General Relativity: Notation and Conventions}\label{app_GRconv}

In this paper the signature of metric is
\be
\eta :=\text{diag}(-1,1,1,1).
\ee
The convention for the completely antisymmetric symbol is
\be
\epsilon^{0123}=1\quad\text{thus}\quad\epsilon_{0123}=-1. 
\ee

The Einstein-Hilbert action with a cosmological constant reads
\be
S_{\text{EH}}:=\frac{1}{2 \kappa}\int_\mathcal{M} (R-2\lambda) \sqrt{-g} \D^4 x + \frac{1}{\kappa} \int_{\partial \mathcal M} K \sqrt{q_3} \D^3 x 
\ee
where $\kappa := 8\pi G=:\ell_P^2/\hbar$ (and $c=1$). 

The Einstein-Hilbert action can be discretized on a 4d simplicial complex with 4-simplices of constant curvature $\l$. This discretization results in the Regge action, \cite{Regge1961,Hartle1981,BarrettFoxon,Bahr2010,Bahr:2009qc,Sorkin:1975ah,eva},
\be
S_{\text{R}} := \frac{1}{\kappa} \left[
 -\sum_{t\ \text{internal}} \text{a}(t)\,\eps(t) - \sum_{t\ \text{boundary}} \text{a}(t)\, \Theta(t)
+\sum_{\sigma} \lambda\, V_4(\sigma) \right],
\ee
here $t$ and $\sig$ denote the triangles and 4-simplices in the simplicial complex and $\text{a}(t)$ is the area of the triangle $t$. $V_4(\sigma)$ is the 4-volume of the 4-simplex $\sig$. $\eps(t)$ is the Lorentzian deficit angle hinged by the internal triangle $t$, and $\Theta(t)$ is the Lorentzian 4d hyper-dihedral angle hinged by the boundary triangle $t$\footnote{Here we follow the same convention of hyper-dihedral angle $\Theta_t(\sig)$ and deficit angle $\eps(t)$ as \cite{BarrettFoxon}. Notice that in Figure 6 of \cite{BarrettFoxon}, the deficit angle $\eps(t)$ is negative.}. 
\be
\eps(t)&=&\sum_{\sig,t\subset\sig}\Theta_t(\sig)\ \ \ \ \text{for internal}\ t ,\nonumber\\
\Theta(t)&=&\sum_{\sig,t\subset\sig}\Theta_t(\sig)\ \ \ \ \text{for boundary}\ t
\ee
$\Theta_t(\sig)$ is the hyper-dihedral (boost) angle in a 4-simplex $\sig$ hinged by $t$, same as $\Theta_{ab}$ in the main text. All the quantities intering $S_{\text{R}}$ are the functions of edge-lengths on the simplicial complex. The first and second terms in $S_{\text{R}}$ are the discretizations of the scalar curvature bulk term $\int_{\cm} R$ and the extrinsic curvature boundary term $\int_{\cm} K$. The third term is the cosmological constant term. 

For a single 4-simplex, the bulk term is absent in Regge action. Thus Regge action reduces to
\be
S_{\text{R}}(\sig):=- \frac{1}{\kappa} \left[\sum_{a<b} \text{a}_{ab} \Theta_{ab}
- \lambda\, V_4 \right].
\ee
where $a,b=1,\cdots,5$ labels the five tetrahedra forming the boundary of 4-simplex. $\text{a}_{ab},\ \Theta_{ab}$ are the area and hyper-dihedral angle of the triangle shared by tetrahedra $a$ and $b$.

Einstein-Hilbert action in Palatini-Cartan formulation reads
\begin{align}
S_{\text{PC}}:=&-\frac{1}{2\kappa}\int_\mathcal{M} \left(\frac{1}{2}\epsilon_{IJKL} e^I\wedge e^J \wedge \F^{KL} - \frac{\lambda}{12} \epsilon_{IJKL} e^I\wedge e^J \wedge e^K \wedge e^L\right)+ \nonumber\\
& +\frac{1}{2\kappa}  \int_{\partial\mathcal M} \epsilon_{IJKL} e^I\wedge e^J \wedge n^K \D_\omega n^L 
\end{align}

Define the 2-forms
\be
B^{IJ} := e^I \wedge e^J,
\ee
implying that
\be
\frac{1}{4}\epsilon^{\mu\nu\rho\sigma} B_{\mu\nu}^{IJ} B_{\rho\sigma}^{KL}= \det(e) \epsilon^{IJKL}. 
\ee
In terms of $B^{IJ}$, Einstein-Hilbert action in Plebanski formulation reads
\begin{align}
S_{\text{Ple}}:=&-\frac{1}{2\kappa} \int_\mathcal{M} \left( \frac{1}{2}\epsilon_{IJKL} B^{IJ}\wedge \F^{KL} -\frac{\lambda}{12}\epsilon_{IJKL} B^{IJ}\wedge B^{KL} + \frac{1}{2}\varphi_{IJKL} B^{IJ}\wedge B^{KL} \right)+ \nonumber\\
& +\frac{1}{2\kappa}  \int_{\partial\mathcal M} \epsilon_{IJKL} B^{IJ} \wedge n^K \D_\omega n^L 
\end{align}
where $\varphi^{IJKL}=-\varphi^{JIKL}=
-\varphi^{IJLK}=\varphi^{KLIJ}$ is a tensor in internal space satisfying $\epsilon_{IJKL}\varphi^{IJKL}=0$. This tensor field serves as Lagrange multiplier for the imposition of the (quadratic form) of the simplicity constraints:
\be
\frac{1}{4}\epsilon^{\mu\nu\rho\sigma}B_{\mu\nu}^{IJ}B_{\rho\sigma}^{KL} \stackrel{!}{=}  ||e|| \epsilon^{IJKL},
\ee
where $||e|| \D^4x := -\frac{1}{4!}\epsilon_{IJKL} B^{IJ}\wedge B^{KL} $. The nontrivial solutions of these equations are
\be
B \stackrel{!}{=}\pm(e\wedge e) \quad\text{and}\quad B\stackrel{!}{=}\pm\star(e\wedge e).
\ee
The first set of solutions reduces from Plebanski action to Palatini-Cartan action. 

Using the $\SLDC$-invariant bilinear forms $\prec\cdot,\cdot\succ$ and $<\cdot,\cdot>$ (see \autoref{app_notations}), this gives
\begin{align}
S_{\text{Ple}}:=&-\frac{1}{2\kappa} \int_\mathcal{M} \left( \prec B\wedge \F \succ -\frac{\lambda}{6} \prec B\wedge B \succ + < (\varphi\triangleright B)\wedge B> \right)+ \nonumber\\
& +\frac{1}{\kappa}  \int_{\partial\mathcal M} \prec B \wedge (n\otimes \D_\omega n)\succ 
\end{align}
where $(\varphi\triangleright B)_{IJ}=\frac{1}{2}\varphi_{IJ}^{\phantom{IJ}KL}B_{KL}$.


\section{Self-dual and Anti-self-dual Decomposition}\label{app_notations}

In our notation for the basis $\cj^{IJ}$ in the Lie algebra $\slc$:
\be
\mathcal{J}^{0i} = K^i\quad\text{and}\quad \mathcal{J}^{ij}=\epsilon^{ij}_{\phantom{ij}k} J^k
\ee
where $J^i,\ K^i$ satisfy
\be
[J^i,J^j]=\epsilon^{ij}_{\phantom{ij}k} J^k \,,\quad 
[K^i,K^j]=-\epsilon^{ij}_{\phantom{ij}k} J^k \,\quad\text{and}\quad
[K^i,J^j]=\epsilon^{ij}_{\phantom{ij}k} K^k
\ee

Given $X$ in the complexification of $\slc$, we define its self-dual and anti-self-dual parts $X_+$ and $X_-$:
\be
X_\pm := \frac{1}{2}(1\mp \I \star) X \ \ \ \ \text{or}\ \ \ \ 
(X_\pm)_{IJ} =\frac{1}{2}\left(X_{IJ}\mp \frac{\I}{2}\epsilon_{IJ}^{\phantom{IJ}KL} X_{KL}\right),
\ee
so that
\be
\star X_\pm = \pm \I X_\pm \ \ \ \ \text{and}\ \ \ \ X=X_+ + X_-.
\ee

We compute
\begin{align}
X_\pm &= \frac{1}{2}(X_{\pm})_{IJ}\mathcal{J}^{IJ} = (X_\pm)_{0k} K^k +   (X_\pm)_{ij} \,\frac{1}{2}\epsilon_{\phantom{ij}k}^{ij} J^k \nonumber\\
&  =  \frac{1}{2}\left(X_{0k}\mp\frac{\I}{2}\epsilon_{0k}^{\phantom{0k}ij} X_{ij}\right) K^k   
+\frac{1}{2}\left(X_{ij}\mp\I\;\epsilon_{ij}^{\phantom{ij}0k'} X_{0k'}\right)\frac{1}{2}\epsilon_{\phantom{ij}k}^{ij} J^k \nonumber\\
& = \frac{1}{2}\left( X_{0k}\pm\frac{\I}{2}\epsilon_{k}^{\phantom{k}ij} X_{ij}\right) K^k 
+\frac{1}{2}\left( \frac{1}{2}\epsilon_{\phantom{ij}k}^{ij} X_{ij}\mp\I X_{0k} \right) J^k\nonumber\\
& = \pm \frac{\I}{2}\left(\frac{1}{2}\epsilon_{\phantom{ij}k}^{ij} X_{ij} \mp \I X_{0k}\right) K^k 
+\frac{1}{2}\left( \frac{1}{2}\epsilon_{\phantom{ij}k}^{ij} X_{ij}\mp\I X_{0k} \right) J^k\nonumber\\
& = \left( \frac{1}{2}\epsilon_{\phantom{ij}k}^{ij} X_{ij}\mp\I X_{0k} \right)
\left(\frac{J^k\pm\I K^k}{2}\right)
\nonumber\\
&=: ( X_{\pm})_kT^k_\pm
\end{align}
where we have defined that 
\be
(X_\pm)_k:= \left( \frac{1}{2}\epsilon_{\phantom{ij}k}^{ij} X_{ij}\mp\I X_{0k} \right)
\quad\text{and}\quad
T^k_\pm :=  \frac{J^k\pm\I K^k}{2}.
\ee
A real element $X\in\slc$ satisfies $\bar{X}_+^k=X^k_-$.

The complexification of $\slc$ is the same as two copies of complexified $\su$, i.e. $\su^\C\times\su^\C$ being the self-dual and anti-self-dual decomposition of complexified $\slc$. $T_+^i$ or $T_-^i$ form a basis in each copy of $\su^\C$, satisfying
\be
[T^i_\pm , T^j_\pm ] = \epsilon^{ij}_{\phantom{ij}k} T^k_\pm 
\quad\text{and}\quad
[T^i_\pm , T^j_\mp] = 0
\ee
The space of real symmetric invariant bilinear forms on $\slc$ is a 2-dimensional vector space. we choose two independent non-degenerate bilinear forms $<\ ,\ >$ and $\prec\ ,\ \succ$ defined by
\begin{align}
<T^i_\pm,T^j_\pm>=\delta^{ij}\;, \quad <T^i_\pm, T^j_\mp > =0\;;\\
\prec T^i_\pm, T^j_\pm \succ = \pm\I \delta^{ij}\;, \quad \prec T^i_\pm, T^j_\mp \succ =0\;.
\end{align}

It is useful to list the spinor representation of $\slc$ basis: In Weyl's left-handed $(\frac{1}{2},0)$ representation, the generators are represented by
\be
J^k \doteq -\frac{\I}{2}\sigma^k \,,
\quad 
K^k \doteq -\frac{1}{2}\sigma^k
\quad\Longrightarrow\quad
T^k_+ \doteq -\frac{\I}{2}\sigma^k =:\tau^k
\ee
In Weyl's right-handed $(0,\frac{1}{2})$ representation, the generators are represented by
\be
J^k \doteq -\frac{\I}{2}\sigma^k \,,
\quad 
K^k \doteq \frac{1}{2}\sigma^k
\quad\Longrightarrow\quad
T^k_- \doteq - \frac{\I}{2}\sigma^k =:\tau^k
\ee
In both Weyl's left- and right-handed representations,
\be
< X_\pm, Y_\pm > \doteq -2 \Tr(X_\pm Y_\pm)  \quad\text{and}\quad \prec X_\pm, Y_\pm \succ \doteq \mp2\I\,\Tr(X_\pm Y_\pm)
\ee

Finally, notice that the right- and left-handed Weyl representations are related at the level of the algebra by the operation 
\be
X_\pm \mapsto X_\mp = -\overline{X_\pm}\,,
\ee
where the overbar stands for complex conjugation. At the level of the group this reads
\be
G_\pm \mapsto G_\mp = [(G_\pm)^{\dagger}]^{-1}
\ee
where $\dagger$ stands for Hermitian conjugation, i.e. for transposition followed by complex conjugation.


\section{Variation of the Action with respect to the Connection}\label{app_variations}

In this appendix we perform the explicit calculation of the variation of the total action $I_{\Gamma_5}$ with respect to the connections $A^i_\mu(x)$ and $\bA^i_\mu(x)$.

We start by discussing the functional derivative of the holonomies $G_{ab}[A]$ and $G^\dagger_{ab}[\bA]$. Such holonomies are defined by
\begin{subequations}\begin{align}
G_\ell[A] &:= \mathbb P \exp   \int_\ell A  \equiv \mathbb P\exp  \int_0^1 A^j_\nu(\ell(s)) \tau_j \frac{\D \ell^\mu}{\D s} \D s   \\
(G_\ell)^\dagger[\bA] &= \mathbb P\exp  \int_{\ell^{-1}} \bA  \equiv \mathbb P \exp \int_0^1 \bA^j_\nu(\ell^{-1}(s)) \tau_j  \frac{\D (\ell^{-1})^\mu}{\D s} \D s   
\end{align}\end{subequations}
where we dropped the edge indices $(ab)$, and we introduced $\ell^{-1}(s):=\ell(1-s)$. Notice that $\bA \equiv \bA^j_\nu(x) \tau_j \D x^\nu := \overline{A^j_\nu(x)}\, \tau_j \D x^\nu$, and the sign difference in the path-ordered exponential comes from the fact that $\tau_j^\dagger = -\tau_j$.

Clearly, if $x\notin \ell$, then $\delta G_\ell / \delta A^i_\mu(x) =0$. Suppose then that there exists and $s_0\in(0,1)$ such that $\ell(s_0) = x$. In this case one can write
\be
G_\ell[A] = \lim_{\epsilon \rightarrow 0} G_{1,s_0+\epsilon} \lt( \mathbb{I} + \int_{s_0-\epsilon}^{s_0+\epsilon} A^j_\nu(\ell(s)) \tau_j \frac{\D \ell^\mu}{\D s} \D s   \rt) G_{s_0-\epsilon,0},
\ee
and therefore
\begin{align}
\frac{\delta G_{ab}[A]}{\delta A_\mu^i(x)}&=\lim_{\epsilon \rightarrow 0} G_{1,(s_0+\epsilon)}\left( \int_{s_0-\epsilon}^{s_0+\epsilon} \delta^{(3)}(x-\ell(s))\delta^j_i \delta^\nu_\mu \tau_j \frac{\D \ell^\nu}{\D s} \D s\right) G_{(s_0-\epsilon), 0} \nonumber\\
&= \left( \int_0^1 \delta^{(3)}(x-\ell(s)) \frac{\D \ell^\mu}{\D s} \D s\right) G_{1, s_0} \tau_i G_{s_0, 0}\;,
\end{align}
We will often write the 2d distribution appearing in this equation symbolically as
\be
\delta_{\ell}^{(2)\;\mu}(x):=  \int_0^1 \delta^{(3)}(x-\ell(s)) \frac{\D \ell^\mu}{\D s} \D s \;.
\ee
For the variation of the hermitian conjugate holonomy $(G_\ell)^\dagger$ with respect to  $\bA^i_\mu(x)$, we find
\be
\frac{\delta (G_{\ell})^\dagger[\bA]}{\delta \bA_\mu^i(x)} = \left( \int_0^1 \delta^{(3)}(x-\ell^{-1}(s)) \frac{\D (\ell^{-1})^\mu}{\D s} \D s\right) (G_{s_0, 0})^\dagger \tau_i (G_{1, s_0})^\dagger  = \delta^{(2)\mu}_{\ell^{-1}}(x) \;(G_{s_0, 0})^\dagger \tau_i (G_{1, s_0})^\dagger\;.
\ee
Notice the absence of the minus sign close to the $\tau_i$.

With this equations we can immediately compute the variation of the spinfoam part of the action $S$ with respect to the connection. For this recall
\be
S = \sum_{ab,a>b} 2j_{ab} \ln \langle J\xi_{ab},\bar z_{ab}\rangle +2j_{ab} \ln\langle \bar z_{ab}, G_{ab}\xi_{ba}\rangle + j_{ab}(\I\gamma-1) \ln \langle \bar z_{ab},  G_{ab}G^\dagger_{ab} \bar z_{ab} \rangle,
\ee
where we used the rescaling symmetry for the variables $\bar z_{ab}$, to fix their norms to be one. Then
\begin{align}
\lt(\frac{\delta S}{\delta A^i_\mu(x)}\rt)_{\Re(I_{\Gamma_5})=0} %
&= \sum_{ab,a>b} j_{ab}  \lt\{ %
2\frac{\langle \bar z_{ab}, [G_{a,s_{ab}}\tau_i G_{s_{ab},b}]\xi_{ba}\rangle}{\langle \bar z_{ab}, G_{ab}\xi_{ba}\rangle} +
(\I\g-1)\frac{\langle \bar z_{ab},  [G_{a,s_{ab}}\tau_i G_{s_{ab},b}]G^\dagger_{ab} \bar z_{ab} \rangle}{\langle \bar z_{ab},  G_{ab}G^\dagger_{ab} \bar z_{ab} \rangle}%
\rt\}  \delta_{\ell_{ab}}^{(2)\;\mu}(x)  \notag\\
& = (1+\I\g) \sum_{ab,a>b} j_{ab}  \langle \xi_{ba}, [(G_{s_{ab},b})^{-1} \tau^i G_{s_{ab},b} ] \xi_{ba} \rangle \;\delta_{\ell_{ab}}^{(2)\;\mu}(x)  
\end{align}
where the variables $\bar z_{ab}$ have been eliminated by use of the equation of motion $\Re(I)=0$, that is $\xi_{ba} \propto_\mathbb{C} G_{ab}^{\dagger} \bar z_{ab}$. Simlarly, one finds
\begin{align}
\lt(\frac{\delta S}{\delta \bA^i_\mu(x)}\rt)_{\Re(I)=0} %
&= \sum_{ab,a>b} j_{ab}  %
(\I\g-1)\frac{\langle \bar z_{ab},G_{ab}  [(G_{s_{ab},b})^\dagger  \tau_i (G_{a,s_{ab}})^\dagger] \bar z_{ab} \rangle}{\langle \bar z_{ab},  G_{ab}G^\dagger_{ab} \bar z_{ab} \rangle}  \delta_{\ell^{-1}_{ab}}^{(2)\;\mu}(x)%
\notag\\
& = (1-\I\g) \sum_{ab,a>b} j_{ab}  \langle [(G_{s_{ab},b})^{-1} \tau_i G_{s_{ab},b} ] \xi_{ba}, \xi_{ba} \rangle \;  \delta_{\ell^{-1}_{ab}}^{(2)\;\mu}(x) \notag\\
& = - (1-\I\g) \sum_{ab,a>b} j_{ab}  \langle [(G_{s_{ab},b})^{-1} \tau_i G_{s_{ab},b} ] \xi_{ba}, \xi_{ba} \rangle \;  \delta_{\ell_{ab}}^{(2)\;\mu}(x),
\end{align}
where in the last step we used $\delta_{\ell^{-1}_{ab}}^{(2)\;\mu}(x)=-\delta_{\ell_{ab}}^{(2)\;\mu}(x)$, since the tangent vector fields along $\ell^{-1}$ and $\ell$ have opposite directions.

We now calculate the variation of the Chern-Simons functional $W[A]$ with respect to $A^i_\mu(x)$:
\begin{align}
W[A] & := \frac{1}{4\pi} \int \Tr\lt( A\wedge \D A + \frac{2}{3} A\wedge A\wedge A \rt) \notag\\
& = -\frac{1}{8\pi} \int \epsilon^{\nu\rho\sigma} \lt(\delta_{jk}A_\nu^j \partial_\rho A_\sigma^k + \frac{1}{3}\epsilon_{jkl} A_\nu^j A_\rho^k A_\sigma^l \rt) \D x^3\,
\end{align}
hence
\begin{align}
\frac{\delta W[A]}{\delta A^i_\mu(x)} & = -\frac{1}{8\pi} \epsilon^{\mu\nu\rho}\lt(\partial_\nu A_\rho^i - \partial_\rho A^i_\nu + \epsilon^{ijk} A^j_\nu A^k_\rho \rt) \notag\\
& = -\frac{1}{8\pi} \epsilon^{\mu\nu\rho} F^i_{\nu\rho}[A].
\end{align}
Clearly $\delta W[\bA] / \delta \bA^i_\mu(x) = c.c.\lt(\delta W[A] / \delta A^i_\mu(x) \rt) $.

Putting all the pieces together, one finally finds that
\begin{align}
\lt(\frac{\delta I}{\delta A^i_\mu(x)}\rt)_{\Re(I)=0} & = -\I \frac{h}{2} \lt(\frac{\delta W[A]}{\delta A^i_\mu(x)}\rt)_{\Re(I)=0} + \lt(\frac{\delta S}{\delta A^i_\mu(x)}\rt)_{\Re(I)=0} \notag\\
& = +\frac{\I h}{16\pi} \epsilon^{\mu\nu\rho} F^i_{\nu\rho}[A] + (1+\I\g) \sum_{ab,a>b} j_{ab}\langle \xi_{ba}, [(G_{s_{ab},b})^{-1} \tau^i G_{s_{ab},b} ] \xi_{ba}\rangle \;  \delta_{\ell_{ab}}^{(2)\;\mu}(x)  
\end{align}
and
\begin{align}
\lt(\frac{\delta I}{\delta \bA^i_\mu(x)}\rt)_{\Re(I)=0} & = -\I \frac{\bar h}{2} \lt(\frac{\delta W[\bA]}{\delta \bA^i_\mu(x)}\rt)_{\Re(I)=0} + \lt(\frac{\delta S}{\delta \bA^i_\mu(x)}\rt)_{\Re(I)=0} \notag\\
& = +\frac{\I \bar h}{16\pi} \epsilon^{\mu\nu\rho} \bar F^i_{\nu\rho}[\bA] - (1-\I\g) \sum_{ab,a>b} j_{ab}  \langle  [(G_{s_{ab},b})^{-1} \tau^i G_{s_{ab},b} ] \xi_{ba}, \xi_{ba}\rangle \; \delta_{\ell_{ab}}^{(2)\;\mu}(x) .
\end{align}
%
%


\section{Lorentzian Gluing}\label{app_Lorentziangluing}

In this section, we shall discuss some details of Lorentzian simplicial geometries. In particular, how one should treat the gluings of the various tetrahedra to one another in a way consistent with the critical point equations discussed in the main text of this paper. Let us assume throughout this section that the reconstruct (i.e. geometrical) cosmological constant is positive.

The first observation is that the holonomy going around a face of a tetrahedron depends only on the bivector $u\wedge  \frak n$, where $u$ is the timelike and $\frak n$ the spacelike normals to the triangle. In future pointing time gauge, i.e.  when the tetrahedron is contained in the spacial slice $t=0$, we obtain $u=(1,0,0,0)$ and $\frak n = (0, \hat{\frak n}$. We have learned that  $\frak n$ should be thought as the outgoing normal to the tetrahedron, if we want the closure equations to be consistent with the structure of the 4-simplex and its orientation. Nonetheless, it is a simple observation that $u\wedge {\frak n} = (-u)\wedge(-{\frak n})=-u\wedge {\frak n}^-$, which tells us that if the tetrahedron is past pointing, we shall interpret the normal $ {\frak n}'$ appearing in the equation as the \textit{inner} pointing one.

To convince ourselves that this is consistent with all the equations, let us focus on the parallel transport equations $J\xi' \propto_\mathbb{C} G \xi$ and $\xi' \propto_\mathbb{C} G J \xi$ (the complex proportionality constants are automatically one the inverse of the other as a consequence of $\det G = 1$). These equations, once read in the vectorial representation, tell us that the proper orthochronus Lorentz transformation associated to $G$ sends the direction $\hat n(\xi):= \langle \xi, \vec \sigma \xi\rangle$ into the direction $\hat n(J\xi') \equiv - \hat n(\xi')$, where $\hat n(\xi)$ and $\hat n(\xi')$ are the spacelike normals to the same triangle as ``seen'' from two neighboring tetrahedra. 

Before starting, notice that an oriented 4-simplex necessary necessarily contains at least one future-pointing and one past-pointing tetrahedron as part of its boundary. Moreover, by convenxity, when all the boosts are smoothly sent to zero, two future pointing (or two past pointing) tetrahedra result glued on opposite sides of the common face, while a couple of oppositely pointing tetrahedra must result glued on the same side of the common face. This is a consequence of the fact that proper orthochronus transformations cannot ``cross the light-cone''. 

\begin{figure}[t]
\centering
\includegraphics[width=.8\textwidth]{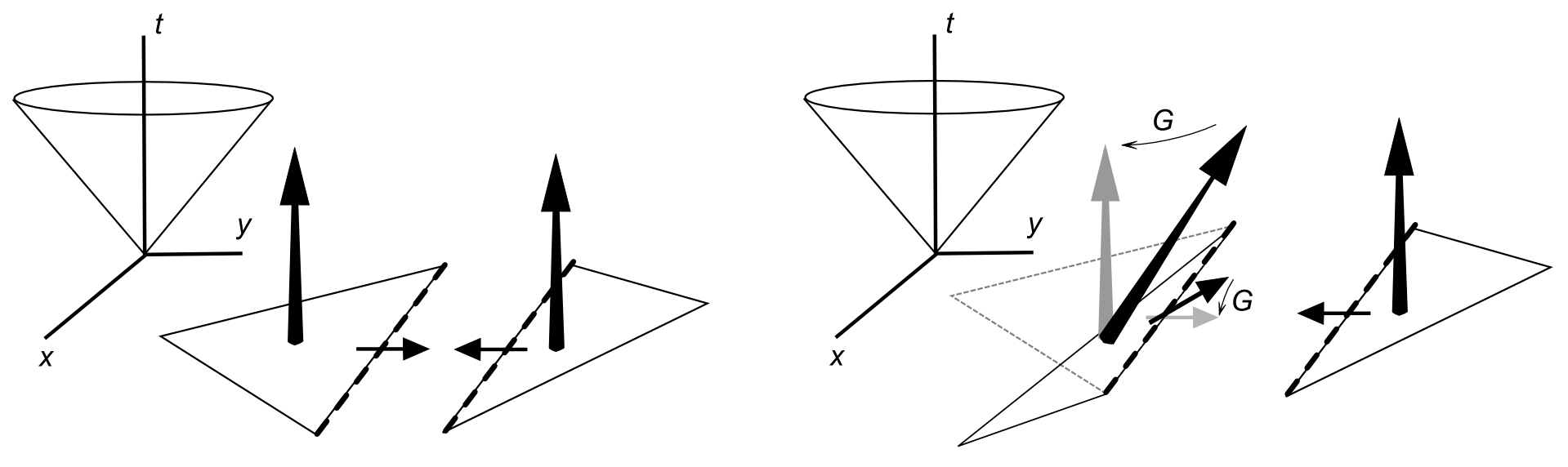}
\caption{The gluing of two future pointing simplices. One spacial dimension has been removed for ease of drawing, and the subsimplex along which the gluing happens is dashed. Notice the Lorentzian nature of the geometry in the way the normals ``rotate'' under the action of boosts.}
\label{fig_ffgluing}
\includegraphics[width=.8\textwidth]{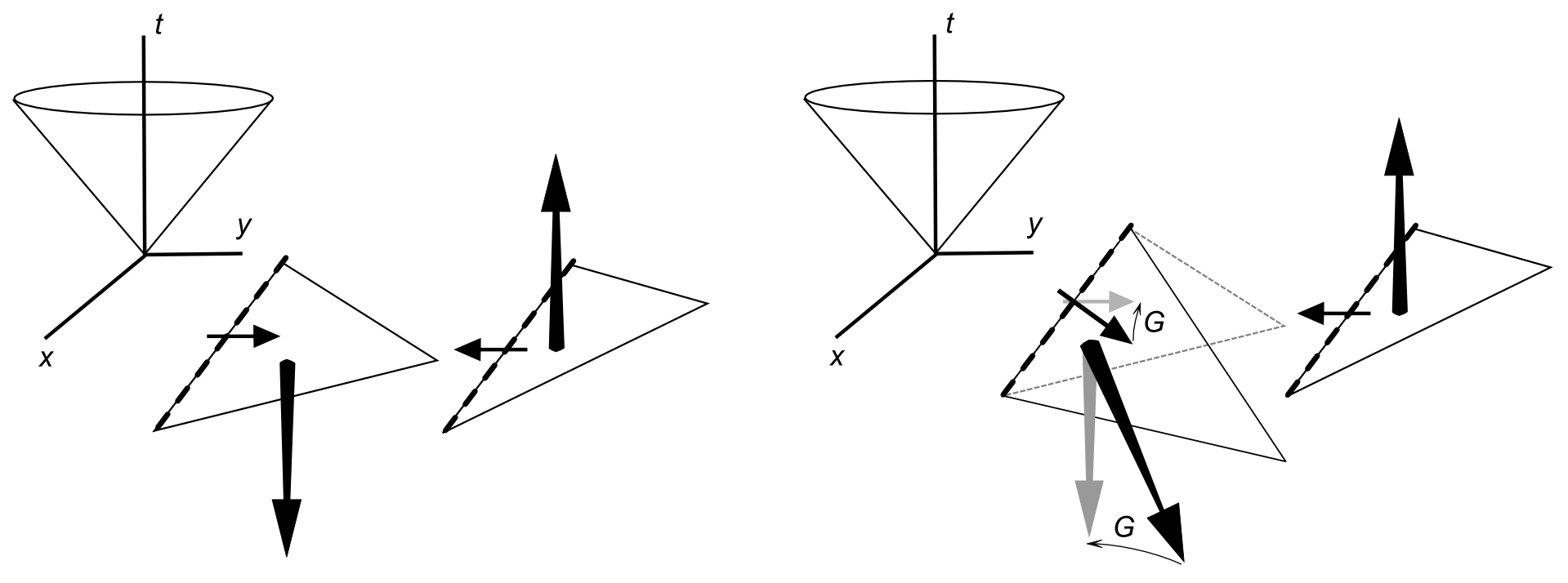}
\caption{The gluing of one future- to one past pointing simplices. One spacial dimension has been removed for ease of drawing, and the subsimplex along which the gluing happens is dashed. Notice the Lorentzian nature of the geometry in the way the normals ``rotate'' under the action of boosts.}
\label{fig_fpgluing}
\end{figure}

Let us consider first the case in which both tetrahedra are future pointing. In this case, the previous statement just mirrors the usual Euclidean result that for gluing two neighboring simplices across a face by preserving orientations their normals should be in opposite directions. This can be seen particularly well in the limit in which $G$ goes to the identity. See \autoref{fig_ffgluing}. 

Let us consider the analogous case in which both tetrahedra are past pointing. In this case, as we previously argued, both spacelike normal must be considered inward pointing. Also in this case, the boost $G$ sends one normal into minus the other. 

We are finally left with the case in which one tetrahedron is future pointing and the other one is past pointing. In this case there is clearly no proper orthochronus Lorentz transformation sending one timelike normal into the other. Similarly, one cannot start with two tetrahedra glued to the exterior of one another to then boost them into their final position. See \autoref{fig_fpgluing}. However, the fact that the two tetrahedra lie on the same side of the common face, is in perfect agreement with the fact that ($i$) one spacelike normal points outwards and the other one inwards and ($ii$) one must have in the limit in which $G$ is the identity $\hat n(\xi') = - \hat n(\xi)$.

Another way to see that all of this is consistent, is the following: instead of speaking about the normal vectors associated to the faces of the tetrahedra, in 4 dimensions it is more appropriate to talk about the bivector associate to them. Call the bivector associated to the face $(ab)$ $B_{ba}$ when it is ``seen'' in the frame of tetrahedron $b$, and $B_{ab}$ when it is ``seen'' in the frame of tetrhadron $a$. Now, to match the orientations, when gluing the two tetrahedra together across the common face, one must always satisfy $B_{ab}=-B_{ba}$. However, $B_{ba} = \tilde N_b\wedge n_{ba}$, where $\tilde N_b$ is the timelike normal to tetrahedron $b$ and $n_{ba}$ is the normal to the face $(ab)$ as seen from tetrahedron $b$. Therefore if $\tilde N_b$ and $\tilde N_a$ have the same time direction, $n_{ba}$ and $n_{ab}$ must have opposite space directions. Similarly, if $\tilde N_b$ and $\tilde N_a$ have opposite time directions, $n_{ba}$ and $n_{ab}$ must have the same space direction.

As a last step in this discussion, we want to relate two different writings of the bivector associated to a triangle. The first way to write the bivector $B_{ba}$ is 
\be
B_{ba} = \frac{\tilde N_b \wedge\tilde N_a}{|\tilde N_b \wedge\tilde N_a|}
\ee
where $\tilde N_b$ is the oriented timelike normal to tetrahedron $b$. When expressing this in the frame of tetrahedron $b$ itself, we find
\be
B_{ba}(b) =\frac{(\pm_b) u \wedge \tilde N_a(b)}{|\tilde N_b(b) \wedge\tilde N_a(b)|},
\ee
where $\pm_b$ depends on whether tetrahedron $b$ is future or past pointing respectively. Now, the time component of $\tilde N_a(b)$ does not matter, because the wedge product is antisymmetric and $u=(1,0,0,0)^T$ has no spacelike component. For what concerns the spacelike part, being $\tilde N_a(b)$ orthogonal to the triangle $(ab)$, it must be proportional to $n_{ba}$. The question is whether it is parallel or antiparallel to it. We see that in both \autoref{fig_ffgluing} and \autoref{fig_fpgluing}, when $\tilde N_b(b) = +u$, the spacial part of $\tilde N_a(b)$ is always antiparallel to $\hat n_{ba}$. It is not hard to see that when $\tilde N_b(b) = -u$,the spacial part of $\tilde N_a(b)$ is always parallel to $\hat n_{ba}$. Hence, the following equality holds
\be
B_{ba}(b) =\frac{\tilde N_b(b) \wedge\tilde N_a(b)}{|\tilde N_b(b) \wedge\tilde N_a(b)|} = -\frac{ u \wedge n_{ba}}{| u \wedge n_{ba}|},
\ee
where in the last equality $n_{ba}$ is understood to be $(0,\hat n_{ba})$.


\section{Perturbative Evaluation of the Critical Chern-Simons Invariant}\label{app_perturbative}

In this appendix we perform a perturbatively evaluation of the Chern-Simons invariant at the critical point. This calculation has the advantage of being completely explicit, and moreover gives a flavor of what the non-Regge contributions might be.

The idea is to consider the graph as ``weak'' source, in the sense that $j/h \sim \Lambda j \sim const \ll 1$, and to expand the Chern-Simons invariant in a formal power series in  $j/h$ around the (vanishing) value it has on the trivial flat connection. Geometrically, this corresponds to evaluating the Chern-Simons functional for a 4-simplex whose physical size is very small with respect to the radius of curvature. We are in other words expanding around a flat solution to the equation of motions. 

At leading order, the transverse holonomies are all trivial, and can be substituted at the first non-trivial order by elements of the Lie algebra (the usual loop quantum gravity fluxes, indeed): 
\be
H_{ab} = \mathbb I + \frac{4\pi}{h}\lt( \frac{1}{\g} + \I \rt) \gamma j_{ab} \hat n_{ab} \cdot \vec\tau + O(j^2/h^2).
\ee 
Thus, the closure equations at the first non-trivial order can be expressed as a linear relations among Lie algebra elements:
\be
\sum_{b, b\neq a}  \gamma j_{ab} \hat n_{ab}\cdot \vec\tau = O(j^2/h^2).
\label{eq_linearclosureeq}
\ee
The latter equation can be interpreted as a consistency equation which involves the boundary data only, which should be imposed also at the purely flat level.

The equations for the longitudinal holonomies also trivialize, and at the leading order they encode the fact that such holonomies are pure gauge and therefore come from a connection which is pure gauge:
\be
G_{ab} = g_a^{-1} g_b \quad \text{for all $(ab)$ except} \quad G_{42} = g_4^{-1} g_2 + \frac{4\pi}{h}\lt( \frac{1}{\g} + \I \rt) \gamma j_{42}\; g_4^{-1}g_3 \hat n_{42} \cdot \vec\tau g_3^{-1} g_2 + O(j^2/h^2).
\ee

To have a clearer notation, let us introduce the pure-gauge connection $(A^0, \bA^0)$ defined on the whole of  $S^3$. Its transverse holonomies $H^0_{ab}:=H_{ab}[A^0] \equiv \mathbb I$ are trivial, and the longitudinal ones $G_{ab}^0:=G_{ab}[A^0]$ are pure gauge, i.e. $G_{ab}^0 = (g_a^0 )^{-1}g_b^0$. This connection clearly solves the leading part of the previous equations. We also introduce a notation for the solution of the flat closure equations $(j_{ab}^0, \xi_{ab}^0$).\footnote{In principle we should also deal with the parallel transport equations for the $\xi_{ab}$ and therefore with the $\mathbb{CP}^1$ variables $z_{ab}$. Since these will not enter explicit the following calculations, we leave them aside.}

Clearly, the contribution of such leading order solutions to the Chern-Simons invariant $W[A^0]$ is not very interesting, since such contribution is just zero modulo $2\pi$. To consider the first non-trivial contribution, we introduce the following notation:
\be
(A,\bA, j_{ab}, \xi_{ab}) = (A^0, \bA^0, j_{ab}^0, \xi^0_{ab}) +  (\delta A, \delta \bA, \delta j_{ab}, \delta \xi_{ab}).
\ee
Clearly the ``$\delta$-variations'' should be considered small, with respect to the leading order solutions, in the sense that such variations are of order $j/h\ll 1$:
\be
(\delta A, \delta \bA, \delta j_{ab}, \delta \xi_{ab}) \sim O(j/h).
\ee

Now, the Chern-Simons functional when evaluated on the connection $A$ can be formally developed in powers of the small parameter $(j/h)$:\footnote{
$W[A^0 + \delta A]  = W[A^0] + \frac{1}{4\pi} \int_{S^3} \Tr \lt[\delta A \wedge D^0 \delta A\rt] + O(\delta A^3)$, where $D^0$ denotes the covariant derivative with respect to $A^0$. Neglecting the contribution of $W[A^0]$ which vanishes modulo $2\pi$, we obtain the sought result by observing that $\frac{1}{4\pi} \int_{S^3} \Tr \lt(\delta A \wedge D^0 \delta A\rt) =\frac{1}{4\pi} \int_{S^3} \Tr \lt(\delta A^{\phantom{0}} \wedge F[A]\rt) + O(\delta A^3)$.
}
\be
W[A] =  \frac{1}{4\pi} \int_{S^3} \Tr \lt(\delta A^{\phantom{0}} \wedge F[A]\rt) + O(j^3/h^3).
\ee
{Notice that in the previous expression, $F\sim O(j/h)$ since it is sourced by the graph, and therefore there is no term of order $O(j/h)$. This had to be expected, since we are perturbing around one of the solutions of the equations of motion of $W[A]$ itself.} Explicitly the curvature $F[A]$ is given by
\be
\epsilon^{\mu\nu\rho} F_{\nu\rho}^i[A(x)] = -\frac{16\pi}{\I h}(1+\I\g) \sum_{a>b} j_{ab} \langle \xi_{ba}, \lt[(G_{s_{ab},b})^{-1} \tau^i G_{s_{ab},b}\rt] \xi_{ba}\rangle \;  \delta_{\ell_{ab}}^{(2)\;\mu}(x) ,
\ee
and by inserting this expression into the perturbative equation for $W[A]$, one obtains\footnote{The mathematical manipulation to get to the following formulas are the same as in the main text, and are therefore skipped.}
\begin{align}
W[A] 
= -\frac{1}{\I h} (1+\I\g) \sum_{a>b} j_{ab}\langle  \xi_{ba}, \lt[G_{ab}^{-1} \delta G_{ab} \rt] \xi_{ba}\rangle.
\end{align}

Let us now evaluate what $\delta G_{ab}$ is in this context. First of all $\delta G_{ab}$ should be understood as $\delta G_{ab} := G_{ab}[A] - G_{ab}[A^0]$, where the equalities holds between $2\times2$ complex matrices. From the equations of motion, we have
\begin{subequations}
\begin{align}
\delta G_{ab} &= (g_a^0)^{-1} (\delta g_b - \delta g_a) g_b^0  \quad \text{for all $(ab)$, except }\\
\delta G_{42} &= (g_4^0)^{-1} (\delta g_2 - \delta g_4) g_2^0 + (g_4^0)^{-1} g_3^0  \lt[ \frac{4\pi}{h}\lt( \frac{1}{\g} + \I \rt) \gamma j_{31} \hat n_{31} \cdot \vec\tau  \rt]  (g_3^0)^{-1} g_2^0.
\end{align}
\end{subequations}
To obtain these expressions, we have parametrized the variations in the $g_a\in\Slc$ by $g_a = (\mathbb I +\delta g_a) g_a^0$. To simplify the notation in what follow, we shall call $F_{31}$ the term appearing in square brackets in the expression of $\delta G_{42}$. Hence,
\begin{align}
W[A] = & -\frac{1}{\I h} \lt(\I+\frac{1}{\g}\rt) \sum_{a>b} \g j_{ab}  \langle\xi_{ba},\lt[g_b^{-1} ( \delta g_b - \delta g_a  )g_b \rt] \xi_{ba}\rangle +\notag\\
&  -\frac{1}{\I h}\lt(\I+\frac{1}{\g}\rt) \g   j_{24} \langle\xi_{24},\lt[g_2^{-1} \lt( g_3 F_{31} g_3^{-1} \rt)g_2 \rt] \xi_{24}\rangle + O(j^3/h^3).
\end{align}
Now, we claim that the first term, in spite of the appearances, is also of order $O(j^3/h^3)$. The reason for this hides in the linearized closure equation (\ref{eq_linearclosureeq}). Indeed, by means of the parallel transport equations we obtain
\begin{align}
 \frac{1}{h}\sum_{a>b} \g j_{ab}\langle\xi_{ba},\lt[g_b^{-1} ( \delta g_b - \delta g_a  )g_b \rt] \xi_{ba}\rangle & =\frac{1}{h}\sum_{a>b} \g j_{ab}\lt \{ \langle\xi_{ba}, g_b^{-1}  \delta g_b g_b  \xi_{ba}\rangle -
\langle J\xi_{ab}, g_a^{-1} \delta g_a g_a  J\xi_{ab}\rangle \rt\}.
\end{align}
Now, the second term on the right-hand side can be manipulated by using the following identities: $\langle J w , z \rangle = -\langle J z , w \rangle$, and $-J g J = (g^{-1})^\dagger$, for any $g\in\Slc$, from which one can deduce\footnote{Indeed, let $g=\exp x$, then $ (g^{-1})^\dagger = (\exp -x)^\dagger = \exp -x^\dagger$. } $-J x J = -x^\dagger$ for any $x \in \slc$. Therefore, recalling that $\delta g\in\slc$, one obtains $\langle J\xi_{ab}, g_a^{-1} \delta g_a g_a  J\xi_{ab}\rangle = - \langle \xi_{ab}, g_a^{-1} \delta g_a g_a  \xi_{ab}\rangle$, and hence
\begin{align}
 \frac{1}{h}\sum_{a>b} j_{ab}\g \langle\xi_{ba},\lt[g_b^{-1} ( \delta g_b - \delta g_a  )g_b \rt] \xi_{ba}\rangle & =\frac{1}{h}\sum_a \sum_{b, b\neq a} \g j_{ab}  \langle\xi_{ab}, g_a^{-1}  \delta g_a g_a  \xi_{ab}\rangle \notag\\
 & = -\frac{\I}{2h}\sum_a \Lambda_{kj}(g_a) \delta g_a^i \sum_{b, b\neq a}\g  j_{ab} \hat n_{ab}^k = O(j^2/h^2),
\end{align}
where we used the fact that $\delta g_a = \delta g_a^k \tau^k = -\frac{\I}{2}\delta g_a^k \tau^k$, and the relation between spinors and face vectors $\hat n_{ab} = \langle \xi_{ab},\vec \sigma \xi_{ab}\rangle$. Also, we made use of the following relation $g^{-1}\sigma^k g = \Lambda(g)^k_{\phantom{k}j} \sigma^j$, which simply follows from the fact that the Pauli matrices are a basis of the complex vector space of $2\times2$ complex matrices of zero trace.

Therefore, we are left with a quite compact expression for the leading order contribution to the Chern-Simons invariant at the critical point:
\be
W[A] =  -\frac{1}{\I h}\lt(\I+\frac{1}{\g}\rt) \g j_{ab}  j_{24} \langle\xi_{24},\lt[g_2^{-1} \lt( g_3 F_{31} g_3^{-1} \rt)g_2 \rt] \xi_{24}\rangle + O(j^3/h^3).
\ee
Notice that this term is actually associated to the presence of a a crossing in the graph $\G_5$. By reinserting the explicit expression for $F_{24}$ one immediately obtains
\begin{align}
W[A] & =  -\frac{4\pi}{\I h^2}\lt(\I+\frac{1}{\g}\rt)^2 (\g  j_{24})(\g j_{31})\hat n_{31}^k \langle\xi_{24},G_{32}^{-1}\tau_k G_{32} \xi_{24}\rangle + O(j^3/h^3) \notag\\
& = \frac{2\pi}{ h^2}\lt(\I+\frac{1}{\g}\rt)^2 (\g  j_{24})(\g j_{31}) \Lambda(G_{32})_{kj} \hat n_{31}^k \hat n_{24}^j + O(j^3/h^3).
\end{align}
And thus the full Chern-Simons invariant relevant for the asymptotics reads
\begin{align}
\text{CS}[A] & = \frac{h}{2}W[A] + \frac{\bar h}{2} W[\bA] \notag\\
& =\frac{\pi}{ h}\lt(\I+\frac{1}{\g}\rt)^2 (\g  j_{24})(\g j_{31}) \Lambda(G_{32})_{kj} \hat n_{31}^k \hat n_{24}^j + c.c.
\end{align}

At this point, one can start drawing a connection with the geometry of a flat 4-simplex. In order to do this, consider a flat 4-simplex, whose sides $\{S_{\bar5\bar1},S_{\bar5\bar2},S_{\bar5\bar3},S_{\bar5\bar4}\}$ all start at vertex $\bar5$ and end at vertex $\{\bar1,\dots,\bar4\}$ respectively. The volume of the four simplex can be then calculated via 
\be
4! \;V_4 = \det (S_{\bar5\bar1},S_{\bar5\bar2},S_{\bar5\bar3},S_{\bar5\bar4}),
\ee
where a certain topological orientation of the 4-simplex has been assumed. This formula can be equally well expressed in terms of the bivectors $\star B_{31}(\bar5) := S_{\bar5\bar2}\wedge S_{\bar5\bar4}$ and $\star B_{24}(\bar5):=S_{\bar5\bar3}\wedge S_{\bar5\bar1}$:
\be
4!\; V_4 = \frac{1}{4}\epsilon_{IJKL}  (\star B_{31})^{IJ} (\star B_{24})^{KL} = \frac{1}{2} \prec \star B_{31}, \star B_{24}  \succ
\ee
where in the last expression we identified the bivector $\star B_{ab}$ with the corresponding $\slc$ element, that is $\frac{1}{2}(\star B_{ab})^{IJ} \mathcal J_{IJ} $. Notice that in this formula it is crucial that all the bivectors are defined at the same point. Though it is not relevant whether such a basepoint is vertex $\bar 5$ or something else. It is now immediate to write the volume as the sum of two piece, each associated to wither the selfdual or anti-selfdual parts of $\slc$:
\be
(2\times 4!) V_4 = \I (\star B_{31}^+)^k (\star B_{24}^+)_k - \I (\star B_{31}^-)^k (\star B_{24}^-)_k
\ee
Now the Lie algebra element $\star B_{ab}(a)$ in the frame of tetrahedron $a$ is given by\footnote{The factor of 2 is due to the fact that $\g j$ is the area of a triangle, and not that of the parallelogram defined by $S_{\bar5\bar2}$ and $S_{\bar5\bar4}$.}
\begin{align}
\star B_{ab}(a) &= 2\gamma j_{ab}(\star u\wedge n_{ab})^{IJ} \mathcal J_{IJ} =2\gamma j_{ab} \hat n_{ab} \cdot \vec J  \notag\\
&=  2\gamma j_{ab} \hat n_{ab}\cdot \vec T_+ +  2\gamma j_{ab} \hat n_{ab}\cdot \vec T_-.
\end{align}
It is not hard to realize that when parallel transported by the holonomy $G_{ca}$, this expression becomes\footnote{Notice that being $\star B_{ab}$ real, its selfdual and anti-selfdual components are related by complex conjugation.}
\begin{align}
\star B_{ab}(c) & =  2\gamma j_{ab} \hat n_{ab} \cdot \lt[ G_{ca} \vec T_+  G_{ca}^{-1}\rt] +  2\gamma j_{ab} \hat n_{ab} \cdot \lt[ G_{ca} \vec T_-  G_{ca}^{-1}\rt]\notag\\
& = 2\gamma j_{ab} \Lambda(G_{ca})_{kj}\hat n_{ab}^k \vec T_+^j   +  2\gamma j_{ab} \overline{\Lambda}(G_{ca})_{kj}\hat n_{ab}^k \vec T_-^j\,.
\end{align}

Going back to the Chern-Simons invariant, the previous results tell us
\begin{align}
\text{CS}[A] & = \frac{\Lambda}{48} \lt(\frac{1}{\g} + \I \rt) (\star B_{31}^+(3))^k(\star B_{24}^+(3))_k +\frac{\Lambda}{48} \lt(\frac{1}{\g} - \I \rt) (\star B_{31}^-(3))^k(\star B_{24}^-(3))_k  + O(j^3/h^3)\notag\\
& =  \frac{\Lambda}{48}\lt[ \I (\star B_{31}^+(3))^k(\star B_{24}^+(3))_k - \I(\star B_{31}^-(3))^k(\star B_{24}^-(3))_k \rt] + \notag\\
& \phantom{=} + \frac{\Lambda }{48\g}\lt[ (\star B_{31}^+(3))^k(\star B_{24}^+(3))_k +(\star B_{31}^-(3))^k(\star B_{24}^-(3))_k \rt] + O(j^3/h^3)\notag\\
& =  \Lambda V_4 + \frac{\Lambda}{48\g} < \star B_{31}(3), \star B_{24}(3)>+ O(j^3/h^3).
\end{align}
Therefore, we obtain the cosmological term plus an extra term which is orientation-reversal invariant. Such a term mirrors the ``twisted volume'' term present in the continuous action of eq. \eqref{HlBF}. 

{
At this point, one might wonder where the sign of the cosmological constant has been fixed in this derivation, where the reconstructed 4-simplex is flat. The point is that by changing the sign of the cosmological constant, one changes the sign of the unit vectors $\hat n_{ab}$ appearing in the holonomies. As a consequence, one is forced to change at the same time the orientation of  the reconstructed 4-simplex, in order for the (curved) closure equations to have a meaning. This change in orientations, flips the sign in the formula relating the 4-volume to the bivectors. Note that to understand this change in sign we need to appeal to the \textit{curved} equations of motion. Indeed, these are the only equations that can be sensitive to the sign of the curvature, since in the flat approximation this information is completely lost.
}

The $j$-independent integration constant $C_{\mathrm{int}}$ in eq. \eqref{eq_CSCint} does not show up at the leading order in the perturbative computation. It cannot appear at higher orders since it is $j$-independent. Therefore $C_{\mathrm{int}}$ is vanishing up to some topological information, e.g. the framing of graph or gauges. So, $C_{\mathrm{int}}$ is a parity invariant contribution (up to an integer multiple of $2\pi  \ell_P^2$).

To conclude, we observe that the volume (as well as the ``twisted volume'') terms can be seen as coming from the crossing in the $\G_5$ graph, and that this result is consistent with the earlier argument of \cite{Han2011}, which used the Vassiliev-Kontsevich invariants.

\bibliographystyle{JHEP}
\bibliography{00biblio_cosmoconstant}

\end{document}